\documentclass[pdflatex,sn-nature]{sn-jnl}

\usepackage{graphicx}%
\usepackage{multirow}%
\usepackage{amsmath,amssymb,amsfonts}%
\usepackage{amsthm}%
\usepackage{mathrsfs}%
\usepackage[title]{appendix}%
\usepackage{textcomp}%
\usepackage{manyfoot}%
\usepackage{booktabs}%
\usepackage{algorithm}%
\usepackage{algorithmicx}%
\usepackage{algpseudocode}%
\usepackage{listings}%

\usepackage[dvipsnames]{xcolor}
\usepackage[table]{xcolor}
\usepackage{tikz}
\usetikzlibrary{tikzmark}
\usetikzlibrary{trees}
\usetikzlibrary{positioning}
\usetikzlibrary{calc}
\usetikzlibrary{fit}
\usetikzlibrary{shapes}
\usetikzlibrary{external}
\usepackage{longtable}
\usepackage{longfigure}
\usepackage{adjustbox}
\usepackage{fullpage}
\usepackage{makecell}
\usepackage{siunitx}
\usepackage{diagbox}
\usepackage[
singlelinecheck=false, justification=raggedright, tableposition=top
]{caption}
\usepackage{multibib}
\newcites{methods}{Supplementary References}

\usepackage{caption}
\DeclareCaptionFont{bold}{\bfseries} 
\usepackage{svg}
\usepackage{bm}
\usepackage{hyperref}

\usepackage[lighttt]{lmodern}

\usepackage{fullpage}
\usepackage[T1]{fontenc}
\usepackage{pgfplots}
\pgfplotsset{compat=1.17}
\usepgfplotslibrary{groupplots}
\usepackage{subcaption}

\usepackage[commentmarkup=todo]{changes}

\makeatletter
\renewcommand\@biblabel[1]{#1.}
\makeatother
\setcitestyle{open={},close={},numbers,super}

\makeatletter
\newcommand*\myfontsize{%
  \@setfontsize\myfontsize{6.7}{8}%
}
\makeatother

\definecolor{myblue}{rgb}{0.2, 0.3, 0.6}
\definecolor{mygreen}{rgb}{0.2, 0.6, 0.3}
\definecolor{myred}{rgb}{0.6, 0.2, 0.3}

\begin{document}

\title[Article Title]{ChemFM as a Scaling Law Guided Foundation Model \\ Pre-trained on Informative Chemicals}


\author[1]{\fnm{Feiyang} \sur{Cai}}
\author [2]{\fnm{Katelin} \sur{Zacour}}
\author[3]{\fnm{Tianyu} \sur{Zhu}}
\author[2]{\fnm{Tzuen-Rong} \sur{Tzeng}}
\author[4]{\fnm{Yongping} \sur{Duan}}
\author[5]{\fnm{Ling} \sur{Liu}}
\author[6]{\fnm{Srikanth} \sur{Pilla}}
\author[7]{\fnm{Gang} \sur{Li}}
\author*[1]{\fnm{Feng} \sur{Luo}}\email{luofeng@clemson.edu}
\affil[1]{\orgdiv{School of Computing}, \orgname{Clemson University}, \orgaddress{\city{Clemson}, \postcode{29634}, \state{SC}, \country{USA}}}
\affil[2]{\orgdiv{Department of Biological Sciences}, \orgname{Clemson University}, \orgaddress{\city{Clemson}, \postcode{29634}, \state{SC}, \country{USA}}}
\affil[3]{\orgdiv{Department of Materials Science and Engineering}, \orgname{Clemson University}, \orgaddress{\city{Clemson}, \postcode{29634}, \state{SC}, \country{USA}}}
\affil[4]{\orgdiv{Horticultural Research Laboratory}, \orgname{USDA}, \orgaddress{\city{Fort Pierce}, \postcode{34945}, \state{FL}, \country{USA}}}
\affil[5]{\orgdiv{College of Computing}, \orgname{Georgia Institute of Technology}, \orgaddress{\city{Atlanta}, \postcode{30332}, \state{GA}, \country{USA}}}
\affil[6]{\orgdiv{Center for Composite Materials}, \orgname{University of Delaware}, \orgaddress{\city{Newark}, \postcode{19716}, \state{DE}, \country{USA}}}
\affil[7]{\orgdiv{Department of Mechanical Engineering}, \orgname{Clemson University}, \orgaddress{\city{Clemson}, \postcode{29634}, \state{SC}, \country{USA}}}
\abstract{

%
Traditional AI methods often rely on task-specific model designs and training, 
which constrain both the scalability of model size and generalization across different tasks.
%
%
Here, we introduce ChemFM, a large foundation model specifically developed for chemicals.  
%
By conducting a series of scaling experiments, we identify UniChem as the informative molecular database for pre-training the foundation model.
%
ChemFM comprises 3 billion parameters and is pre-trained on 178 million molecules using self-supervised causal language modeling
to extract generalizable molecular representations.
%
This model can be adapted to diverse downstream chemical applications using either full-parameter or parameter-efficient fine-tuning methods.
ChemFM consistently outperforms state-of-the-art task-specific AI models across all tested tasks. 
%
%
%
Notably, it achieves up to 67.48\% performance improvement across 34 property prediction benchmarks, 
up to 33.80\% reduction in mean average deviation between conditioned and actual properties of generated molecules in 
conditional molecular generation tasks, 
and up to 3.7\% top-1 accuracy improvement across 4 reaction prediction datasets. 
Moreover, ChemFM demonstrates its superior performance in predicting antibiotic activity and cytotoxicity, highlighting its potential to advance the discovery of novel antibiotics.
Furthermore, we demonstrate that, as a foundation model, ChemFM exhibits strong data efficiency, requiring significantly fewer labeled training samples to achieve state-of-the-art performance.
%
We anticipate that ChemFM will significantly advance chemistry research by providing a foundation model capable of effectively generalizing across a broad range of tasks with minimal additional training. 
}

\maketitle
\section{Introduction}\label{sec:main}

Over the past decade, 
artificial intelligence 
has revolutionized research methodologies across scientific disciplines~\cite{ai4science1, ai4science2}, including chemistry.
The prevailing AI-based paradigm in computational chemistry focuses on developing models for specific tasks.
For example, deep learning models are often trained using pre-calculated molecular descriptors or fingerprints~\cite{ecfp}, molecular graph representations~\cite{graph_conv}, or serialization format representations~\cite{smiles}. 
These models excel at tasks such as predicting molecular properties\cite{MMNB, dmpnn}, designing and optimizing molecules\cite{molgpt, molecular_design}, and forecasting chemical synthesis and retro-synthesis\cite{chemformer, rsmiles}.
Despite their advancements, these task-specific models have limitations.
First, training a high-performing task-specific model often requires large amounts of high-quality data.
Annotating chemical data is typically costly and time-consuming, and may involve extensive laboratory experiments.
Second, the models struggle to capture general patterns that reflect the inherent structural dependencies and contextual relationships within molecules, leading to over-fitting and poor generalization to novel molecular features.
Furthermore, the diversity of chemical tasks and datasets makes it impractical to annotate comprehensive chemical datasets and train large-scale models for every individual application. However, as suggested by the success in other domains such as computer vision~\cite{dino} and natural language processing~\cite{gpt3,llama3}, scaling model sizes could unlock new possibilities in chemical AI models.
%

One promising approach to addressing these challenges is the development of foundation models. 
These models are pre-trained on large unannotated datasets, often using weakly supervised or unsupervised methods to extract complex, general-domain features, enabling them to be fine-tuned for various downstream tasks with minimal additional training. 
%
Initially pioneered in language and image modalities, foundation models have demonstrated substantial performance improvements in other scientific domains, such as retinal imaging\cite{retinalFM}, single-cell transcriptomics\cite{RNAFM}, and histopathology imaging~\cite{pathologyFM}.
Existing efforts to construct large chemical models can be broadly categorized into two directions.
The first focuses on pre-training models exclusively on chemical data.
Early works, such as developing unified pre-trained models~\cite{molebert,smilestransformer,chemberta2,chemformer}, have been limited by the scale of both model architectures and pre-training datasets (Table~\ref{table:pre-trained_model_comparison}). 
Most of these studies directly used public molecular databases~\cite{zinc20, pubchem}, without additional preprocessing or quality assessment.
As a result, the datasets often contained redundant or noisy molecules and lacked systematic evaluation of chemical diversity, hindering meaningful scaling analysis and limiting generalization across tasks.
Moreover, no prior work has systematically analyzed how large a chemical foundation model should be or how its performance scales with model size, except for a preliminary study by Frey et al.\cite{mit_pretrain} that explored scaling laws during pre-training. 
However, it did not demonstrate whether scaling leads to improved performance on downstream tasks.
%
%
The second reframes chemical tasks as natural language problems, fine-tuning large language models to augment chemical knowledge~\cite{chemllm,chemdfm}. 
Although this approach benefits from the capabilities of large pre-trained language models, it lacks fundamental alignment between linguistic tokens and molecular representations. 
As a result, such models struggle even with basic molecule recognition and manipulation tasks\cite{mollangbench}, and therefore cannot be expected to outperform chemical-specific models, as has been empirically observed.

In this work, we introduced ChemFM, 
a $3$-billion-parameter foundation model 
designed for chemicals that can be fine-tuned for various chemical design and property prediction tasks. 
Our study primarily focuses on two key aspects often overlooked in prior works: (1) analyzing the scaling behavior of chemical foundation models across different model and dataset sizes, and (2) preparing and evaluating large-scale chemical datasets to understand how data diversity and quality affect model performance.
%
By conducting a series of scaling experiments, we demonstrated that UniChem~\cite{unichem}, while being relatively smaller with 178 million molecules, is more diverse and information richer than the 
 much larger ZINC20 database~\cite{zinc20} of 1.8 billion molecules.
 Therefore, we trained ChemFM on SMILES strings~\cite{smiles} from $178$ million molecules in UniChem database~\cite{unichem} (Fig.~\ref{fig:pretrain_pipeline}\textbf{a} and~\ref{fig:pretrain_pipeline}\textbf{b}). 
By leveraging the paradigm of casual language modeling~\cite{gpt1},  
ChemFM effectively learned SMILES syntax as well as the molecular internal relationships between atoms and bonds, 
enabling its adaptation for various downstream tasks (Fig.~\ref{fig:finetune_pipeline}).
We first validated ChemFM on $34$ property prediction datasets from domains including pharmaceutical, physicochemical, and bioactivity, 
showing consistent outperformance over existing approaches across all datasets.
%
%
%
Moreover, ChemFM demonstrated superior performance for potential antibiotic screening, highlighting its potential to advance real-world drug discovery.
ChemFM also exhibited flexibility and versatility in conditional molecular generation tasks.
Unlike previous approaches that required training separate models for each condition or condition combination, 
ChemFM allowed the training of a single unified model capable of handling all variations of condition combinations. 
The unified model not only achieved strong generative performance but also enabled effective control and matching of flexible desired conditions.
Furthermore, we demonstrated that ChemFM can be seamlessly integrated with existing sequence editing-based methods for reaction prediction~\cite{rsmiles}, resulting in state-of-the-art performance on $4$ reaction prediction tasks, including both forward synthesis and retro-synthesis.
Furthermore,
ChemFM also exhibited remarkable training and data efficiency.
By leveraging parameter-efficient fine-tuning~\cite{lora}, ChemFM can be fine-tuned within a single moderate GPU machine,
making it broadly accessible for research applications.
ChemFM achieved state-of-the-art results while using significantly less data than existing task-specific methods.
%
ChemFM can be leveraged for diverse chemical research endeavors and may significantly advance chemistry research.

\section{Results}\label{sec:results}

\subsection{Pre-training dataset selection}
Traditionally, ZINC20~\cite{zinc20} and UniChem~\cite{unichem}, containing $1.8$ billion and $178$ million molecules, respectively, are suitable candidates for pre-training ChemFM due to their extensive molecular coverage.
We first conducted a series of scaling experiments to access the informativeness of both datasets (\hyperref[sec:pre-training_data_selection]{Methods} and Supplementary Fig.~\ref{fig:scaling_law}). 
The scaling laws of neural models
reveals that model performance follows a power-law relationship between model size or dataset size, 
provided the other is not bottlenecked~\cite{scalinglaw}.
Our scaling experiments evaluated models ranging from approximately 10 million to 200 million parameters.
The results showed that, for UniChem, the model performance (measured by validation loss) closely followed a power-law scaling trend with model size, 
with no signs of performance saturation.
%
In contrast, the ZINC20 dataset exhibited performance saturation when the model size reached $60$ million parameters.
This suggests that the knowledge contained within ZINC20 becomes a bottleneck, limiting further gains from increasing model size.
This may be due to the fact that ZINC20 is primarily designed for ligand discovery, enriching commercially available compounds with mainstream medicinal chemistry scaffolds. 
Many molecules share core structures with minor variations, limiting structural diversity and reducing the dataset’s informativeness for large-scale molecular representation learning.
Based on these findings, we selected UniChem as our pre-training dataset, as it offers a broader chemical space and greater structural diversity.
{Additionally, we experimentally demonstrated that the model pre-trained on the UniChem dataset outperforms the one pre-trained on ZINC20 in downstream tasks (\hyperref[sec:comparison_pretraining]{Methods}; Supplementary Table~\ref{table:comparison_pretraining}.)}

\subsection{Training of ChemFM}
%
%
By leveraging self-supervised causal language modeling, we developed two model variants: ChemFM-1B and ChemFM-3B, comprising approximately $970$ million and $3.0$ billion trainable parameters, respectively.
Both models were trained for one epoch on $1.78$ billion SMILES strings, augmented from $178$ million molecules from UniChem dataset (\hyperref[sec:pre-traing]{Methods}).
%
%
Throughout pre-training, the validation perplexity for both ChemFM-1B and ChemFM-3B steadily decreased 
(Fig.~\ref{fig:pretrain_pipeline}\textbf{c}), 
showing no signs of saturation until processing $818$ billion tokens.
%
ChemFM-3B achieved a lower final validation perplexity compared to ChemFM-1B.
%
Moreover, the final validation losses of both models start to deviate from the predicted scaling law (Supplementary Fig.~\ref{fig:scaling_law}), suggesting that under the current data regime, further model scaling may lead to diminishing returns as the loss approaches a plateau.

\subsection{Unconditional molecule generation using pre-trained ChemFM}

We evaluated ChemFM-3B in unconditional molecule generation by randomly generating $\num{100000}$ molecules and benchmarking validity, uniqueness, novelty, internal diversity, and distribution similarity with the training dataset (Fig.~\ref{fig:pretrain_pipeline}\textbf{d}).
%
ChemFM-3B achieved a remarkable validity score of $0.996$ without additional constraints during tokenization, model training, or generation. 
The uniqueness score was perfect ($1.0$), indicating no duplicate canonical SMILES strings among generated molecules. 
High internal diversity scores (IntDiv$_1$ of $0.904$ and IntDiv$_2$ of $0.896$) demonstrated the diversity of molecular structures of generations.
%
By comparing various physicochemical descriptors---such as molecular complexity, weight, and structural characteristics---and \texttt{ECFP} fingerprints between the training and generated molecules, we observed that ChemFM faithfully captured the distribution of molecules in the training data without overfitting to a narrow subset
(\hyperref[sec:methods_unconditional_benchmark]{Methods} and Supplementary Fig.~\ref{fig:unconditional_generation_desc_dist} and~\ref{fig:tsne_embedding}). 
More importantly, over half ($55.8\%$) of the generated molecules were entirely novel, not found in the extensive training dataset, highlighting the potential of these models for exploring chemical space, discovering new molecules, and optimizing molecular structures.

\subsection{Molecular property prediction}
We evaluated the adaptability of ChemFM for molecular property prediction using two widely-used benchmarks: MoleculeNet \cite{wu2018moleculenet} and ADMET \cite{admet}, covering a total of $34$ datasets across diverse domains, including pharmaceutical, physicochemical, and bioactivity applications. 
Across all evaluated datasets from both MoleculeNet and ADMET benchmarks, ChemFM models consistently outperformed existing state-of-the-art methods.

The MoleculeNet benchmark consists of $4$ regression datasets ($4$ properties in total) 
and $8$ classification datasets ($189$ properties in total) (Supplementary Table~\ref{table:dataset_molecule_net}). 
Comparisons with the methods in the literature for MoleculeNet datasets are often challenging due to varying dataset splitting strategies and random seed choices.
To ensure comprehensive evaluation,
we compared ChemFM models with different sets of methods using the same splitting methods and random seeds. 
%
We first compared the fine-tuned ChemFM-3B models on the standard MoleculeNet datasets 
against {SMILES Transformer~\cite{smilestransformer}}, MoleculeNet models~\cite{moleculenet}, directed message passing neural networks (D-MPNN or Chemprop)~\cite{dmpnn}, MolMapNet OOTB (MMNB)~\cite{MMNB}, and {Chemformer\cite{chemformer}} (Fig.~\ref{fig:results_prop_pred}; full comparison results are provided in Supplementary Table~\ref{table:rst_prop_prediction}). 
For classification tasks, ChemFM-3B demonstrated a consistent performance advantage, with improvements in the area under the receiver operating curve (ROC-AUC) of $0.012$ on BBBP, $0.034$ on BACE, $0.030$ on HIV, $0.018$ on Tox21, $0.029$ on SIDER, and $0.030$ on ClinTox. 
Additionally, ChemFM-3B showed improvements of $0.026$ and $0.010$ on the MUV and PCBA datasets, respectively, in the area under the precision-recall curve (PRC-AUC). 
For regression tasks, ChemFM-3B reduced root mean squared errors (RMSE) by $0.039$ on ESOL, $0.245$ on FreeSolv, $0.010$ on Lipophilicity, and $0.024$ on PDBbind.

We also compared ChemFM against methods that use different dataset splits, including {Pretrain GNNs\cite{pretraingnns}, ChemBERTa-2\cite{chemberta2}}, AttentiveFP\cite{attentiveFP}, 3D InfoMax\cite{3dinfomax}, Mole-BERT\cite{molebert}, GraphMVP\cite{graphmvp}, and MoleculeSDE\cite{moleculeSDE}
(\hyperref[sec:methods_moleculenet]{Methods} and Supplementary Table~\ref{table:rst_prop_prediction_attentive_fp} and~\ref{table:rst_prop_prediction_molebert} ). 
Across all comparison settings, ChemFM consistently delivered better results than the other methods did.
Additionally, we observed that ChemFM-3B generally outperformed ChemFM-1B (Fig.~\ref{fig:results_prop_pred}) {and its non-pre-trained counterpart} (\hyperref[sec:methods_effect_pre-training]{Methods} and Supplementary Table~\ref{table:without_pretraining}), underscoring the benefits of larger model sizes {and pre-training}.
%

%

On the ADMET benchmark, which includes $13$ classification and $9$ regression datasets (each representing a single property; Supplementary Table~\ref{table:dataset_admet}),
ChemFM again achieved superior performances across all datasets (quantitative results in Supplementary Table~\ref{table:rst_prop_prediction_admet}), with an average improvement of approximately $7.09\%$. 
The improvements ranged from a minimum of $0.11\%$ on the DILI dataset to a maximum of $67.48\%$ on the Half\_Life\_Obach dataset.

{
For a wider application, we also evaluated ChemFM on two additional datasets beyond the MoleculeNet and ADMET benchmarks: an odor prediction dataset~\cite{odor} and a chromatographic retention time prediction dataset~\cite{smrt}. 
Across both tasks, ChemFM-3B outperformed the specialized baselines (\hyperref[sec:addtional_property_prediction]{Methods} and Supplementary Table~\ref{table:additional_property_prediction}), with particularly large improvements ($31.6\%$ reduction in mean absolute error) on retention time prediction, 
demonstrating that our approach generalizes effectively to broader chemical property prediction tasks.
}

\subsection{Potential discovery of novel antibiotics}
Recent study~\cite{antibiotics} has leveraged multiple Chemprop~\cite{dmpnn} models to predict antibiotic activity and human cell cytotoxicity, successfully screening over $10$ million molecules to identify novel antibiotic candidates with high antibiotic activity and low cytotoxicity. 
Here, we fine-tuned ChemFM model for the same tasks of predicting antibiotic activity and cytotoxicity across different cell types (Supplementary Fig.~\ref{fig:precision-recall-antibiotics}). 
Specifically, ChemFM has significantly improved the performance with the PRC-AUC values increasing from $0.364$ to $0.428$ for antibiotic activity, $0.176$ to $0.461$ for cytotoxicity in human liver carcinoma cells (HepG2), $0.168$ to $0.459$ for human primary skeletal muscle cells (HSkMC), and $0.335$ to $0.414$ for cytotoxicity in human lung fibroblasts cells (IMR-90).
%

To further evaluate ChemFM's predictive capabilities, we applied both ChemFM and Chemprop to an antibiotic library of $\num{1173}$ molecules~\cite{medchemexpress_antibiotic}, 
which consists of real antibiotics but differs significantly from the positive samples in the training dataset. 
ChemFM labeled $\num{149}$ molecules as positives, whereas Chemprop labeled only $\num{29}$ (Supplementary Data 1), 
suggesting that ChemFM has a higher true positive rate in discovering antibiotics.

These results highlight ChemFM's potential to significantly improve the screening process for novel antibiotics by providing more accurate predictions for both antibiotic activity and toxicity.

\subsection{Conditional molecule generation}


Conditional molecule generation is critical for designing molecules to meet specific property criteria or incorporate particular scaffold structures. 
We fine-tuned two separate ChemFM-3B models: 
one on the GuacaMol~\cite{guacamol} dataset for property-based generation and another on the MOSES~\cite{moses} dataset for scaffold and property-based generation. 
%
For each dataset, we considered four continuous properties: octanol-water partition coefficient (logP), synthetic accessibility score (SAS), topological polar surface area (TPSA), and quantitative estimate of drug-likeness (QED). 
{Traditional methods such as cRNN~\cite{crnn} and MolGPT~\cite{molgpt} require a separate model for each property combination, resulting in $15$ models to cover all four conditions for each dataset. 
In contrast, by carefully designing the input condition to the model
(\hyperref[sec:conditonal_generation_technique]{Methods})
ChemFM can handle all combinations within a single unified model.}


We first evaluated the property-based generation model trained on the GuacaMol dataset. 
For each property combination, we generated $\num{10000}$ molecules at different sample points and evaluated their validity, uniqueness, novelty, and mean absolute deviation (MAD) between the conditioned and computed properties (Table~\ref{table:rst_condition_generation} and Supplementary Fig.~\ref{fig:rst_mol_gen_prop_dist_1d}). 
ChemFM outperformed MolGPT in validity, uniqueness, and novelty across all conditioned properties, whether for individual properties or multiple combined conditions (Table~\ref{table:rst_condition_generation}; {comparison with cRNN~\cite{crnn} is given in Supplementary Table~\ref{table:result_conditional_generation_full}}). 

On average, ChemFM achieved improvements of $0.0079$ in validity, $0.0104$ in uniqueness, and $0.0151$ in novelty over MolGPT.
Furthermore, ChemFM demonstrated stronger adherence to the desired property values, with an average percentage reduction in MAD across all four properties of $21.19\%$, ranging from $7.70\%$ for SAS to $33.80\%$ for TPSA.

%

Next, we evaluated conditional generation based on both scaffold and property on the MOSES dataset. 
%
Using the same $5$ test scaffolds as MolGPT, 
we generated $\num{10000}$ molecules at each sample point across different scaffold and property combinations (Supplementary Table~\ref{table:rst_conditional_molecular_generation} and Supplementary Fig.~\ref{fig:rst_mol_gen_scaffold_prop_dist_1d}).
ChemFM consistently outperformed MolGPT across all scaffold and property combinations,
generating more valid, unique, and novel molecules, 
with average improvements of 
$1.93\%$, $26.69\%$, and $26.69\%$, respectively.
%
Moreover, ChemFM showed a stronger alignment with the desired conditions by: 1) generating more molecules that shared the same scaffold as the conditioned scaffold, with an average improvement of $25.73\%$ over MolGPT, and 2) achieving an average reduction in MAD across all four properties, with reductions of $15.31\%$, $9.63\%$, $13.35\%$, and $1.96\%$ for logP, SAS, TPSA, and QED, respectively.

\subsection{Reaction prediction}

We fine-tuned ChemFM-3B model for both reaction synthesis and retro-synthesis tasks using three USPTO benchmark datasets: USPTO-Full\cite{uspto_full}, USPTO-MIT\cite{usptomit}, and USPTO-50K\cite{uspto50k}. 
These datasets, comprising organic chemical reactions extracted from US patents and applications, are widely used for evaluating reaction prediction tasks (Supplementary Table~\ref{table:dataset_uspto}).
%
%
We compared ChemFM with existing methods in the literature, employing the same data splitting methods for training and evaluation~\cite{at,rsmiles}. Table~\ref{table:rst_rct} presents a comparison between ChemFM and previous best and second-best performing models, while complete results comparing ChemFM with other methods are available in Supplementary Table~\ref{table:rst_rct_prediction_detail}.  
%

For the retro-synthesis task, 
%
ChemFM consistently achieved higher top-$1$, top-$3$, and top-$5$ accuracies compared to previous best methods. 
Our experiments also highlight the training efficiency of the ChemFM foundation model. 
%
%
For instance, on the USPTO-50K dataset, we achieved state-of-the-art results after just one epoch ($\num{25000}$ steps) of training on the augmented dataset (equivalent to five epochs due to five-fold augmentation),
already surpassing the performance of R-SMILES~\cite{rsmiles}, which used approximately ten times the number of training steps. 
With additional training, the top-$1$ accuracy could be further improved (while top-$5$ accuracy may decrease).
Moreover, for USPTO-50K and USPTO-MIT, top-$1$ accuracies further reach $59.7\%$ and $62.4\%$, respectively. 
The top-$1$ accuracy improvements over the previous best methods were $3.7\%$ for USPTO-50K, $2.1\%$ for USPTO-MIT, and $2.3\%$ for USPTO-Full.


%

For the reaction synthesis task, we focused on the more challenging setting where reactants and reagents are mixed, evaluating ChemFM on the USPTO-MIT dataset. 
ChemFM demonstrated competitive performance, surpassing the previous best method (AT\cite{at}) by $0.1\%$ on both top-1 and top-5 accuracies. 
\subsection{Training and data efficiency}

ChemFM supported 
parameter-efficient fine-tuning methods, such as the low-rank adaptation (LoRA) technique\cite{lora}, which significantly reduces the number of trainable parameters and GPU memory requirements (\hyperref[sec:methods_lora]{Methods}). 
For example, with a LoRA rank of 4 in ChemFM-3B (using 32-bit float precision), the number of trainable parameters is reduced by $460\times$, from 3 billion to 6.5 million. 
This reduction lowers the GPU memory required during training from $\qty{51}{\giga\byte}$ to $\qty{20}{\giga\byte}$, making fine-tuning feasible on a single moderate GPU machine. 
Additionally, the checkpoint size is reduced from $\qty{12}{\giga\byte}$ to $\qty{26}{\mega\byte}$, allowing for minimal storage for adapters on each dataset.


{
Furthermore, ChemFM also demonstrates strong data efficiency. We evaluated its performance on a classification task (CYP2D6\_Substrate\_CarbonMangels) and a regression task (Half\_Life\_Obach). In the classification task, ChemFM outperformed previous state-of-the-art methods even when fine-tuned on just $50\%$ of the training data. Similarly, in the regression task, fine-tuning with only $10\%$ of the data was sufficient to achieve state-of-the-art results on the test set (\hyperref[sec:methods_data_efficient]{Methods} and Supplementary Fig.~\ref{fig:data_efficiency}).
This suggests that ChemFM effectively learns transferable molecular representations, reducing the need for large labeled datasets and making it well-suited for small-data scenarios. This is an important advantage in chemistry research, where collecting and measuring data can be costly and labor-intensive.
}

\section{Discussion}\label{sec:discussion}
The tasks in computational chemistry are complex and diverse,
and training specific models for each task is both resource-intensive and time-consuming. 
In this work, we introduced ChemFM, a general-purpose foundation model specifically designed for chemicals. 
By leveraging the causal language modeling framework and extensive self-supervised training on $178$ million molecules, ChemFM has successfully learned the molecular structures represented by SMILES, as well as the contextual relationships of atoms and bonds within molecules.
%
%

%
ChemFM effectively characterized the structures of molecules, 
helping to establish structure-property relationships. 
%
%
Evaluated against 34 molecular property prediction datasets from the MoleculeNet and ADMET benchmarks, ChemFM achieved an average $6.98\%$ improvement over previous state-of-the-art methods. 
Moreover, in an antibiotic discovery application, ChemFM substantially outperformed models used in a prior study~\cite{antibiotics}.
In the conditional molecule generation, 
%
we showed that a single unified ChemFM model can generate molecules given an arbitrary combination of property conditions with high validity, uniqueness, and novelty while precisely matching desired properties or scaffold structures.
We further demonstrated ChemFM's ability to improve both accuracy and computational efficiency in predicting chemical reactions. 
ChemFM integrated seamlessly with SMILES sequence editing-based methods designed for reaction prediction, such as the root-aligned SMILES (R-SMILES) technique. 
Requiring fewer training steps, ChemFM consistently achieved higher prediction accuracy than existing models on synthesis and retro-synthesis tasks across USPTO benchmark datasets.
{Beyond the tasks studied in this work, we believe ChemFM can be effectively extended to broader downstream tasks, including molecular optimization~\cite{molopt,ttt}, which represent promising directions for future exploration.}

%

This work has a few limitations. 
ChemFM was fully trained on the most informative dataset available to our knowledge, 
which makes the distribution of generated molecules closely mirror the training dataset, limiting exploration of the potentially broader chemical space. 
While fine-tuning ChemFM is efficient with the LoRA technique, its inference time is not yet comparable to smaller-scale models, particularly when screening large amounts of data. 
Distilling smaller, cost-efficient models from ChemFM could improve evaluation efficiency.

In conclusion, ChemFM demonstrates its capability as a versatile chemical foundation model, which can efficiently be adapted to diverse tasks and improve upon state-of-the-art performance. 
The success of ChemFM in unifying various chemical tasks under a single model architecture highlights the capability of foundation models in computational chemistry, potentially significantly advancing drug discovery, molecule optimization, and chemical synthesis planning.

\clearpage
\section*{Figures}
\noindent	
	\begin{minipage}{1.0\textwidth}
	\tikzsetnextfilename{pretrain}
	\input{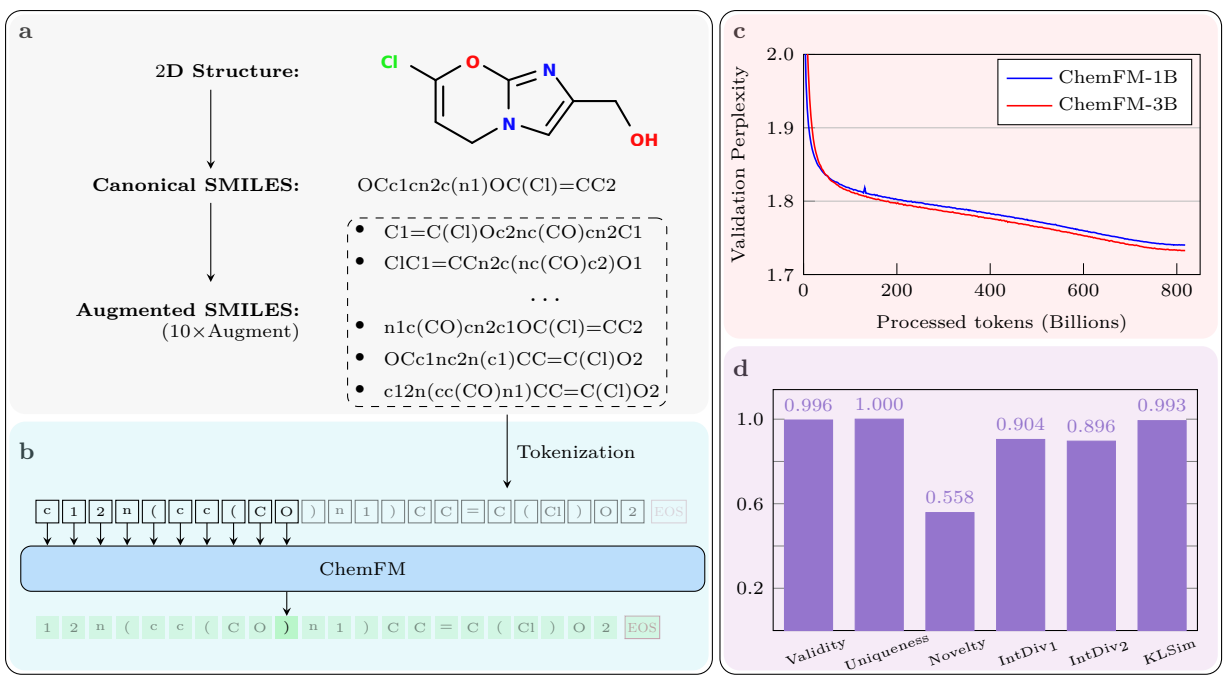}
	\captionof{figure}{\textbf{Pre-training and unconditional molecular generation benchmarking of ChemFM models.}
		\textbf{a},
		Pre-processing pipeline for ChemFM's pre-training dataset. 
		The pipeline starts with $178$ million molecules from the UniChem database, initially represented by International Chemical Identifier (InChI)~\cite{inchi}. These InChIs are converted into canonical SMILES strings using \texttt{RDKit}\cite{rdkit}. The SMILES strings are then augmented tenfold through the SMILES enumeration technique\cite{smilesaug}, resulting in approximately $1.78$ billion SMILES strings for use as the pre-training dataset.
		\textbf{b}, 
		Pre-training process for ChemFM. 
		SMILES strings are segmented, tokenized, and terminated with an end token. These tokens are fed into ChemFM, a causal decoder-only transformer. Pre-training uses self-supervised causal language modeling, where the task is to predict each token based on preceding tokens.
		\textbf{c}, Pre-training performance of ChemFM-1B and ChemFM-3B models, measured by perplexity (exponentiated average negative log-likelihood) on the validation set. Models are trained through $818$ billion tokens, slightly exceeding one epoch.
		\textbf{d}, 
		Unconditional generation benchmarking for ChemFM-3B. A total of $\num{100000}$ molecules are generated randomly using a temperature setting of $1.0$. 
		The validity, uniqueness, and novelty scores of the generated molecules are reported. Additionally, internal diversity metrics (IntDiv$_1$, IntDiv$_2$) assess the diversity of the generated molecules, while KL similarity (KLSim) evaluates how closely the distribution of generated molecules aligns with that of the training dataset.
	}
	\label{fig:pretrain_pipeline}
\end{minipage}

\noindent	
\begin{minipage}{1.0\textwidth}
	\tikzsetnextfilename{finetune}
		\begin{tikzpicture}
		
		
		\definecolor{color_chemFM}{HTML}{BBDEFB}
		\definecolor{color_activate_token}{HTML}{B9F6CA}
		\definecolor{color_tokenize_example}{HTML}{F44336}
		\definecolor{color_property_prediction_bg}{HTML}{B2EBF2}
		\definecolor{color_molecular_generation_bg}{HTML}{E1BEE7}
		\definecolor{color_reaction_bg}{HTML}{FFCDD2}

		\input{figures/property_prediction}
		\input{figures/reaction_prediction}
		\input{figures/conditional_generation}
		\draw [rounded corners=6] (0,0.25) -- (0,8.05) -- (16,8.05) 
		-- (16, 0.25) -- cycle;

	\end{tikzpicture}
	
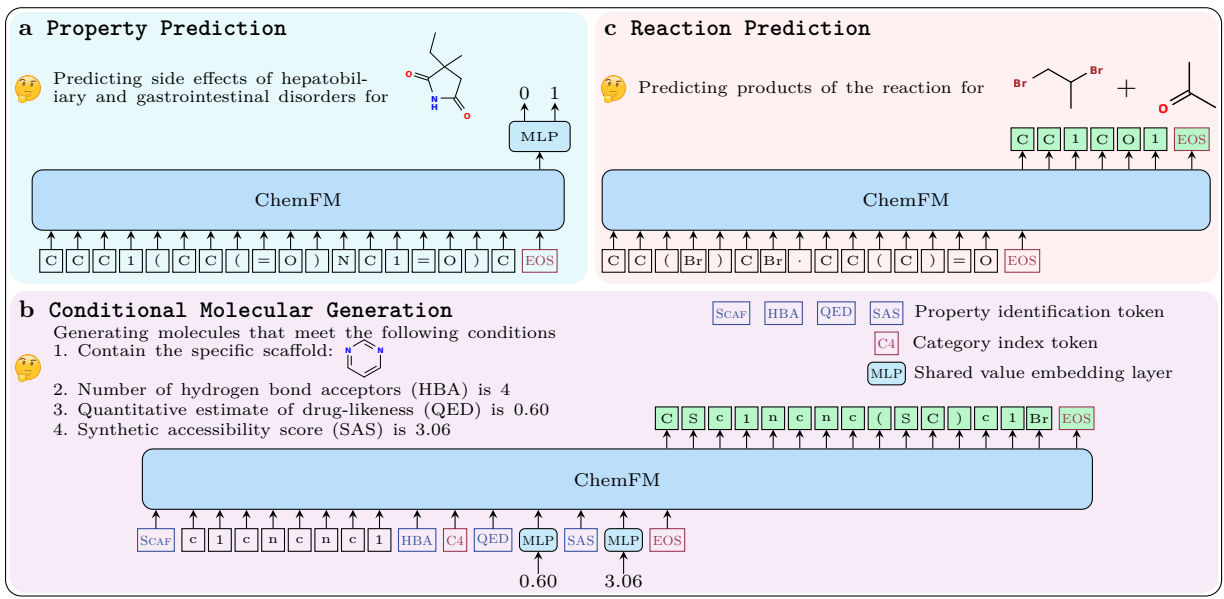
\captionof{figure}{\textbf{Illustrations of fine-tuning ChemFM model for downstream tasks.}
		\textbf{a}, 
		Property prediction fine-tuning. 
		During fine-tuning, the SMILES strings of molecules are augmented with a probability of $1.0$ and tokenized before input to ChemFM.
		An MLP layer is added to the final token's hidden state in the final layer to handle single or multiple regression or classification tasks.
		For inference, the canonical SMILES is input into ChemFM to predict the desired properties.
		\textbf{b},
		Conditional molecular generation fine-tuning. 
		This task is also framed as a sequence-to-sequence problem. 
		The input comprises a sequence of conditions, each initiated by a unique property identification token followed by single or multiple tokens representing the property values.
		Classification values are encoded as special tokens corresponding to their class indices, 
		continuous values are normalized and mapped into the embedding space using a shared MLP,
		and scaffolds are represented by their SMILES and tokenized into sequences.
		During fine-tuning, the target molecules 
		are augmented with a probability of $1.0$.
		\textbf{c},
		Reaction prediction fine-tuning for both forward synthesis and retro-synthesis.
		These tasks are approached as sequence-to-sequence problems, where the model predicts the product (or reactant) sequence based on the reactant (or product) sequence. 
		The root-aligned SMILES technique~\cite{rsmiles} is employed, aligning both sequences using the same root atom and augmenting them by enumerating different atoms as roots.
		%
	}
	\label{fig:finetune_pipeline}
\end{minipage}

\clearpage
\input{figures/results_PropPred}

\clearpage
\section*{Tables}

\noindent
\begin{minipage}{1.0\textwidth}
	{
		\captionof{table}{\textbf{
				Comparison of representative molecular pre-trained models with ChemFM.}}
		\tablebodyfont
		\label{table:pre-trained_model_comparison}
		\begin{tabular}{p{2.4cm} p{1.9cm} p{1.5cm} p{3.6cm} p{4.4cm}}
			\toprule
			\textbf{Model} & \textbf{Pre-training Data} & \textbf{Parameter Size} & \textbf{Pre-training Strategies} & \textbf{Downstream Tasks} \\
			\midrule
			Mole-BERT\cite{molebert} & ZINC15\cite{zinc15} \ \ \ \ (2M) & 1.86M & Masked atom modeling; contrastive learning & Property prediction \\
			\midrule
			SMILES \ \ \ \ \ \ \ Transformer~\cite{smilestransformer} & ChEMBL24\cite{chembl} (861K) & 4.26M & Reconstruction & Property prediction \\
			\midrule
			ChemBERTa-2~\cite{chemberta2} & PubChem\cite{pubchem} (77M) & up to 46M & Masked language modeling; multi-task regression & Property prediction \\
			\midrule
			Chemformer~\cite{chemformer}& ZINC-15\cite{zinc15} (100M) & up to 230M & Masked language modeling; SMILES canonicalization & Property prediction; reaction prediction; molecular optimization \\
			\midrule
			ChemFM & UniChem\cite{unichem} (178M) & up to 3B & Next token prediction & Property prediction; reaction prediction; molecular generation \\
			\bottomrule
		\end{tabular}
		\captionof*{tabledescription}{We compare ChemFM with other representative chemical pre-trained models reported in the literature in terms of model size, pre-training data and scale, pre-training strategies, and downstream tasks. 
				Models that are not open-sourced or whose reported performance could not be reliably validated through our replication efforts are excluded from this comparison. 
				The downstream task performance of ChemFM relative to these pre-trained models is presented in Tables~\ref{table:rst_prop_prediction}  (SMILES Transformer and Chemformer on property prediction), \ref{table:rst_prop_prediction_molebert} (Mole-BERT and ChemBERTa-2 on property prediction), and \ref{table:rst_rct_prediction_detail} (Chemformer on reaction prediction), in Supplementary Information. 
		}
	}
\end{minipage}

\clearpage
\noindent
\begin{minipage}{0.9\textwidth}
	\captionof{table}{\textbf{Performance comparison for conditional molecule generation on the GuacaMol\cite{guacamol} dataset.}}
		\tablebodyfont

	\begin{tabular}{cccccc}
	\toprule
	Property & Model & Validity $\uparrow$ & Uniqueness $\uparrow$ & Novelty $\uparrow$ & \makecell[c]{Mean average  deviation \\ (MAD) $\downarrow$} \\   
	
	\midrule
	\multirow{2}{*}{logP}      
	&  MolGPT     & $0.971$            &  $0.969$         &  $0.947$             &   $0.230$ \\
	&  ChemFM-3B    & $\mathbf{0.981}$            &  $\mathbf{0.981}$         &  $\mathbf{0.966}$             &   $\mathbf{0.182}$  \\
	\midrule
	\multirow{2}{*}{TPSA}      &  MolGPT     & $0.971$            &  $0.969$         &  $0.945$             &   $3.562$ \\
	&  ChemFM-3B    & $\mathbf{0.979}$            &  $\mathbf{0.979}$         &  $\mathbf{0.963}$             &   $\mathbf{2.466}$  \\
	\midrule
	\multirow{2}{*}{SAS}       &  MolGPT     & $0.978$            &  $0.974$         &  $0.941$             &   $0.133$ \\
	&  ChemFM-3B    & $\mathbf{0.986}$            &  $\mathbf{0.985}$         &  $\mathbf{0.957}$             &   $\mathbf{0.126}$  \\
	\midrule
	\multirow{2}{*}{QED}       & MolGPT     & $0.974$            &  $0.971$         &  $0.940$             &   $0.056$ \\
	&  ChemFM-3B    & $\mathbf{0.982}$            &  $\mathbf{0.982}$         &  $\mathbf{0.963}$             &   $\mathbf{0.045}$  \\
	\midrule
	\multirow{2}{*}{SAS + logP}      &  MolGPT     & $0.972$            &  $0.963$         &  $0.947$             &   $0.147 / 0.253$ \\
	&  ChemFM-3B    & $\mathbf{0.980}$            &  $\mathbf{0.975}$         &  $\mathbf{0.960}$             &   $\mathbf{0.137 / 0.195}$  \\
	\midrule
	\multirow{2}{*}{SAS + TPSA}      &  MolGPT     & $0.971$            &  $0.960$         &  $0.944$             &   $0.155 / 3.785$ \\
	&  ChemFM-3B    & $\mathbf{0.980}$            &  $\mathbf{0.971}$         &  $\mathbf{0.956}$             &   $\mathbf{0.138} / \mathbf{2.659}$ \\
	\botrule
	\multirow{2}{*}{TPSA + logP}      &   MolGPT     & $0.964$            &  $0.958$         &  $0.947$             &   $3.715 / 0.243$ \\
	&  ChemFM-3B    & $\mathbf{0.973}$            &  $\mathbf{0.970}$         &  $\mathbf{0.962}$             &   $\mathbf{2.415 / 0.184}$  \\
	\botrule
	\multirow{2}{*}{TPSA + logP + SAS}      &  MolGPT     & $0.972$            &  $0.942$         &  $0.931$             &   $3.797 / 0.268 / 0.180$  \\
	&  ChemFM-3B    & $\mathbf{0.975}$            &  $\mathbf{0.946}$         &  $\mathbf{0.936}$             &   $\mathbf{2.289}/ \mathbf{0.191} / \mathbf{0.166}$  \\						
	\bottomrule
\end{tabular}

	\label{table:rst_condition_generation}
	\captionof*{tabledescription}{
		Molecules were generated based on desired property values, with a performance comparison between ChemFM-3B, which uses a single model, and MolGPT~\cite{molgpt}, which uses $8$ separate models. Metrics include validity, uniqueness, novelty, and mean absolute deviation (MAD) between the conditioned and actual properties of the generated molecules. 
		\textbf{Bold} values indicate the best performance for each metric. 
		It should be noted that validity, uniqueness, and novelty are computed against the total number of generated molecules, rather than only the valid ones, to more accurately reflect model performance (\hyperref[sec:conditonal_generation_evaluation]{Methods}). 
	}
\end{minipage}

\clearpage
\noindent
\begin{minipage}{0.815\textwidth}
	\captionof{table}{\textbf{Performance comparison of ChemFM with the best and second-best models on standard USPTO benchmarks for synthesis and retro-synthesis reaction prediction tasks, showing top-1, top-3, and top-5 accuracies (in percentages).}}
	\tablebodyfont
	\begin{tabular}{lllccc}
		\toprule
		Task category & Dataset & Model & Top-$1$ & Top-$3$ & Top-$5$ \\ 
		\midrule
		Synthesis & USPTO-MIT & Prev. best: AT\cite{at} & 90.4 & - & 96.5 
														 \\
		                                                 &                            & Prev. second-best: R-SMILES\cite{rsmiles} & 90.0 & 95.6 & 96.4 
		                                                 \\
		                                                 &                            & ChemFM & $\mathbf{90.5}$ & $\mathbf{95.7}$ & $\mathbf{96.6}$ 
		                                                 \\                                                
		\midrule
		Retro-synthesis & USPTO-50K & Prev. best: R-SMILES\cite{rsmiles} & 56.0 & 79.0 & 86.1 
																		\\
														 &                            & Prev. second-best: Graph2Edits\cite{graph2edits} & 55.1 & 77.3 & 83.4
														 \\
														 &                            & ChemFM & $58.0$ & $\mathbf{80.0}$ & $\mathbf{86.3}$  \\
														 &                            & ChemFM$^\star$ & $\mathbf{59.7}$ & $79.2$ & $84.2$ 															 
														 \\
														 \cmidrule(r){2-6}
														& USPTO-MIT & Prev. best: R-SMILES\cite{rsmiles} & 60.3 & 77.9 & 82.8 
														\\
														&                            & Prev. second-best: RetroTRAE\cite{retroTRAE} & 60.3 & 77.9 & 82.8 
														\\
														&                            & ChemFM & $61.6$ & $\mathbf{78.7}$ & $\mathbf{83.0}$ \\
																																								&                      	& ChemFM$^\star$ & $\mathbf{62.4}$ & $78.5$ & $82.5$ 														
														\\
														\cmidrule(r){2-6}
														& USPTO-Full & Prev. best: RetroXpert\cite{retroxpert} & 49.4 & 63.6 & 67.6 
														\\
														&                            & Prev. second-best: R-SMILES\cite{rsmiles} & 48.9 & 66.6 & 72.0 
														\\
														&                            & ChemFM & $\mathbf{51.7}$ & $\mathbf{68.0}$ & $\mathbf{72.5}$ 
														\\
		
\bottomrule
\end{tabular}

	\captionof*{tabledescription}{The best and second-best models are determined based on top-$1$ performance.
		\textbf{Bold} values indicate the best performance for each metric. 
		A hyphen ``-'' indicates that the value was not reported in the original paper.
		Results for R-SMILES are obtained through our replication using publicly available models\cite{rsmiles}.
		ChemFM$^\star$ denotes ChemFM with further pre-training, which achieves better top-$1$ results but shows a decrease in top-$3$ and $5$ performance.
	}
	\label{table:rst_rct}
\end{minipage}
\clearpage

%


\newpage 
\section{Methods}
\label{sec:methods}
%

\subsection{Chemical language modeling}
Molecular serialization systems, such as the simplified molecular input line entry system (SMILES)~\cite{smiles} or self-referencing embedded strings (SELFIES)~\cite{selfies},
represent molecules as linear sequences. 
This linearization enables the use of sequence-based models to effectively model chemical language.
Formally, consider a corpus of molecules $\mathcal{C} = \{\bm{s}_1, \bm{s}_2, \ldots, \bm{s}_m\}$.
Each molecule $\bm{s}$ is represented as a sequence of tokens (subwords), $\bm{s} = (t_1, t_2, \ldots, t_n)$, using a serialization system.
The chemical language model is tasked with computing the joint probability of the sequence:
\begin{equation*}
	P(\bm{s}) = P(t_1, t_2, \ldots, t_n).
\end{equation*}
ChemFM extends the principles of causal language modeling\cite{gpt1} by employing a unidirectional transformer decoder, also known as a causal decoder-only transformer, to model chemical language in an auto-regressive manner. 
Within this framework, each token in the sequence is predicted based solely on its preceding tokens, allowing the joint probability of the sequence to be factorized as a Markov chain:
\begin{equation*}
	P(\bm{s}) = \prod_{i=1}^{n} p(t_i | t_1, \ldots, t_{i-1}).
\end{equation*}
By pre-training on a large corpus of molecules, ChemFM learned the syntactic rules of serialization systems and the sequential dependencies inherent in molecular structures. These capabilities in representation learning can then be adapted to a wide range of chemical tasks.
\subsection{Model architecture}\label{sec:models}

The ChemFM models were based on TinyLlama\cite{tinyllama}, a parameter-compact version of the Llama 2 architecture\cite{llama2}, which employed a causal decoder-only transformer. 
In this work, we presented two model variations, 
ChemFM-1B, with $970$ million trainable parameters, and ChemFM-3B, with $3.0$ billion trainable parameters.
These variations differ in the number of hidden layers, the number of attention heads, the dimension of the hidden representations, and the dimension of the multi-layer perceptron (MLP) representations.
Detailed architectures for both models are outlined in Supplementary Table~\ref{table:model_architecture}.
%
%
\subsection{Molecule representation and tokenization}\label{sec:molecule_representation}

We utilized SMILES, a serialization format widely used in computational chemistry\cite{attentiveFP,molgpt,rsmiles}, to represent molecular structures.
Molecules were first transformed into SMILES strings using the \texttt{RDKit} library\cite{rdkit}.
%
The resulting SMILES strings were segmented and tokenized using a sub-word tokenizer\cite{bpe} with a predetermined vocabulary of $266$ tokens. 
The vocabulary includes both uppercase and lowercase representations of the $118$ elements from the periodic table, numerical digits from $0$ to $9$, $19$ special symbols as specified by SMILES syntax\cite{smiles}, and a special end token indicating the termination of a SMILES string. {It should be noted that our vocabulary includes a small number of redundant tokens (e.g., lowercase element symbols that rarely or never appear in aromatic form). However, these have no impact on training efficiency or model performance.}. 
These tokens form the foundational vocabulary used during the pre-training phase, while additional special tokens are introduced during fine-tuning to address specific task requirements, as detailed in subsequent sections.

\subsection{Pre-training dataset selection}
\label{sec:pre-training_data_selection}

Public chemical databases like ZINC\cite{zinc15, zinc20}, PubChem\cite{pubchem}, ChemBL\cite{chembl}, and UniChem\cite{unichem} contain billions of molecules and are commonly used in pre-training chemical models. 
For example, 
Chemformer\cite{chemformer} utilized $100$ million molecules randomly sampled from ZINC15, Grover\cite{grover} was developed on $11$ million molecules sampled from ZINC15 and ChemBL, 
and MolFormer\cite{molformer} employed a combination of ZINC15 and PubChem.
However, the literature often lacks justification for specific dataset selections. 

Considering the large scale of ChemFM and in order to avoid performance saturation,
ZINC20 and UniChem (which encompasses most of the molecules from PubChem)---housing $1.8$ billion and $178$ million molecules respectively---are well-suited candidates for pre-training ChemFM.
However, a large dataset alone does not guarantee sufficient information richness. 
Given the computational intensity of pre-training large models, 
careful dataset assessment is crucial prior to pre-training.

Therefore, we conducted a series of scaling experiments to evaluate the information content in the UniChem and ZINC20 datasets.
The scaling laws of neural language models\cite{scalinglaw} reveal that model performance strongly depends on the scale of the model's non-embedding parameters and the dataset size. 
Empirical evidence shows that performance (as measured by loss) follows a power-law relationship with each of these factors, provided the other is not bottlenecked.
Our scaling experiments utilized causal decoder-only transformers from the ChemFM family, with models ranging from approximately $10$M to $200$M parameters. 
Each model was trained using cross-entropy loss, with a fixed data budget of $\num{250000}$ steps and a consistent batch size of $\num{1024}$ across all runs. 
The detailed architectures for the models used in these experiments are provided in Supplementary Table~\ref{table:model_architecture}. 
The loss, measured on a validation dataset, was recorded at the end of each run.

For the UniChem dataset, we observed that the validation loss closely followed a power-law scaling with respect to the number of non-embedding parameters, showing no sign of performance saturation as the model size increased. 
%
%
In contrast, the ZINC20 dataset exhibited performance saturation when the model size reached $60$M parameters.
This suggests that the knowledge contained within ZINC20 becomes a bottleneck, limiting the model's ability to benefit from increased parameter size.

{To directly evaluate the impact of the pre-training dataset, we fully pre-trained a 1B-parameter model on ZINC20 and compared its downstream property prediction performance with the UniChem-pretrained counterpart, with details provided in Section~\ref{sec:comparison_pretraining}.
}

%
%

\subsection{Pre-training details}
\label{sec:pre-training}

We selected the UniChem dataset for pre-training ChemFM. Using the SMILES data enumeration technique\cite{smilesaug}, we augmented the dataset tenfold, resulting in a final pre-training dataset comprising $1.78$ billion molecules, with $90\%$ allocated for training and $10\%$ for validation.

Both model variants were trained using the 
AdamW\cite{adamw} optimizer.
The learning rate was initially warmed up to $4\times10^{-4}$ over $\num{2000}$ steps and then decayed following a cosine scheduler down to $4\times10^{-5}$.
Each sequence was truncated to $\num{512}$ tokens, and a batch size of $\num{1024}$ sequences was utilized. 
The models were trained for one epoch, processing a total of $818$ billion tokens.
The two ChemFM variants were pre-trained in a distributed manner on different hardware configurations: 
ChemFM-1B was trained across eight NVIDIA A100 nodes, each with $2\times$A100 80GB GPUs, while ChemFM-3B was trained across two NVIDIA HGX H100 nodes, each with $8\times$H100 80GB GPUs.
The pre-training required $23.2$ days for ChemFM-1B and $27.6$ days for ChemFM-3B.
Both pre-trained models and the training codes are publicly available (\hyperref[sec:code]{Code availability}).

\subsection{Benchmarking unconditional generation for pre-trained models}
\label{sec:methods_unconditional_benchmark}
We evaluated the unconditional generation capability of the pre-trained ChemFM model by generating 
$\num{100000}$ molecules.
For each molecule, the generation process began by sampling a start token according to its frequency distribution in the training dataset.
%
The model then autoregressively generated tokens until producing an end token, thus completing the molecule.
%
A temperature of $1.0$ was applied to the $\mathtt{SoftMax}$ during generation.

We assessed the generated molecules using established metrics from molecule generation benchmarks such as GuacaMol~\cite{guacamol} and MOSES~\cite{moses}.
Specifically, we measured  
the validity, uniqueness, novelty, and internal diversity (IntDiv$_1$, IntDiv$_2$, {and Sphere Exclusion Diversity (SEDiv)}).
We also compared the distributions of $9$ physicochemical descriptors (computed using the \texttt{RDKit} library) between $\num{100000}$ generated molecules and $\num{100000}$ randomly sampled from the training dataset (Supplementary Fig.~\ref{fig:unconditional_generation_desc_dist}). 
We quantified the similarity between these distributions by computing Kullback-Leibler (KL) divergence for each descriptor and aggregating them into a final KL similarity (KLSim) score.
Additionally, \texttt{ECFP4}\cite{ecfp} fingerprints were computed for both sets, and their 2D t-SNE mapping was visualized to further evaluate how well the generated molecules aligned with the training data (Supplementary Fig.~\ref{fig:tsne_embedding}). 
{We also compared internal diversity between the generated molecules and the pre-training dataset (UniChem) to further evaluate their alignment (Supplementary Table~\ref{table:diveristy_comparison}).}

Details on these metrics can be found in the \hyperref[sup:generation_metrics]{Supplementary Information}.
%

{
It is worth noting that the ChemFM models were trained on a dataset more than $100\times$ larger than those used in the GuacaMol and MOSES benchmarks, which limits the validity of direct performance comparisons. 
This is also why the novelty of molecules generated by ChemFM is lower than values typically reported in the literature.
To verify this, we compared ChemFM-3B with MolGPT by generating $\num{100000}$ molecules each and computing novelty separately against the GuacaMol and MOSES benchmarks. The results are reported in Supplementary Table~\ref{table:novelty comparison}.
In this setting, ChemFM achieves higher novelty than MolGPT, confirming that the lower novelty observed with respect to UniChem reflects the scale of the reference database rather than a limitation of the model.
}

\subsection{Training objective for property prediction}
\label{sec:methods_lora}

Fine-tuning ChemFM for supervised molecular property prediction tasks 
follows the framework of sequence classification and regression in causal language models~\cite{gpt1}. 
Given a labeled dataset $\mathcal{C}$, each sample consists of a molecule $\bm{s}$ represented as a SMILES string and its corresponding label set $\bm{y} = (y_1, \ldots, y_m)$  for $m$ prediction tasks.
These labels represent either regression or binary classification tasks but do not mix the two, following the settings in the MoleculeNet\cite{moleculenet} and ADMET\cite{admet} benchmarks.
The SMILES string $\bm{s}$ is tokenized into a sequence of tokens $t_1, t_2, \ldots, t_n$, terminated with a special end token. 
This tokenized sequence is processed by ChemFM, from which the hidden state $h_l^n$ 
from the last layer $l$ corresponding to the final token $t_n$ is extracted.
%
A linear layer, $W_{\bm{y}} \in \mathbb{R}^{d_\text{model} \times m}$, where $d_\text{model}$ is the dimension of the model’s hidden representations, is applied to this hidden state, to make the predictions $\hat{\bm{y}} = (\hat{y}_1, \ldots, \hat{y}_m)$ for the $m$ tasks, as shown in Fig.~\ref{fig:finetune_pipeline}\textbf{a}.
For regression tasks, the model minimizes the mean square error (MSE) loss over the dataset:
\begin{equation*}
	\mathcal{L}_{\text{regression}} = \frac{1}{|\mathcal{C}|}\sum_{(\bm{s}, \bm{y})\in\mathcal{C}} \frac{1}{m} \sum_{i=1}^{m}(\hat{y}_i - y_i)^2.
\end{equation*}
For binary classification tasks, the model computes a probability distribution for each task using a $\mathtt{Sigmoid}$ activation function, and minimizes the binary cross-entropy loss:
\begin{equation*}
	\mathcal{L}_{\text{classification}} = -\frac{1}{|\mathcal{C}|}\sum_{(\bm{s}, \bm{y})\in\mathcal{C}} \frac{1}{m} \sum_{i=1}^{m} P_i(y_i|\bm{s}),
\end{equation*}
where $P_i(y_i|\bm{s}) = \mathtt{Sigmoid}(\hat{y}_i)$ represents the predicted probability for task $i$.

\subsection{Parameter efficient fine-tuning}
Adapting all parameters of ChemFM is resource-intensive, requiring substantial GPU memory and storage. 
We utilized Low-Rank Adaptation (LoRA)~\cite{lora}, a popular parameter-efficient fine-tuning technique that reduces the number of trainable parameters by introducing low-rank decomposition matrices for each layer instead of updating the full parameter set.
We applied LoRA across all linear layers in the transformer blocks of ChemFM, while freezing the embedding layer, as no task-specific tokens are introduced for molecular property prediction tasks. 
The prediction head $W_{\bm{y}}$ is fully adapted to predict labels. 

%

The number of trainable parameters is controlled by adjusting the rank $r$ of the  decomposition matrices,
and we report the number of trainable parameters for each task in Supplementary Table~\ref{table:hyperparameter_moleculenet}, \ref{table:hyperparameter_admet},
and \ref{table:hyperparameter_reaction}.
For instance, with $r=4$ in ChemFM-3B (using $32$-bit float precision), the number of trainable parameters is reduced by $460\times$, from $3$ billion to $6.5$ million. 
This reduces video RAM requirements during training from $\qty{51}{\giga\byte}$ to $\qty{20}{\giga\byte}$ and checkpoint size from $\qty{12}{\giga\byte}$ to $\qty{26}{\mega\byte}$.

\subsection{Data pre-processing and training setups for property prediction}
SMILES augmentation during training has been shown to improve molecular property prediction\cite{CLM}.
During training, we applied SMILES enumeration\cite{smilesaug} with probability $p = 1.0$, while during inference, we used only canonical SMILES strings. 
Though synthesizing results from multiple augmentations could potentially improve performance, this approach was not explored in our experiments.

%
Through the reduction in GPU memory achieved by using the LoRA technique, 
both ChemFM-1B and ChemFM-3B can be fine-tuned on a single GPU.
While our experiments were conducted on a single H100 $\qty{80}{\giga\byte}$ GPU, 
the fine-tuning process is feasible on more modest hardware setups.

\subsection{Experimental setting on MoleculeNet benchmark for property prediction}
\label{sec:methods_moleculenet}

We began by fine-tuning the ChemFM model on datasets from the MoleculeNet benchmark\cite{moleculenet} (dataset descriptions are provided in Supplementary Table~\ref{table:dataset_molecule_net}). 
{
While many methods have been developed and evaluated on the MoleculeNet benchmark, 
comparisons between them are often problematic due to variations in dataset splitting strategies
and random seed selections across studies.
This issue is particularly exacerbated by the ``scaffold'' split, which partitions molecules based on structural scaffolds.
Although this method creates more structurally diverse and challenging train/validation/test folds than random splitting,
it can result in significantly different test sets across experiments, complicating cross-study comparisons.}

{
To ensure a comprehensive evaluation of ChemFM, we conducted three distinct sets of comparisons with existing methods, all using the same train/validation/test datasets. 
We excluded methods that are not open-sourced since verifying their dataset splits is not possible.}

\noindent \textbf{Comparison set 1:}
We first fine-tuned both ChemFM-1B and ChemFM-3B models on standard MoleculeNet datasets\cite{moleculenet}, as provided in the MoleculeNet paper. 
Different datasets used different scaffold methods and are detailed in  Supplementary Table~\ref{table:dataset_molecule_net}.
The methods we compared against include {SMILES Transformer~\cite{smilestransformer}}, MoleculeNet models\cite{moleculenet}, Direct Message Passing Neural Networks (D-MPNN or Chemprop)\cite{dmpnn}, MolMapNet (MMNB)\cite{MMNB}, and {Chemformer\cite{chemformer}}.
We conducted a random search for the training hyperparameters and LoRA configurations for each dataset, with the selected hyperparameters detailed in Supplementary Table~\ref{table:hyperparameter_moleculenet}.
\emph{Importantly, 
hyperparameter tuning was based solely on validation performance, 
and no tuning was performed on the test datasets.}
We evaluated our models across three folds and reported the average performance, along with the corresponding split method and evaluation metrics, in  Supplementary Table~\ref{table:rst_prop_prediction}.

\noindent \textbf{Comparison set 2:}
We then compared ChemFM with the AttentiveFP method\cite{attentiveFP}, 
which used random splitting methods but with different seeds than the standard MoleculeNet benchmark. 
MMNB provided a direct comparison with AttentiveFP (as shown in Table 2 of the MMNB paper\cite{MMNB}), 
where MMNB outperformed AttentiveFP on most datasets. 
Since ChemFM models consistently outperformed MMNB, we can reasonably infer that ChemFM is also superior to AttentiveFP on these datasets. 
However, on four specific datasets---Tox21, ESOL, FreeSolv, and Lipophilicity---AttentiveFP outperformed MMNB. 
For these datasets, we conducted a direct comparison by reevaluating AttentiveFP across three folds split by different random seeds and fine-tuning ChemFM using identical data splits. 
The results are presented in  Supplementary Table~\ref{table:rst_prop_prediction_attentive_fp}.

\noindent \textbf{Comparison set 3:}
The third comparison set focused on methods using a deterministic scaffold split to generate a single fold of train/validation/test sets, including {Pretrain GNNs\cite{pretraingnns}, ChemBERTa-2\cite{chemberta2}}, 3D InfoMax\cite{3dinfomax}, Mole-BERT\cite{molebert}, GraphMVP\cite{graphmvp}, and MoleculeSDE\cite{moleculeSDE}. 
We adopted the same settings as these methods---training on the same data fold with three different random training seeds (which only affect the training procedure like the network weights initialization and network dropout, but not dataset splitting)---and reported the average performance in  Supplementary Table~\ref{table:rst_prop_prediction_molebert}.

It is important to highlight that, despite differences in splitting methods and random seeds, no additional hyperparameter tuning was performed for comparison sets 2 and 3. 
We reuse the hyperparameters optimized for the standard MoleculeNet datasets (shown in Supplementary Table~\ref{table:hyperparameter_moleculenet}). 
For each dataset, we first fine-tuned the ChemFM-3B model.
%
If ChemFM-3B did not outperform all other methods (specifically, the ESOL dataset in comparison set 2 and the MUV dataset in comparison set 3), we proceeded to fine-tune ChemFM-1B.
The results indicated that at least one of our ChemFM models outperformed all other compared methods across the evaluated datasets, even without additional hyperparameter tuning on these specific data folds.

\subsection{Experimental setting on ADMET benchmark for property prediction}
\label{sec:methods_admet}

We compared ChemFM with methods on the leaderboard of the ADMET benchmark\cite{admet},
which comprises $22$ datasets and provides standard data splits and performance evaluation metrics.
The leaderboard facilitates cross-method comparisons on these datasets. 
However, 
not all methods on the leaderboard are evaluated correctly.
For example, common reasons for mis-evaluation included optimizing hyperparameters using the test datasets and combining the training and validation datasets for model training (practices that can improve performance, especially in the ADMET benchmarks, where most datasets contain fewer than 1,000 instances).
We carefully reviewed the public codes of the methods on the leaderboard and excluded those that were mis-evaluated. 
The methods and corresponding reasons for exclusion are listed in the Supplementary Table~\ref{tab:exclusion_reasons}.

For each dataset, we first conducted a hyperparameter search for the ChemFM-3B model. 
The adapted ChemFM-3B model outperforms the best models on the leaderboard for 20 out of the 22 datasets, with the exceptions of the Caco2\_Wang and HIA\_Hou datasets. 
For these two datasets, we then performed a hyperparameter search for the ChemFM-1B model,
which can achieve state-of-the-art results. 
The comparisons between ChemFM and the previous best models are presented in  Supplementary Table~\ref{table:rst_prop_prediction_admet}, with the hyperparameters used detailed in Supplementary Table~\ref{table:hyperparameter_admet}.

\subsection{Effect of pre-training on ChemFM performance}
\label{sec:methods_effect_pre-training}
{To directly assess whether ChemFM's improvements stem from pre-training rather than simply from model size, we fine-tuned ChemFM-3B on the ADMET dataset with the same hyperparameters as before but initialized the model from scratch (random initialization) instead of using pre-trained weights. The results are reported in the  Supplementary Table~\ref{table:without_pretraining} (A similar comparison between pre-training and non-pre-training was conducted for conditional molecular generation, as described in \hyperref[sec:conditonal_generation_evaluation]{Methods}).  }

{	
	As shown, models trained without pre-training perform substantially worse than the pre-trained ChemFM model across all property prediction tasks, demonstrating that the observed improvements are indeed due to large-scale pre-training rather than model size alone.
}

\subsection{Effect of pre-training dataset on ChemFM performance}
\label{sec:comparison_pretraining}
{
To further evaluate the impact of the pre-training dataset, we conducted a direct comparison between ChemFM models pre-trained on UniChem and ZINC20. Specifically, we pre-trained a 1B-parameter model on ZINC20 using the same configuration and training steps as the UniChem-pretrained counterpart. Both models were then fine-tuned on the MoleculeNet datasets for molecular property prediction.}

{
The results, summarized in Supplementary Table~\ref{table:comparison_pretraining}, show that the UniChem-pretrained model consistently outperforms the ZINC20-pretrained model on 9 out of 11 datasets, by a large margin, with the remaining two datasets yielding nearly identical performance. This demonstrates that pre-training on the more informative UniChem dataset leads to stronger molecular representations and superior downstream performance.
}

\subsection{Additional experiments on property prediction tasks}
\label{sec:addtional_property_prediction}

{
We further evaluated ChemFM on two molecular property prediction tasks extending to applications relevant for a broader community of chemists: odor prediction\cite{odor} and chromatographic retention time prediction\cite{smrt}.}

{
For odor prediction~\cite{odor}, we used a dataset of approximately $\num{5000}$ molecules annotated with $138$ odor labels, where each molecule may have multiple labels. 
This is formulated as a multi-label classification task. 
We compared ChemFM-3B with the open-source reproduction of the original MPNN-based approach (\href{https://github.com/BioMachineLearning/openpom}{OpenPOM})~\cite{OpenPOM}, ensuring identical 5-fold cross-validation splits (Supplementary Table~\ref{table:additional_property_prediction}).
}

{
For chromatographic retention time prediction, we used the METLIN small molecule retention time (SMRT) dataset~\cite{smrt}, which contains experimentally acquired reverse-phase chromatography measurements for up to $\num{80038}$ molecules. 
ChemFM-3B was compared against the baseline regression neural network built on molecular fingerprints and descriptors, using the same $75\%/25\%$ train/test random split (Supplementary Table~\ref{table:additional_property_prediction}).
}

\subsection{Experimental setting for potential antibiotics screening}

We fine-tuned ChemFM-1B on a dataset used for screening potential antibiotics~\cite{antibiotics}. 
This dataset contains $\num{39312}$ compounds, with measurements of antibiotic activity based on RN4220 growth inhibition and cytotoxicity data across three human cell types: liver carcinoma cells (HepG2), primary skeletal muscle cells (HSkMC), and lung fibroblast cells (IMR-90). 
For a fair comparison, we adhered to the data split protocol from previous work, using $80\%$ of the dataset for training and $20\%$ for testing. 
While exact train-test splits from the original study were not available, we ensured that active compounds in our train and test sets reflected a similar distribution to the full dataset ($1.3\%$ for antibiotic activity, $8.5\%$ for HepG2 cytotoxicity, $3.8\%$ for HSkMC cytotoxicity, and $8.8\%$ for IMR-90 cytotoxicity), consistent with the original paper.
We used an empirical hyperparameter setup (detailed in Supplementary Table~\ref{table:hyperparameter_antibiotics}) without additional hyperparameter tuning. 
For each task, in contrast to the previous study, which employed an ensemble of $20$ Chemprop models\cite{dmpnn}, we trained a single ChemFM model. 
Evaluation employed bootstrapping, with $100$ resampled test sets generated by repeatedly drawing samples of equal size to the original test set. 
This approach allowed us to compute $95\%$ confidence intervals for the PRC-AUC and capture the variability in precision-recall curves.

We further evaluated both ChemFM and Chemprop on an antibiotic library containing $\num{1994}$ real antibiotics~\cite{medchemexpress_antibiotic}. 
To focus on structurally novel molecules, we deduplicated the library and excluded antibiotics with Tanimoto similarity scores below $0.5$ to any known antibiotics in the training dataset, resulting in $\num{1173}$ novel molecules.
Since both models were trained as classifiers, we applied a threshold of $0.5$ to the prediction scores to distinguish positives from negatives.
ChemFM labeled \num{149} molecules as positives, whereas Chemprop labeled only \num{29}. 
Even when lowering the threshold to $0.4$—following the approach used in Wong et al. \cite{antibiotics} to identify antibiotic activity hits—Chemprop labeled only \num{42}, still far fewer than ChemFM.
Details of this antibiotics dataset and prediction scores from both models are provided in Supplementary Data 1 in a separate file.

\subsection{Experimental setting for data efficiency}
\label{sec:methods_data_efficient}
To evaluate the data efficiency of ChemFM, we conducted experiments on both a regression and a classification task. Specifically, we selected the CYP2D6\_Substrate\_CarbonMangels classification task and the Half\_Life\_Obach regression task.
For each task, we randomly sampled $10\%$, $20\%$, and $50\%$ of the original training dataset to create reduced training subsets. 
To ensure robustness, we generated five independent subsets for each ratio. 
ChemFM-3B was fine-tuned on these subsets and compared against the previous best methods: Chemprop-RDKit~\cite{chemprop-rdkit} for CYP2D6\_Substrate\_CarbonMangels and DeepPurpose~\cite{deeppurpose} for Half\_Life\_Obach.
We used the full test set for evaluation.
The final reported results for each ratio are the average over five runs, providing a comparison of model performance under different training data constraints (Supplementary Fig~\ref{fig:data_efficiency}).
\subsection{Training objective for conditional generation}
Conditional molecular generation tasks aim to produce molecules that meet specified criteria, such as desired molecular properties or structural constraints like scaffold fragments, and can be formalized as a sequence-to-sequence problem, where the goal is to generate a target sequence conditioned on a given input sequence.
%
%
Let $\mathcal{C}$ denote a dataset where each instance includes an input sequence, $\bm{s}_i$, representing the desired molecular characteristics or structural constraints (details on condition representation are provided in the next section), and a corresponding target molecular SMILES sequence, $\bm{s}_o$. 
These sequences are tokenized into series of tokens: $\bm{s}_i = (t_{-m}, t_{-m+1}, \ldots, t_0)$ for the input and $\bm{s}_o = (t_1, t_2, \ldots, t_n)$ for the target sequence.
The ChemFM model generates the output sequence autoregressively, conditioned on the input sequence and previously generated tokens, which is illustrated in Fig.~\ref{fig:finetune_pipeline}\textbf{b}.
The training objective is to maximize the conditional probability distribution $P(\bm{s}_o | \bm{s}_i)$:
\begin{equation}
	\label{eq:seq2seq}
	P(\bm{s}_o | \bm{s}_i) =  \prod_{i=1}^{n} p(t_i | t_{-m}, \ldots, t_{0}, \ldots, t_{i-1}).
\end{equation} 


\subsection{Condition representation for conditional generation}
In conditional molecular generation tasks, the input sequence can consist of multiple conditions, 
each represented by two components: a property name and a property value.
A concrete example is shown in the Fig.~\ref{fig:finetune_pipeline}\textbf{b}.
The property name serves as a unique identifier and is denoted by a special token 
indicating the specific molecular property being conditioned upon. 
%
%
The property value can take one of three forms:
\begin{description}
	\item[\textbf{\upshape Continuous values}] Represented by a special placeholder token. 
	These values are normalized before being processed by the model. 
	They are then mapped into the embedding space through a shared linear layer, which is applied to all real-valued properties, allowing the model to capture the continuous nature of the property.
	%
	%
	\item[\textbf{\upshape Classification values}]  
	Encoded as special tokens that correspond to specific class indices. 
	For example, a property ``isRing'' followed by a classification token ``C1'' indicates that the molecule should contain a ring structure, where the class index ``1'' denotes the presence of a ring.	
	\item[\textbf{\upshape String representations}] Used in cases such as scaffold-conditioned generation, where the scaffold fragment is represented as a SMILES string.
\end{description}


\subsection{Data pre-processing and training details for conditional generation}
\label{sec:conditonal_generation_technique}

We followed the experimental setup of MolGPT\cite{molgpt} to evaluate conditional molecular generation using the GuacaMol\cite{guacamol} and MOSES\cite{moses} datasets. 
The GuacaMol dataset focuses on generation based on molecular properties, while for the MOSES dataset includes both scaffold and molecular properties as generation conditions.
For both datasets, we considered four continuous molecular properties: $\log$P, synthetic accessibility score (SAS), topological polar surface area (TPSA), and quantitative estimate of drug-likeness (QED). 
These properties can be directly computed via \texttt{RDKit}\cite{rdkit}, enabling automatic performance evaluation of conditional molecular generation models.
Unlike MolGPT, which requires $15$ separate models to cover all property combinations for each dataset, 
we developed a single unified model for each dataset. 
This approach allows our models to handle multiple property combinations more flexibly. 
While it is feasible to train a single model that combines both datasets, we maintained separate models for GuacaMol and MOSES to ensure a fair comparison with MolGPT.


During training, we applied a probabilistic property selection strategy: one property was selected with a probability of $0.1$, two properties with $0.2$, three with $0.3$, and four with $0.4$. 
The order of properties was randomized. 
Additionally, we used the SMILES enumeration technique~\cite{smilesaug} with a probability of $1.0$ to augment target SMILES strings .
Both models underwent full-parameter fine-tuning using the AdamW optimizer with a weight decay of $0.01$. 
The learning rate was initialized at $6\times10^{-4}$, with a warm-up phase spanning $0.1$ epochs, and was decayed using a cosine schedule to a minimum of $6\times10^{-5}$. 
Fine-tuning was conducted on an NVIDIA HGX H100 node with 8$\times$80GB GPUs for $10$ epochs, using a batch size of $384$.

%

\subsection{Evaluation details for conditional generation}
\label{sec:conditonal_generation_evaluation}
Our evaluation followed the setup of MolGPT to ensure a fair comparison.
For property-based generation (GuacaMol dataset), we evaluated $8$ distinct property combinations {and compared performance with both cRNN~\cite{crnn} and MolGPT}. 
For each combination, multiple sample points (representing specific property values) were tested, and for each point, we generated $\num{10000}$ molecules with the temperature setting to $1.0$. 
The distribution of the generated molecules' properties across sample points for each combination is presented in Supplementary Fig.~\ref{fig:rst_mol_gen_prop_dist_1d}. 
To assess the basic generation capabilities of the models, we reported the validity, uniqueness, and novelty scores for each property combination. 
%
%
Additionally, to evaluate how well the model adheres to the property conditions, we computed the mean absolute deviation (MAD) between the conditioned property and the computed property. 
These results are summarized and compared with {cRNN and} MolGPT in Supplementary Table~\ref{table:result_conditional_generation_full}. 
{
The MolGPT results were obtained by re-running the published checkpoints, while cRNN results are based on our reimplementation and training using the published code, since the original paper did not perform the same experiments as conducted here.}

{
To demonstrate that pre-training benefits conditional molecular generation tasks, we additionally trained a randomly initialized ChemFM-3B model for this task. 
During these experiments, we observed that the model without pre-training was unstable to train and often diverged, producing almost entirely invalid molecules. 
After  hyperparameter tuning, we were able to obtain a trained model and report the results in Supplementary Table~\ref{table:result_conditional_generation_full}. The results indicate that simply increasing model size does not yield performance improvements; in fact, the randomly initialized ChemFM-3B often performs worse than the much smaller MolGPT model. These findings confirm that pre-training is crucial for effective conditional molecular generation.
}

Instead of computing the novelty, and uniqueness scores against on the valid generated molecules, we compute these scores against the total number of generations. This is due to the fact that the standard uniqueness and novelty computation cannot effectively reflect model performance when validity is low. 
For example, assuming two models generating $\num{10000}$ molecules each, one generates $\num{5000}$ unique molecules out of $\num{9000}$ valid, while the other generates $\num{5300}$ unique molecules out of $\num{9800}$ valid.
Although the second model exhibits better uniqueness performance, calculating uniqueness as a ratio would yield $0.56$ for the first model ($\num{5000} / \num{9000}$) and $0.54$ for the second ($\num{5300} / \num{9800}$). 

For scaffold and property-based generation (MOSES dataset), we evaluated the model conditioned on five testing scaffolds and 8 different property combinations. 
For each sample point, $\num{10000}$ molecules were generated, and 
the distribution of the generated molecules' properties is presented in Supplementary Fig.~\ref{fig:rst_mol_gen_scaffold_prop_dist_1d}.
Here, a valid molecule is defined by two criteria: 1) the SMILES string is syntactically correct and represents a feasible molecular structure, and 2) the scaffold of the generated molecule has a Tanimoto similarity of at least $0.8$ to the desired scaffold.
Instead of reporting the validity, novelty, and uniqueness scores, we directly presented the counts of valid, unique, and novel molecules generated. 
%
%
We also evaluated the count of molecules that retained the same scaffold as the conditioned scaffold and computed the MAD between the conditioned property and the generated property values.
The results of this evaluation, compared with MolGPT, are presented in Supplementary Table~\ref{table:rst_conditional_molecular_generation}.

{In the scaffold-conditioned generation experiments, we observed that although ChemFM substantially outperforms baseline methods in both generation and matching metrics, a non-negligible fraction of generated molecules still failed to include the specified scaffold. Scaffold-constrained generation techniques such as PromptSMILES\cite{promptsmiles} are fully compatible with ChemFM: by rooting the desired scaffold at the beginning of the SMILES string, PromptSMILES ensures that the generated molecules contain the specified scaffold. Importantly, this approach does not require retraining a scaffold-specific conditioned model and can be directly applied to the property-conditioned ChemFM models for both property- and scaffold-conditioned generation.}

\subsection{Training objective for reaction prediction}
We focused on both forward synthesis and retro-synthesis reaction prediction tasks, 
which leverage the same training objective used in conditional molecular generation as sequence-to-sequence problems.
%
%
Let $\mathcal{C}$ denote a reaction dataset, where each instance consists of an input sequence, $\bm{s}_i$, and a corresponding target sequence, $\bm{s}_o$. 
In the forward synthesis task, $\bm{s}_i$ represents the reactants (and possibly includes reagents), while $\bm{s}_o$ denotes the products. 
In retro-synthesis, the roles are reversed. Both input and target sequences are represented as SMILES strings.
When multiple compounds appear in either the reactants or products, they are separated by a predefined delimiter, ``.'', in the SMILES representation. 
These sequences are then tokenized into series of tokens: $\bm{s}_i = (t_{-m}, t_{-m+1}, \ldots, t_0)$ for the input and $\bm{s}_o = (t_1, t_2, \ldots, t_n)$ for the target sequence.
The ChemFM model generates the output sequence autoregressively, conditioned on the input sequence and previously generated tokens, as illustrated in Fig.~\ref{fig:finetune_pipeline}\textbf{c}.
The training objective is the same as in conditional molecular generation, defined in Eq.~(\ref{eq:seq2seq}).

%

\subsection{Datasets and pre-processing for reaction prediction}
For reaction prediction tasks, we fine-tuned ChemFM-3B on widely-used USPTO-series datasets, including USPTO-50K\cite{uspto50k}, USPTO-MIT\cite{usptomit}, and USPTO-Full\cite{uspto_full}, commonly employed for benchmarking both forward synthesis and retro-synthesis tasks. 
Detailed statistics for these datasets are provided in Supplementary Table~\ref{table:dataset_uspto}.
In the forward synthesis task, we focused on the USPTO-MIT dataset with the setting where reactants and reagents are mixed in the input sequence.
For retro-synthesis prediction, we conducted experiments on USPTO-50K, USPTO-MIT, and USPTO-Full datasets, 
focusing on the challenging setting where the reaction class is not provided.
For the USPTO-Full dataset, following previous work\cite{at, rsmiles}, we removed invalid reactions, such as those containing no products or just single ions as reactants.

Typically, input and output SMILES strings in reaction tasks vary significantly as they are pre-processed independently\cite{at} and no inherent relationship between them is considered.
Root-aligned SMILES (R-SMILES)\cite{rsmiles}, however, defined a tight, one-to-one mapping between reactant and product SMILES by aligning the same atom as the root in both strings, making them more similar and improving the efficiency of reaction prediction.
Following Zhong et al. \cite{rsmiles}, we augmented the training data by enumerating different root atoms, generating $n$ augmented input-output pairs for each reaction.
The augmentation folds for each dataset are determined based on dataset size: USPTO-50K was augmented 20-fold, and USPTO-MIT and USPTO-Full were augmented 5-fold.
During inference, test data were also augmented to generate multiple input sequences. 
We employed beam search to generate $m$ predictions for each augmented input sequence, yielding $n \times m$ predictions.
Both beam size and the number of generations were are to $10$ for all experiments.
Final predictions were selected based on the scores of these generations, 
following the scoring strategy of Zhong et al. \cite{rsmiles}.

\subsection{Training details for reaction prediction}
Both input and output sequences are represented as R-SMILES and tokenized, with reactant and product sequences truncated at $512$ tokens each. 
%
%
The context length of ChemFM is increased from $512$ tokens (used in pre-training) to $\num{1024}$ tokens to accommodate these longer sequences.
Given the large size of the reaction datasets (e.g., USPTO-Full contains approximately 1 million reactions), 
fine-tuning was performed on an NVIDIA HGX H100 node with $8\times \qty{80}{\giga\byte}$ GPUs.
We used empirical hyperparameter settings without conducting hyperparameter searches. 
The detailed hyperparameter settings are provided in Supplementary Table~\ref{table:hyperparameter_reaction}. {{We should note that, in forward reaction prediction, we only reported the results with full-parameter fine-tuning. However, LoRA fine-tuning remains a viable option. While its performance was slightly below that of full-parameter fine-tuning, it was still strong and competitive, so we report the latter as the main result.}}.

\subsection{Comparison with state-of-the-art methods for reaction prediction}
We compared the performance of our adapted ChemFM models with various sequence- and graph-based reaction prediction methods reported in the literature, using the same datasets and splits for a fair comparison.
Methods that are not open-sourced or cannot be reproduced were excluded from the comparison.
For the retro-synthesis task on the USPTO-50K, USPTO-MIT, and USPTO-Full datasets, our model outperformed existing methods by a significant margin (complete results are shown in Supplementary Table~\ref{table:rst_rct_prediction_detail}). 
For forward synthesis on the USPTO-MIT dataset, our initial results were just below the best-reported performance of Chemformer\cite{chemformer}. 
We observed that Chemformer simplified $903$ reactions with multiple products into single-product reactions, which is inconsistent with the original USPTO-MIT dataset. 
When we excluded this portion of the test data to ensure a fair comparison, the top-1 accuracy of our model reached $91.4\%$, surpassing previously reported results of Chemformer.

\section*{Data availability}
\label{sec:data}
The pre-training datasets for ChemFM are sourced from the UniChem database\cite{unichem} (\url{https://ftp.ebi.ac.uk/pub/databases/chembl/UniChem/data/}).
The molecular property prediction datasets are derived from the MoleculeNet\cite{moleculenet} (\url{https://github.com/shenwanxiang/ChemBench}) and ADMET\cite{admet} (\url{https://tdcommons.ai/benchmark/admet_group/overview}) benchmarks. 
%
Datasets for molecular conditional generation tasks are sourced from the GuacaMol\cite{guacamol} database (\url{https://github.com/BenevolentAI/guacamol}) and the MOSES\cite{moses} database (\url{https://github.com/molecularsets/moses}).
Datasets for reaction prediction tasks involving the USPTO series are formatted into Root-aligned SMILES\cite{rsmiles} and are available at \url{https://github.com/otori-bird/retrosynthesis}.

\section*{Code availability}
\label{sec:code}
The ChemFM-1B and ChemFM-3B models are publicly available on the Hugging Face Model Hub at \url{https://huggingface.co/ChemFM}. 
Additionally, the source code for pre-training and fine-tuning these models, along with the model checkpoints, can be accessed on our GitHub repository at \url{https://github.com/TheLuoFengLab/ChemFM} and has been archived on Zenodo at \url{https://zenodo.org/records/17450883}.

\section*{Acknowledgements}
This work was supported as part of the AIM for Composites, an Energy Frontier Research Center funded by the U.S. Department of Energy, Office of Science, Basic Energy Sciences at Clemson University under award \#DE-SC0023389. We would also like to thank Clemson University's Palmetto Cluster team for their invaluable support with cloud computing resources and maintenance.

\section*{Author contributions}
FC designed all models and experiments, performed all experiments and analyses, and drafted the initial version of the paper. 
KZ, TT, and YD assisted with the potential antibiotic screening testing. 
TZ, SP, and GL provided suggestions and feedback on the chemical aspects of this research. 
LL provided suggestions and feedback on the model learning aspects of this research.
FL conceived the study and directed and supervised the whole study. 
All authors contributed to manuscript editing.

\section*{Competing interests}
The authors declare no competing interests.

\bibliography{sn-bibliography}

\setcounter{NAT@ctr}{0} 

\newpage
\begin{appendices}
	\setcounter{page}{1} 


\section*{Supplementary Information}\label{sec:supple_info}

\newcounter{suptable}
\newcounter{supfigure}

\makeatletter
\renewcommand{\thesubsection}{S\arabic{subsection}}
\renewcommand{\thetable}{S\arabic{subsection}.\arabic{suptable}}
\renewcommand{\thefigure}{S\arabic{subsection}.\arabic{supfigure}}
\renewcommand{\tablename}{Table \@gobble}
\renewcommand{\figurename}{Fig. \@gobble}

\makeatother

\newenvironment{sup_table}[1][]{%
	\refstepcounter{suptable} 
	\renewcommand{\thetable}{S\arabic{subsection}.\arabic{suptable}}
	\begin{table}[#1]
		\centering
	}{%
	\end{table}
}

\newenvironment{sup_figure}[1][]{%
	\refstepcounter{supfigure} 
	\renewcommand{\thefigure}{S\arabic{subsection}.\arabic{supfigure}}
	\begin{figure}[#1]
		\centering
	}{%
	\end{figure}
}

\subsection{Supplementary information for pre-training benchmarking}
\label{sup:generation_metrics}

{
\subsubsection*{Definition of benchmark metrics for unconditional molecular generation}
\begin{description}
	\item[\textbf{\upshape Validity:}] 
	The validity score measures the proportion of generated SMILES strings that are valid.
	A SMILES string is considered valid if it is syntactically correct and represents a feasible molecular structure, such as correct atom valency and consistent bond arrangements in aromatic rings. 
	In our experiments, validity is implicitly checked by the \texttt{RDKit} parser when converting a SMILES string into an \texttt{RDKit} molecule object\citemethods{rdkit_methods}.
	
	\item[\textbf{\upshape Uniqueness:}]
	To ensure the model does not collapse into generating a subset of repetitive molecules, 
	we measure the uniqueness score, defined as the proportion of unique SMILES strings among the total generated SMILES strings.
	
	\item[\textbf{\upshape Novelty:}] 
	The novelty score evaluates the model's ability to explore chemical space by generating new molecules that are not present in the training dataset,
	%
	which can be defined as the proportion of novel molecules among generations: 
	
	\item[\textbf{\upshape Internal diversity:}] 
	Generative model often encounter the issue of mode collapse, where the generated molecules are concentrated in a small region of chemical space.
	Internal diversity measures how well the model generates diverse molecules by penalizing high similarity between molecules pairs within the generated set, and it is defined as: 
	\begin{equation*}
		\text{IntDiv}_p (G) =  1- \sqrt[p]{\frac{1}{|G|^2} \sum_{m_1, m_2 \in G}{S(m_1, m_2)^p}},
	\end{equation*} 
	where $S(\cdot)$ is the Tanimoto Similarity between molecule pair $m_1$ and $m_2$ in the generated set $G$.
	We evaluate both $\text{IntDiv}_1$ and  $\text{IntDiv}_2$ in our experiment.

	\item[\textbf{\upshape Sphere exclusion diversity:}]
	{Sphere exclusion diversity is a quantitative measure of molecular diversity derived from the principle of sphere exclusion clustering. 
	In this approach, molecules are embedded in a chosen chemical space (e.g., using structural fingerprints), and representative compounds are sequentially selected while all others within a predefined similarity radius are excluded.
	This process continues until no unassigned molecules remain, ensuring that the resulting set consists of compounds that are mutually dissimilar by at least the specified threshold. 
	Consequently, sphere exclusion diversity provides an effective means of capturing the breadth of chemical space in a compound library, reducing redundancy, and enhancing coverage of structurally distinct scaffolds.
	In our evaluation, we employed the MolScore~\citemethods{molscore_methods} framework with Morgan fingerprints as the representation space, and we set a Tanimoto distance threshold of $0.65$.
	}
	
	\item[\textbf{\upshape  KL similarity:}]
	Kullback-Leibler similarity (KLSim) measures how closely the distribution of generated molecules matches that of the training dataset.
	We compare the distributions of various physicochemical descriptors between the generated set $G$ and the training set $T$ using KL divergence, and the KL similarity score is computed as:
	\begin{equation*}
		\text{KLSim} (G, T; \text{Desp}) = \frac{1}{|\text{Desp}|} \sum_{i=1}^{|\text{Desp}|}  e^{-D_{\text{KL}}(\text{Desp}_i(G),\text{Desp}_i(T))},
	\end{equation*} 
	where Desp is a set of the descriptors measured using \texttt{RDKit} toolbox, and $D_{\text{KL}}$ is the KL divergence between the two distributions.
	We follow the settings established by the GuacaMol\citemethods{guacamol_methods} benchmarking platform, which measures $9$ different molecular descriptors, including molecular complexity, molecular weight, etc. 
	The comparison of these descriptor distributions between the generated set and the training set is illustrated in Supplementary Fig.~\ref{fig:unconditional_generation_desc_dist}.
	Due to the computational expense of using the entire training dataset, we use a representative subset of $\num{100000}$ molecules from the training dataset. 
	Additionally, given the high diversity of the training dataset, a high KL similarity score also suggests that the generated molecules maintain a high degree of diversity\citemethods{guacamol_methods}.
\end{description}
}

\setcounter{suptable}{0}
\setcounter{supfigure}{0}
\input{figures/scaling_law}
\newpage
\noindent
\refstepcounter{supfigure}
\begin{minipage}{1.0\textwidth}
	\tikzsetnextfilename{unconditional_generation_descriptor_distribution}
	\begin{tikzpicture}
		\definecolor{clr1}{RGB}{255, 251, 180}
		\definecolor{clr2}{RGB}{178, 178, 255}
		\definecolor{darkgray176}{RGB}{255, 251, 180}
		\definecolor{darkorange25512714}{RGB}{255,127,14}
		\definecolor{lightgray204}{RGB}{204,204,204}
		\definecolor{train}{HTML}{000075}
		\definecolor{chemfm3b}{HTML}{469990}
		
		\begin{groupplot}[group style={group size=3 by 3, 		vertical sep=40pt,
				horizontal sep=60pt,}, 	width=0.30\textwidth,
			height=0.30\textwidth,
			clip=false,
			every axis/.append style={
				font=\scriptsize
			},
			 legend style={at={(3.1,1.2)}, anchor=south, legend columns=2, /tikz/every even column/.append style={column sep=0.5cm}}, 
			legend cell align={left},
			]
			
\nextgroupplot[
tick pos=left,
xlabel={\texttt{BertzCT}},
xmin=0, xmax=4000,
ylabel={Density},
ymin=0, ymax=0.0012,
xtick={0,2000,4000},
xticklabels={0,2000,4000},
ytick={0,0.0005,0.0010},
yticklabels={0.0000,0.0005,0.0010},
scaled y ticks=false,
]
\addplot [semithick, train]
table {%
	58.3863749710494 8.49667953099538e-05
	168.798164506824 0.000266523070707314
	279.209954042599 0.000497804222957572
	389.621743578374 0.000809770268939356
	500.033533114149 0.00101742755558333
	610.445322649924 0.00112527191324201
	720.857112185699 0.00109092705554265
	831.268901721474 0.000991579455393484
	941.680691257249 0.000805206204765247
	1052.09248079302 0.00060924161961594
	1162.5042703288 0.000456409016702484
	1272.91605986457 0.000340461506429034
	1383.32784940035 0.000242542892850588
	1493.73963893612 0.000166403192011435
	1604.1514284719 0.00011911762986731
	1714.56321800767 8.70251881496963e-05
	1824.97500754345 6.23557886124767e-05
	1935.38679707922 4.31083003002233e-05
	2045.798586615 3.13003654871762e-05
	2156.21037615077 2.64493064104627e-05
	2266.62216568655 1.74798560145793e-05
	2377.03395522232 1.6035999161918e-05
	2487.4457447581 1.55122004770955e-05
	2597.85753429387 1.34535579561907e-05
	2708.26932382965 1.18011615458336e-05
	2818.68111336542 1.16913130354559e-05
	2929.0929029012 8.42221548692937e-06
	3039.50469243697 7.86332860653809e-06
	3149.91648197275 8.92480542093778e-06
	3260.32827150852 7.1287938498407e-06
	3370.7400610443 5.91624443444367e-06
	3481.15185058007 5.49540757761331e-06
	3591.56364011585 4.24228442207718e-06
	3701.97542965162 4.27460689462095e-06
	3812.3872191874 4.15796998979976e-06
	3922.79900872317 2.89345138269849e-06
};\addlegendentry{{Training molecules}};
\addplot [semithick, chemfm3b]
table {%
	32.4163106830544 7.31381466500037e-05
	89.7039607042157 0.000207332773632991
	146.991610725377 0.000394398328350917
	204.279260746538 0.000554269377768445
	261.566910767699 0.000683520101123696
	318.854560788861 0.00083844486263362
	376.142210810022 0.00099193204038395
	433.429860831183 0.00110048421565077
	490.717510852344 0.00115917807631892
	548.005160873505 0.00117667480938708
	605.292810894667 0.00115953602385569
	662.580460915828 0.00110487672291166
	719.868110936989 0.0010377507714333
	777.15576095815 0.000964488682431818
	834.443410979311 0.000869985633186332
	891.731061000473 0.000764606331728405
	949.018711021634 0.000667118841940394
	1006.3063610428 0.000567436032077079
	1063.59401106396 0.000476665987048101
	1120.88166108512 0.00040296138964079
	1178.16931110628 0.00034088473987525
	1235.45696112744 0.000286759038460279
	1292.7446111486 0.000241388322670324
	1350.03226116976 0.000202629719981384
	1407.31991119092 0.000169660916855601
	1464.60756121209 0.000143845780994374
	1521.89521123325 0.000119917186779833
	1579.18286125441 9.79193729879717e-05
	1636.47051127557 8.00894015195449e-05
	1693.75816129673 6.72216064430997e-05
	1751.04581131789 5.94234463879328e-05
	1808.33346133905 5.14102167810462e-05
	1865.62111136021 4.24919676369151e-05
	1922.90876138137 3.45186906289037e-05
	1980.19641140254 2.97290516927e-05
	2037.4840614237 2.69677425064973e-05
	2094.77171144486 2.38652530417235e-05
	2152.05936146602 2.01111465882279e-05
	2209.34701148718 1.6182422379891e-05
	2266.63466150834 1.40075753118825e-05
	2323.9223115295 1.3290286495474e-05
	2381.20996155066 1.28831614159799e-05
	2438.49761157183 1.19638773968855e-05
	2495.78526159299 9.78380125633186e-06
	2553.07291161415 7.77793021654612e-06
	2610.36056163531 6.9383127898672e-06
	2667.64821165647 6.77091754550361e-06
	2724.93586167763 6.81361037989887e-06
	2782.22351169879 7.40414562067726e-06
	2839.51116171995 7.06792936617619e-06
	2896.79881174112 5.62838141233943e-06
	2954.08646176228 5.26757305160714e-06
	3011.37411178344 5.02548405350876e-06
	3068.6617618046 4.52330442449171e-06
	3125.94941182576 4.20531571664521e-06
	3183.23706184692 3.50305236193818e-06
	3240.52471186808 3.03435572877166e-06
	3297.81236188924 3.00274875526781e-06
	3355.10001191041 3.08360390017164e-06
	3412.38766193157 3.10162465828632e-06
	3469.67531195273 3.11833330063157e-06
	3526.96296197389 2.91245095267001e-06
	3584.25061199505 2.27363646404059e-06
	3641.53826201621 1.77140545309699e-06
	3698.82591203737 1.74809992764213e-06
	3756.11356205853 1.86406892777646e-06
	3813.4012120797 1.6824373348871e-06
	3870.68886210086 1.25815103489411e-06
	3927.97651212202 1.33530677008237e-06
	3985.26416214318 1.22344338059425e-06
};\addlegendentry{{Generated molecules}};
\node[anchor=north west, font=\small] at (rel axis cs:-0.18,1.1) {\textbf{a}};	

\nextgroupplot[
tick pos=left,
xlabel={\texttt{MolLogP}},
xmin=-10, xmax=30,
ylabel={Density},
ymin=0, ymax=0.25,
xtick={-10,0,10, 20},
xticklabels={-10,0,10, 20},
ytick={0.0, 0.1, 0.2},
yticklabels={0.0, 0.1, 0.2},
]
\addplot [semithick, train]
table {%
	-9.05636986668415 7.90326189432731e-05
	-7.01382676925084 0.000306636728071298
	-4.97128367181753 0.000370459139940249
	-2.92874057438422 0.00172738881172345
	-0.886197476950912 0.011056847486164
	1.1563456204824 0.0995066584508187
	3.19888871791571 0.240418819278794
	5.24143181534902 0.0955351939269426
	7.28397491278233 0.0229600353720184
	9.32651801021566 0.00642696670164988
	11.369061107649 0.00340426308551175
	13.4116042050823 0.00190922283725036
	15.4541473025156 0.00140899542996832
	17.4966903999489 0.00112014204161213
	19.5392334973822 0.000792800581088616
	21.5817765948155 0.000563194719158335
	23.6243196922488 0.00040569371808851
	25.6668627896821 0.000200705001008782
	27.7094058871155 0.000122772953432671
	29.7519489845488 7.49377758091542e-05
};
\addplot [semithick, chemfm3b]
table {%
	-9.85962372057378 8.49280185531938e-05
	-8.90131662993808 0.000177953969995139
	-7.94300953930239 0.000132891067528396
	-6.9847024486667 0.000311077241698615
	-6.02639535803101 0.000293554113429794
	-5.06808826739531 0.000337717095621018
	-4.10978117675963 0.000846035852531015
	-3.15147408612393 0.00163310220031674
	-2.19316699548824 0.00334761360836032
	-1.23485990485256 0.0092360573302838
	-0.276552814216856 0.0291987341033867
	0.681754276418829 0.0796602704526305
	1.64006136705453 0.162179672561712
	2.59836845769021 0.232642673442962
	3.55667554832591 0.224410934397444
	4.5149826389616 0.140892588770347
	5.4732897295973 0.0691079916018798
	6.43159682023298 0.0343017289447343
	7.38990391086867 0.018681095426507
	8.34821100150437 0.0103877974196837
	9.30651809214005 0.00598772740015513
	10.2648251827758 0.00377090872392304
	11.2231322734114 0.00284868391247056
	12.1814393640471 0.00215293869978497
	13.1397464546828 0.00161859904534316
	14.0980535453185 0.00159458660794134
	15.0563606359542 0.000932816406337381
	16.0146677265899 0.00091941139416748
	16.9729748172256 0.000827780704122291
	17.9312819078613 0.00119637831774608
	18.889588998497 0.000868962162907247
	19.8478960891327 0.000664231422623953
	20.8062031797684 0.000716922796341542
	21.764510270404 0.000428828151509751
	22.7228173610397 0.0004558209799157
	23.6811244516754 0.000323343883662108
	24.6394315423111 0.000277339777558066
	25.5977386329468 0.000204794863453813
	26.5560457235825 0.000104915526455347
	27.5143528142182 0.000138595618052164
	28.4726599048539 0.000100932553692659
	29.4309669954896 3.71063320680328e-05
};
\node[anchor=north west, font=\small] at (rel axis cs:-0.18,1.1) {\textbf{b}};	
			\nextgroupplot[
tick pos=left,
xlabel={\texttt{MolWt}},
xmin=0, xmax=1500,
ylabel={Density},
ymin=0, ymax=0.0045,
xtick={0, 500, 1000, 1500},
xticklabels={0, 500, 1000, 1500},
ytick={0.0, 0.002, 0.004},
yticklabels={0.000, 0.002, 0.004},
scaled y ticks=false
]
\addplot [semithick, train]
table {%
	14.2440627154104 4.73263785660996e-07
	49.7401077639851 4.05635374330358e-06
	85.2361528125599 2.80125329921427e-05
	120.732197861135 0.000145983028624879
	156.228242909709 0.000456859215010919
	191.724287958284 0.000991463850704675
	227.220333006859 0.00181989686035718
	262.716378055433 0.0029259472427132
	298.212423104008 0.00355813511332315
	333.708468152583 0.00395666778975166
	369.204513201158 0.0041567902454994
	404.700558249732 0.00290380459364541
	440.196603298307 0.00184525573988901
	475.692648346882 0.00144037222822711
	511.188693395457 0.000986199133600209
	546.684738444031 0.000670955877971879
	582.180783492606 0.000463688309600701
	617.676828541181 0.000342194167897619
	653.172873589756 0.000258645500536258
	688.66891863833 0.000203697966564245
	724.164963686905 0.000169382904687804
	759.66100873548 0.000129060830303546
	795.157053784054 0.0001007187614517
	830.653098832629 8.40122776082338e-05
	866.149143881204 6.91745867553996e-05
	901.645188929779 6.06249370996719e-05
	937.141233978353 5.31192685912784e-05
	972.637279026928 4.37420575996702e-05
	1008.1333240755 3.33194911729278e-05
	1043.62936912408 2.5890247266631e-05
	1079.12541417265 2.55907852631041e-05
	1114.62145922123 2.45019048020471e-05
	1150.1175042698 1.91057640178007e-05
	1185.61354931838 1.67681213351679e-05
	1221.10959436695 1.59166696031115e-05
	1256.60563941553 1.43345511618647e-05
	1292.1016844641 1.09885879431179e-05
	1327.59772951268 1.10421440350978e-05
	1363.09377456125 8.74158376204271e-06
	1398.58981960982 7.86057349660319e-06
	1434.0858646584 8.01432446795147e-06
	1469.58190970697 8.11897879658771e-06
};
\addplot [semithick, chemfm3b]
table {%
	40.7322322114266 3.7498763437403e-06
	116.922377062072 0.000200942419731991
	193.112521912717 0.00141055996744659
	269.302666763363 0.00345741625942231
	345.492811614008 0.00392410592695418
	421.682956464654 0.00202610410932082
	497.873101315299 0.000946329098152425
	574.063246165945 0.000420988515488514
	650.25339101659 0.000224487251010373
	726.443535867236 0.000144113810378387
	802.633680717881 7.74358815233819e-05
	878.823825568527 6.32547487095283e-05
	955.013970419172 4.59401665625851e-05
	1031.20411526982 2.72360139178728e-05
	1107.39426012046 2.28891252158291e-05
	1183.58440497111 1.63685927681785e-05
	1259.77454982175 1.29314493366892e-05
	1335.9646946724 9.20923903214287e-06
	1412.15483952304 8.21682801936852e-06
	1488.34498437369 4.833741800333e-06
};
\node[anchor=north west, font=\small] at (rel axis cs:-0.18,1.1) {\textbf{c}};	
			
\nextgroupplot[
tick pos=left,
xlabel={\texttt{TPSA}},
xmin=-0.5, xmax=400,
ylabel={Density},
ymin=0, ymax=0.015,
xtick={0, 100, 200, 300, 400},
xticklabels={0, 100, 200, 300, 400},
ytick={0.0, 0.005, 0.010},
yticklabels={0.000, 0.005, 0.010},
scaled y ticks=false
]
\addplot [semithick, train]
table {%
	0.0 0.0
	14.4608444404427 0.00280388618798715
	45.7830693699814 0.00970478886067264
	77.10529429952 0.0121622516140683
	108.427519229059 0.00468033131082766
	139.749744158597 0.00123666844610901
	171.071969088136 0.00047479396507795
	202.394194017675 0.000218301529742714
	233.716418947213 0.000137721583497339
	265.038643876752 8.23981364920266e-05
	296.360868806291 3.6197814100439e-05
	327.683093735829 3.50958063914001e-05
	359.005318665368 2.39938459052715e-05
	390.327543594907 2.00065511424991e-05
};
\addplot [semithick, chemfm3b]
table {%
	0.0 0.0
	14.2321674122915 0.00312965862054021
	28.7805041567808 0.0061889236843255
	43.3288409012701 0.00980610900727886
	57.8771776457594 0.0127345318660753
	72.4255143902488 0.0125645189569629
	86.9738511347381 0.0091837431361874
	101.522187879227 0.00547302702896558
	116.070524623717 0.00300348062378635
	130.618861368206 0.00163619716744007
	145.167198112695 0.000986770092701091
	159.715534857185 0.000678154064351381
	174.263871601674 0.000410278175835009
	188.812208346163 0.000280966621228728
	203.360545090653 0.000211067785580785
	217.908881835142 0.000157435941276726
	232.457218579631 0.00016215250278408
	247.005555324121 9.30008103276893e-05
	261.55389206861 7.87651679291549e-05
	276.102228813099 5.97698242317605e-05
	290.650565557589 5.53439820982549e-05
	305.198902302078 3.67349482733835e-05
	319.747239046567 3.60437102601468e-05
	334.295575791057 2.29975626542687e-05
	348.843912535546 2.53910097724858e-05
	363.392249280035 3.23875954882261e-05
	377.940586024525 1.97305926006978e-05
	392.488922769014 2.02673159363716e-05
};

\node[anchor=north west, font=\small] at (rel axis cs:-0.18,1.1) {\textbf{d}};	
			\nextgroupplot[
tick pos=left,
xlabel={\texttt{NumHAcceptors}},
xmin=-0.5, xmax=20,
ylabel={Density},
ymin=0, ymax=0.3,
xtick={0, 5, 10, 15, 20},
xticklabels={0, 5, 10, 15, 20},
ytick={0.0, 0.1, 0.2, 0.3},
yticklabels={0.0, 0.1, 0.2, },
scaled y ticks=false
]
\draw[draw=none,fill=train] (axis cs:-0.4,0) rectangle (axis cs:0,0.0116956881405487);
\draw[draw=none,fill=train] (axis cs:0.6,0) rectangle (axis cs:1,0.0344189338678318);
\draw[draw=none,fill=train] (axis cs:1.6,0) rectangle (axis cs:2,0.110735495435981);
\draw[draw=none,fill=train] (axis cs:2.6,0) rectangle (axis cs:3,0.197341114137997);
\draw[draw=none,fill=train] (axis cs:3.6,0) rectangle (axis cs:4,0.221303113715766);
\draw[draw=none,fill=train] (axis cs:4.6,0) rectangle (axis cs:5,0.170631609320622);
\draw[draw=none,fill=train] (axis cs:5.6,0) rectangle (axis cs:6,0.11423138156307);
\draw[draw=none,fill=train] (axis cs:6.6,0) rectangle (axis cs:7,0.0615560381791949);
\draw[draw=none,fill=train] (axis cs:7.6,0) rectangle (axis cs:8,0.0333469631071508);
\draw[draw=none,fill=train] (axis cs:8.6,0) rectangle (axis cs:9,0.0169381497481941);
\draw[draw=none,fill=train] (axis cs:9.6,0) rectangle (axis cs:10,0.00932931277093828);
\draw[draw=none,fill=train] (axis cs:10.6,0) rectangle (axis cs:11,0.00497249790611766);
\draw[draw=none,fill=train] (axis cs:11.6,0) rectangle (axis cs:12,0.00372072240978161);
\draw[draw=none,fill=train] (axis cs:12.6,0) rectangle (axis cs:13,0.00223054759122533);
\draw[draw=none,fill=train] (axis cs:13.6,0) rectangle (axis cs:14,0.001789796742485);
\draw[draw=none,fill=train] (axis cs:14.6,0) rectangle (axis cs:15,0.00195877924410558);
\draw[draw=none,fill=train] (axis cs:15.6,0) rectangle (axis cs:16,0.00122621156362192);
\draw[draw=none,fill=train] (axis cs:16.6,0) rectangle (axis cs:17,0.000775574640747933);
\draw[draw=none,fill=train] (axis cs:17.6,0) rectangle (axis cs:18,0.00070195637962942);
\draw[draw=none,fill=train] (axis cs:18.6,0) rectangle (axis cs:19,0.00109611353499139);
\draw[draw=none,fill=chemfm3b] (axis cs:-2.77555756156289e-17,0) rectangle (axis cs:0.4,0.0150336596999306);
\draw[draw=none,fill=chemfm3b] (axis cs:1,0) rectangle (axis cs:1.4,0.0408243356108556);
\draw[draw=none,fill=chemfm3b] (axis cs:2,0) rectangle (axis cs:2.4,0.127111907181742);
\draw[draw=none,fill=chemfm3b] (axis cs:3,0) rectangle (axis cs:3.4,0.21655916358916);
\draw[draw=none,fill=chemfm3b] (axis cs:4,0) rectangle (axis cs:4.4,0.21920565120702);
\draw[draw=none,fill=chemfm3b] (axis cs:5,0) rectangle (axis cs:5.4,0.157833301468147);
\draw[draw=none,fill=chemfm3b] (axis cs:6,0) rectangle (axis cs:6.4,0.100616842931463);
\draw[draw=none,fill=chemfm3b] (axis cs:7,0) rectangle (axis cs:7.4,0.0524668685913239);
\draw[draw=none,fill=chemfm3b] (axis cs:8,0) rectangle (axis cs:8.4,0.028869859222959);
\draw[draw=none,fill=chemfm3b] (axis cs:9,0) rectangle (axis cs:9.4,0.0155066061563541);
\draw[draw=none,fill=chemfm3b] (axis cs:10,0) rectangle (axis cs:10.4,0.00825140626100607);
\draw[draw=none,fill=chemfm3b] (axis cs:11,0) rectangle (axis cs:11.4,0.00464896303973756);
\draw[draw=none,fill=chemfm3b] (axis cs:12,0) rectangle (axis cs:12.4,0.00329049981384022);
\draw[draw=none,fill=chemfm3b] (axis cs:13,0) rectangle (axis cs:13.4,0.00226410537649557);
\draw[draw=none,fill=chemfm3b] (axis cs:14,0) rectangle (axis cs:14.4,0.00173078277669883);
\draw[draw=none,fill=chemfm3b] (axis cs:15,0) rectangle (axis cs:15.4,0.00219366654256015);
\draw[draw=none,fill=chemfm3b] (axis cs:16,0) rectangle (axis cs:16.4,0.00114714672409109);
\draw[draw=none,fill=chemfm3b] (axis cs:17,0) rectangle (axis cs:17.4,0.000694325648791974);
\draw[draw=none,fill=chemfm3b] (axis cs:18,0) rectangle (axis cs:18.4,0.000623886814856556);
\draw[draw=none,fill=chemfm3b] (axis cs:19,0) rectangle (axis cs:19.4,0.00112702134296668);
\node[anchor=north west, font=\small] at (rel axis cs:-0.18,1.1) {\textbf{e}};	
			\nextgroupplot[
tick pos=left,
xlabel={\texttt{NumHDonors}},
xmin=-0.5, xmax=10,
ylabel={Density},
ymin=0, ymax=0.4,
xtick={0, 5, 10, 15},
xticklabels={0, 5, 10, 15},
ytick={0.0, 0.1, 0.2, 0.3},
yticklabels={0.0, 0.1, 0.2, 0.3},
scaled y ticks=false
]
\draw[draw=none,fill=train] (axis cs:-0.4,0) rectangle (axis cs:0,0.227855546183172);
\draw[draw=none,fill=train] (axis cs:0.6,0) rectangle (axis cs:1,0.385825464256486);
\draw[draw=none,fill=train] (axis cs:1.6,0) rectangle (axis cs:2,0.244476166492262);
\draw[draw=none,fill=train] (axis cs:2.6,0) rectangle (axis cs:3,0.0886658923188362);
\draw[draw=none,fill=train] (axis cs:3.6,0) rectangle (axis cs:4,0.02896265355713);
\draw[draw=none,fill=train] (axis cs:4.6,0) rectangle (axis cs:5,0.0102750593747566);
\draw[draw=none,fill=train] (axis cs:5.6,0) rectangle (axis cs:6,0.0048919631227104);
\draw[draw=none,fill=train] (axis cs:6.6,0) rectangle (axis cs:7,0.00249086719833017);
\draw[draw=none,fill=train] (axis cs:7.6,0) rectangle (axis cs:8,0.00181673553249295);
\draw[draw=none,fill=train] (axis cs:8.6,0) rectangle (axis cs:9,0.00108678458496023);
\draw[draw=none,fill=train] (axis cs:9.6,0) rectangle (axis cs:10,0.000825018777303541);
\draw[draw=none,fill=chemfm3b] (axis cs:-2.77555756156289e-17,0) rectangle (axis cs:0.4,0.232383612288774);
\draw[draw=none,fill=chemfm3b] (axis cs:1,0) rectangle (axis cs:1.4,0.382929132592578);
\draw[draw=none,fill=chemfm3b] (axis cs:2,0) rectangle (axis cs:2.4,0.242309469750246);
\draw[draw=none,fill=chemfm3b] (axis cs:3,0) rectangle (axis cs:3.4,0.0887902107737748);
\draw[draw=none,fill=chemfm3b] (axis cs:4,0) rectangle (axis cs:4.4,0.028069681930519);
\draw[draw=none,fill=chemfm3b] (axis cs:5,0) rectangle (axis cs:5.4,0.0113524483111977);
\draw[draw=none,fill=chemfm3b] (axis cs:6,0) rectangle (axis cs:6.4,0.00486246458638912);
\draw[draw=none,fill=chemfm3b] (axis cs:7,0) rectangle (axis cs:7.4,0.00257188209528019);
\draw[draw=none,fill=chemfm3b] (axis cs:8,0) rectangle (axis cs:8.4,0.00194900440032952);
\draw[draw=none,fill=chemfm3b] (axis cs:9,0) rectangle (axis cs:9.4,0.0011151520022504);
\node[anchor=north west, font=\small] at (rel axis cs:-0.18,1.1) {\textbf{f}};	
			\nextgroupplot[
tick pos=left,
xlabel={\texttt{NumRotatableBonds}},
xmin=-0.5, xmax=20,
ylabel={Density},
ymin=0, ymax=0.18,
xtick={0, 5, 10, 15, 20},
xticklabels={0, 5, 10, 15, 20},
ytick={0.0, 0.05, 0.1, 0.15},
yticklabels={0.00, 0.05, 0.10, 0.15 },
scaled y ticks=false
]
\draw[draw=none,fill=train] (axis cs:-0.4,0) rectangle (axis cs:0,0.0164495650571445);
\draw[draw=none,fill=train] (axis cs:0.6,0) rectangle (axis cs:1,0.0381931850290499);
\draw[draw=none,fill=train] (axis cs:1.6,0) rectangle (axis cs:2,0.0822455606255476);
\draw[draw=none,fill=train] (axis cs:2.6,0) rectangle (axis cs:3,0.125141404008787);
\draw[draw=none,fill=train] (axis cs:3.6,0) rectangle (axis cs:4,0.155804302296173);
\draw[draw=none,fill=train] (axis cs:4.6,0) rectangle (axis cs:5,0.158299751930497);
\draw[draw=none,fill=train] (axis cs:5.6,0) rectangle (axis cs:6,0.137903267351048);
\draw[draw=none,fill=train] (axis cs:6.6,0) rectangle (axis cs:7,0.1007553373694);
\draw[draw=none,fill=train] (axis cs:7.6,0) rectangle (axis cs:8,0.0672294911459565);
\draw[draw=none,fill=train] (axis cs:8.6,0) rectangle (axis cs:9,0.0406058887844146);
\draw[draw=none,fill=train] (axis cs:9.6,0) rectangle (axis cs:10,0.0247540167293878);
\draw[draw=none,fill=train] (axis cs:10.6,0) rectangle (axis cs:11,0.0150589264577457);
\draw[draw=none,fill=train] (axis cs:11.6,0) rectangle (axis cs:12,0.0108283239355274);
\draw[draw=none,fill=train] (axis cs:12.6,0) rectangle (axis cs:13,0.00705415337761026);
\draw[draw=none,fill=train] (axis cs:13.6,0) rectangle (axis cs:14,0.00524129290776144);
\draw[draw=none,fill=train] (axis cs:14.6,0) rectangle (axis cs:15,0.00378475040974336);
\draw[draw=none,fill=train] (axis cs:15.6,0) rectangle (axis cs:16,0.003175379538745);
\draw[draw=none,fill=train] (axis cs:16.6,0) rectangle (axis cs:17,0.00233668323086394);
\draw[draw=none,fill=train] (axis cs:17.6,0) rectangle (axis cs:18,0.00211152682748468);
\draw[draw=none,fill=train] (axis cs:18.6,0) rectangle (axis cs:19,0.00302719298711148);
\draw[draw=none,fill=chemfm3b] (axis cs:-2.77555756156289e-17,0) rectangle (axis cs:0.4,0.021165917445754);
\draw[draw=none,fill=chemfm3b] (axis cs:1,0) rectangle (axis cs:1.4,0.042864310802093);
\draw[draw=none,fill=chemfm3b] (axis cs:2,0) rectangle (axis cs:2.4,0.0874284485495151);
\draw[draw=none,fill=chemfm3b] (axis cs:3,0) rectangle (axis cs:3.4,0.12898204941786);
\draw[draw=none,fill=chemfm3b] (axis cs:4,0) rectangle (axis cs:4.4,0.153414501776626);
\draw[draw=none,fill=chemfm3b] (axis cs:5,0) rectangle (axis cs:5.4,0.152994665000973);
\draw[draw=none,fill=chemfm3b] (axis cs:6,0) rectangle (axis cs:6.4,0.132238344409515);
\draw[draw=none,fill=chemfm3b] (axis cs:7,0) rectangle (axis cs:7.4,0.0966034180857491);
\draw[draw=none,fill=chemfm3b] (axis cs:8,0) rectangle (axis cs:8.4,0.0665902085871981);
\draw[draw=none,fill=chemfm3b] (axis cs:9,0) rectangle (axis cs:9.4,0.0394134573046479);
\draw[draw=none,fill=chemfm3b] (axis cs:10,0) rectangle (axis cs:10.4,0.0239614159763253);
\draw[draw=none,fill=chemfm3b] (axis cs:11,0) rectangle (axis cs:11.4,0.0155442006205392);
\draw[draw=none,fill=chemfm3b] (axis cs:12,0) rectangle (axis cs:12.4,0.0110795949087111);
\draw[draw=none,fill=chemfm3b] (axis cs:13,0) rectangle (axis cs:13.4,0.00693242675896249);
\draw[draw=none,fill=chemfm3b] (axis cs:14,0) rectangle (axis cs:14.4,0.00558075713978517);
\draw[draw=none,fill=chemfm3b] (axis cs:15,0) rectangle (axis cs:15.4,0.00404476893617457);
\draw[draw=none,fill=chemfm3b] (axis cs:16,0) rectangle (axis cs:16.4,0.00332797444115629);
\draw[draw=none,fill=chemfm3b] (axis cs:17,0) rectangle (axis cs:17.4,0.00248830088984917);
\draw[draw=none,fill=chemfm3b] (axis cs:18,0) rectangle (axis cs:18.4,0.00199678466469377);
\draw[draw=none,fill=chemfm3b] (axis cs:19,0) rectangle (axis cs:19.4,0.0033484542838711);

\node[anchor=north west, font=\small] at (rel axis cs:-0.18,1.1) {\textbf{g}};	
			
			\nextgroupplot[
tick pos=left,
xlabel={\texttt{NumAliphaticRings}},
xmin=-0.5, xmax=10,
ylabel={Density},
ymin=0, ymax=0.4,
xtick={0, 5, 10, 15, 20},
xticklabels={0, 5, 10, 15, 20},
ytick={0.0, 0.1, 0.2, 0.3},
yticklabels={0.0, 0.1, 0.2, 0.3},
scaled y ticks=false
]
\draw[draw=none,fill=train] (axis cs:-0.4,0) rectangle (axis cs:0,0.393354803988705);
\draw[draw=none,fill=train] (axis cs:0.6,0) rectangle (axis cs:1,0.37309321348474);
\draw[draw=none,fill=train] (axis cs:1.6,0) rectangle (axis cs:2,0.159790556597282);
\draw[draw=none,fill=train] (axis cs:2.6,0) rectangle (axis cs:3,0.0444229187911918);
\draw[draw=none,fill=train] (axis cs:3.6,0) rectangle (axis cs:4,0.0178901471088157);
\draw[draw=none,fill=train] (axis cs:4.6,0) rectangle (axis cs:5,0.00629838207312465);
\draw[draw=none,fill=train] (axis cs:5.6,0) rectangle (axis cs:6,0.00257624020798017);
\draw[draw=none,fill=train] (axis cs:6.6,0) rectangle (axis cs:7,0.00102432709797388);
\draw[draw=none,fill=train] (axis cs:7.6,0) rectangle (axis cs:8,0.000661577270732181);
\draw[draw=none,fill=train] (axis cs:8.6,0) rectangle (axis cs:9,0.000298310081010923);
\draw[draw=none,fill=train] (axis cs:9.6,0) rectangle (axis cs:10,0.000208789501532104);
\draw[draw=none,fill=chemfm3b] (axis cs:-2.77555756156289e-17,0) rectangle (axis cs:0.4,0.349961341888323);
\draw[draw=none,fill=chemfm3b] (axis cs:1,0) rectangle (axis cs:1.4,0.387886455603418);
\draw[draw=none,fill=chemfm3b] (axis cs:2,0) rectangle (axis cs:2.4,0.177556204877951);
\draw[draw=none,fill=chemfm3b] (axis cs:3,0) rectangle (axis cs:3.4,0.0499944774126176);
\draw[draw=none,fill=chemfm3b] (axis cs:4,0) rectangle (axis cs:4.4,0.0214477211796247);
\draw[draw=none,fill=chemfm3b] (axis cs:5,0) rectangle (axis cs:5.4,0.00773162233535159);
\draw[draw=none,fill=chemfm3b] (axis cs:6,0) rectangle (axis cs:6.4,0.00305248466226868);
\draw[draw=none,fill=chemfm3b] (axis cs:7,0) rectangle (axis cs:7.4,0.00105431213663885);
\draw[draw=none,fill=chemfm3b] (axis cs:8,0) rectangle (axis cs:8.4,0.000582381942143366);
\draw[draw=none,fill=chemfm3b] (axis cs:9,0) rectangle (axis cs:9.4,0.000311273107007661);

\node[anchor=north west, font=\small] at (rel axis cs:-0.18,1.1) {\textbf{h}};	
			\nextgroupplot[
tick pos=left,
xlabel={\texttt{NumAromaticRings}},
xmin=-0.5, xmax=10,
ylabel={Density},
ymin=0, ymax=0.4,
xtick={0, 5, 10, 15},
xticklabels={0, 5, 10, 15},
ytick={0.0, 0.1, 0.2, 0.3},
yticklabels={0.0, 0.1, 0.2, 0.3},
scaled y ticks=false
]
\draw[draw=none,fill=train] (axis cs:-0.4,0) rectangle (axis cs:0,0.123699811403878);
\draw[draw=none,fill=train] (axis cs:0.6,0) rectangle (axis cs:1,0.280533241251715);
\draw[draw=none,fill=train] (axis cs:1.6,0) rectangle (axis cs:2,0.327803436429225);
\draw[draw=none,fill=train] (axis cs:2.6,0) rectangle (axis cs:3,0.177570044102871);
\draw[draw=none,fill=train] (axis cs:3.6,0) rectangle (axis cs:4,0.0603517318962644);
\draw[draw=none,fill=train] (axis cs:4.6,0) rectangle (axis cs:5,0.0136109239142557);
\draw[draw=none,fill=train] (axis cs:5.6,0) rectangle (axis cs:6,0.00513285510146778);
\draw[draw=none,fill=train] (axis cs:6.6,0) rectangle (axis cs:7,0.00226191279225979);
\draw[draw=none,fill=train] (axis cs:7.6,0) rectangle (axis cs:8,0.00213003982048584);
\draw[draw=none,fill=train] (axis cs:8.6,0) rectangle (axis cs:9,0.00166972171259798);
\draw[draw=none,fill=train] (axis cs:9.6,0) rectangle (axis cs:10,0.00160609671559701);
\draw[draw=none,fill=chemfm3b] (axis cs:-2.77555756156289e-17,0) rectangle (axis cs:0.4,0.204801494065908);
\draw[draw=none,fill=chemfm3b] (axis cs:1,0) rectangle (axis cs:1.4,0.330341185211961);
\draw[draw=none,fill=chemfm3b] (axis cs:2,0) rectangle (axis cs:2.4,0.286272265397514);
\draw[draw=none,fill=chemfm3b] (axis cs:3,0) rectangle (axis cs:3.4,0.127286784344438);
\draw[draw=none,fill=chemfm3b] (axis cs:4,0) rectangle (axis cs:4.4,0.0366989979316023);
\draw[draw=none,fill=chemfm3b] (axis cs:5,0) rectangle (axis cs:5.4,0.00784183786171858);
\draw[draw=none,fill=chemfm3b] (axis cs:6,0) rectangle (axis cs:6.4,0.00273108821816575);
\draw[draw=none,fill=chemfm3b] (axis cs:7,0) rectangle (axis cs:7.4,0.00137558487459084);
\draw[draw=none,fill=chemfm3b] (axis cs:8,0) rectangle (axis cs:8.4,0.000923750426732534);
\draw[draw=none,fill=chemfm3b] (axis cs:9,0) rectangle (axis cs:9.4,0.000451834447858305);

\node[anchor=north west, font=\small] at (rel axis cs:-0.18,1.1) {\textbf{i}};	
		\end{groupplot}
	\end{tikzpicture}
	\\
	\\
	\captionof{figure}{\textbf{Comparison of physicochemical descriptor distributions between training and generated molecules.}
	The descriptors were computed for 178 million molecules in the training dataset and \num{100000} molecules randomly sampled from the ChemFM-3B model, using \texttt{RDKit}~\protect\citemethods{rdkit_methods}. 
	The descriptors are: 
	\textbf{a}, \texttt{BertzCT},  a topological index quantifying molecular complexity;
	\textbf{b}, \texttt{MolLogP},  the octanol-water partition coefficient;
	\textbf{c}, \texttt{MolWt}, molecular weight;
	\textbf{d}, \texttt{TPSA}, topological polar surface area;
	\textbf{e}, \texttt{NumHAcceptors}, number of hydrogen bond acceptors;
	\textbf{f}, \texttt{NumHDonors}, number of hydrogen bond donors;
	\textbf{g}, \texttt{NumRotatableBonds}, number of rotatable single bonds;
	\textbf{h}, \texttt{NumAliphaticRings}, number of aliphatic (non-aromatic) rings;
	\textbf{i}, \texttt{NumAromaticRings}, number of aromatic rings.
	}
	\label{fig:unconditional_generation_desc_dist}
\end{minipage}
\newpage
\input{figures/tnse_embedding}
\newpage
\noindent
\refstepcounter{suptable}
	\begin{minipage}{0.67\textwidth}
	\captionof{table}{\textbf{Architectures of the ChemFM models.}} 	 \label{table:model_architecture}
	\tablebodyfont
	\begin{tabular}{lcccccc}
		\toprule
		\makecell[c]{Model} &
		$n_\text{params}$ & $n_\text{layers}$ & $n_\text{heads}$  & \makecell[c]{$n_\text{ctx}$ \\ (pre-training)} & $d_\text{model}$ & $d_\text{ff}$ \\
		\midrule
		ChemFM-1B & $970$M & $22$ & $32$  & $512$ & $2048$ & $5632$ \\
		ChemFM-3B & $3.0$B & $30$ & $48$  & $512$ & $3072$ & $8640$ \\
		\midrule
		ChemFM-10M & $9.8$M & $3$ & $8$  & $512$ & $512$ & $1408$ \\
		ChemFM-20M & $20.3$M & $4$ & $10$  & $512$ & $640$ & $1760$ \\
		ChemFM-30M & $29.2$M & $4$ & $12$  & $512$ & $768$ & $2112$ \\
		ChemFM-40M & $39.6$M & $4$ & $14$  & $512$ & $896$ & $2464$ \\
		ChemFM-50M & $49.5$M & $5$ & $14$  & $512$ & $896$ & $2464$ \\
		ChemFM-60M & $56.8$M & $5$ & $15$  & $512$ & $960$ & $2640$ \\
		ChemFM-100M & $97.9$M & $6$ & $18$ & $512$ & $1152$ & $3168$ \\
		ChemFM-200M & $201.1$M & $10$ & $20$  & $512$ & $1280$ & $3520$ \\
		\botrule
	\end{tabular}
	\captionof*{tabledescription}{
		ChemFM-1B and ChemFM-3B are the primary models, while ChemFM-10M to ChemFM-200M are the models used for the pre-training dataset selection experiments.
		$n_\text{params}$ is the actual number of non-embedding trainable parameters, $n_\text{layers}$ is the number of hidden layers in the Transformer decoder, $n_\text{heads}$ is the number of attention heads for each attention layer in the Transformer decoder, 
		$n_\text{ctx}$ is the context length, $d_\text{model}$ is the dimension of the hidden representations, and $d_\text{ff}$ is the dimension of the MLP representations. %
		It should be noted that, for training efficiency, the context lengths during fine-tuning may differ from those used in pre-training, depending on the maximum length of the dataset.}
\end{minipage}
\newpage
\noindent
\refstepcounter{suptable}
\begin{minipage}{0.53\textwidth}
	{
	\captionof{table}{{\textbf{Comparison of molecular diversity between ChemFM-3B generated molecules and the UniChem training dataset.}}}	\label{table:diveristy_comparison}
		\small
	\begin{tabular}{cccc}
	\toprule
	 & IntDiv1 & IntDiv2 & SEDiv \\ 
	\midrule
	UniChem & $0.903/0.903$ & $0.895/0.894$ & $0.697/0.377$ \\
	ChemFM-3B & $0.904/0.904$ & $0.895/0.896$ & $0.701/0.388$ \\
	\botrule
	\end{tabular}
}
		\tablebodyfont
		\captionof*{tabledescription}{{We report three diversity metrics---IntDiv1, IntDiv2, and Sphere Exclusion Diversity (SEDiv) (computed at sample sizes of 10,000/100,000 molecules).}}
\end{minipage}

\newpage
\noindent
\refstepcounter{suptable}
\begin{minipage}{0.58\textwidth}
	{
	\captionof{table}{{\textbf{Novelty of generated molecules for ChemFM-3B and MolGPT.}}} 	\label{table:novelty comparison}
		\small
    \begin{tabular}{lccc}
	\toprule
	Benchmark & Model & Valid / Unique & Novelty (\%) \\
	\midrule
	\multirow{2}{*}{MOSES} 
	& ChemFM-3B & 99,596 / 99,585 & \textbf{100.0} \\
	& MolGPT    & 99,737 / 98,865 & 99.1 \\
	\midrule
	\multirow{2}{*}{GuacaMol} 
	& ChemFM-3B & 99,596 / 99,585 & \textbf{100.0} \\
	& MolGPT    & 93,254 / 87,439 & 93.8 \\
	\botrule
\end{tabular}
}
		\tablebodyfont
		\captionof*{tabledescription}{{Both models generated $\num{100000}$ molecules, and novelty was evaluated separately with respect to MOSES and GuacaMol.}}
\end{minipage}

\newpage

\newpage
\setcounter{suptable}{0}
\setcounter{supfigure}{0}
\subsection{Supplementary information for property prediction tasks}

\noindent
\refstepcounter{supfigure}
\begin{minipage}{0.92\textwidth}
	\tikzsetnextfilename{antibiotics}
	\definecolor{c2}{HTML}{9ecbe2}
	\definecolor{c4}{HTML}{fd9272}
	\begin{tikzpicture}
			\begin{groupplot}[group style={group size=2 by 2, 		vertical sep=60pt,
				horizontal sep=60pt,}, 	width=0.45\textwidth,
			height=0.45\textwidth,
			every axis/.append style={
				font=\small
			},
		clip=false,
		]	
			\nextgroupplot[
			legend cell align={left},
			legend style={fill opacity=0.8, draw opacity=1, text opacity=1, at={(1,1)},
				anchor=north east,
				legend columns=1},
			tick pos=left,
			xlabel={Recall},
			xmin=-0.00, xmax=1.00,
			xtick={0.0, 0.2, 0.4, 0.6, 0.8, 1.0},
			xticklabels={0.0, 0.2, 0.4, 0.6, 0.8, 1.0},
			ylabel={Precision},
			ymin=-0.00, ymax=1.01,
			ytick={0.2, 0.4, 0.6, 0.8, 1.0},
			yticklabels={0.2, 0.4, 0.6, 0.8, 1.0},
			title={Antibiotic activity}
			]
			\foreach \i in {0,...,99}{
					\addplot[blue,mark=none,opacity=0.1] table[x index=1, y index=0, col sep=comma] {./figures/new_antibiotics_data/data_\i.csv};
			}	
			\addplot[black,mark=none, line width=1pt] table [x index=1, y index=0, col sep=comma] {./figures/new_antibiotics_data/data_main.csv};
			\node[anchor=north east, align=right, font=\footnotesize] at (rel axis cs:1,1) 
			{PRC-AUC: $0.428$ \\ $95\%$ CI: $[0.331, 0.551]$};
			\node[anchor=north west, font=\small] at (rel axis cs:-0.15,1.12) {\textbf{a}};	
		\nextgroupplot[
		legend cell align={left},
		legend style={fill opacity=0.8, draw opacity=1, text opacity=1, at={(1,1)},
			anchor=north east,
			legend columns=1},
		tick pos=left,
		xlabel={Recall},
		xmin=-0.00, xmax=1.00,
		xtick={0.0, 0.2, 0.4, 0.6, 0.8, 1.0},
		xticklabels={0.0, 0.2, 0.4, 0.6, 0.8, 1.0},
		ylabel={Precision},
		ymin=-0.00, ymax=1.01,
		ytick={0.2, 0.4, 0.6, 0.8, 1.0},
		yticklabels={0.2, 0.4, 0.6, 0.8, 1.0},
		title={HepG2 cytotoxicity}
		]
		\foreach \i in {0,...,99}{
			\addplot[blue,mark=none,opacity=0.1] table[x index=1, y index=0, col sep=comma] {./figures/new_hepg2_data/data_\i.csv};}
		\addplot[black,mark=none,line width=1pt] table [x index=1, y index=0, col sep=comma] {./figures/new_hepg2_data/data_main.csv};
		\node[anchor=north east, align=right, font=\footnotesize] at (rel axis cs:1,1) 
		{PRC-AUC: $0.461$ \\ $95\%$ CI: $[0.421, 0.500]$};
		\node[anchor=north west, font=\small] at (rel axis cs:-0.15,1.12) {\textbf{b}};	
	
\nextgroupplot[
	legend cell align={left},
	legend style={fill opacity=0.8, draw opacity=1, text opacity=1, at={(1,1)},
		anchor=north east,
		legend columns=1},
	tick pos=left,
	xlabel={Recall},
	xmin=-0.00, xmax=1.00,
	xtick={0.0, 0.2, 0.4, 0.6, 0.8, 1.0},
	xticklabels={0.0, 0.2, 0.4, 0.6, 0.8, 1.0},
	ylabel={Precision},
	ymin=-0.00, ymax=1.01,
	ytick={0.2, 0.4, 0.6, 0.8, 1.0},
	yticklabels={0.2, 0.4, 0.6, 0.8, 1.0},
	title={HSkMC cytotoxicity}
	]
	\foreach \i in {0,...,99}{
		\addplot[blue,mark=none,opacity=0.1] table[x index=1, y index=0, col sep=comma] {./figures/new_hskmc_data/data_\i.csv};
	}	
	\addplot[black,mark=none, line width=1pt] table [x index=1, y index=0, col sep=comma] {./figures/new_hskmc_data/data_main.csv};
\node[anchor=north east, align=right, font=\footnotesize] at (rel axis cs:1,1) 
{PRC-AUC: $0.459$ \\ $95\%$ CI: $[0.395, 0.527]$};
\node[anchor=north west, font=\small] at (rel axis cs:-0.15,1.12) {\textbf{c}};	

\nextgroupplot[
legend cell align={left},
legend style={fill opacity=0.8, draw opacity=1, text opacity=1, at={(1,1)},
	anchor=north east,
	legend columns=1},
tick pos=left,
xlabel={Recall},
xmin=-0.00, xmax=1.00,
xtick={0.0, 0.2, 0.4, 0.6, 0.8, 1.0},
xticklabels={0.0, 0.2, 0.4, 0.6, 0.8, 1.0},
ylabel={Precision},
ymin=-0.00, ymax=1.01,
ytick={0.2, 0.4, 0.6, 0.8, 1.0},
yticklabels={0.2, 0.4, 0.6, 0.8, 1.0},
title={IMR-90 cytotoxicity}
]
\foreach \i in {0,...,99}{
	\addplot[blue,mark=none,opacity=0.1] table[x index=1, y index=0, col sep=comma] {./figures/new_imr90_data/data_\i.csv};
}	
\addplot[black,mark=none,line width=1pt] table [x index=1, y index=0, col sep=comma] {./figures/new_imr90_data/data_main.csv};
\node[anchor=north east, align=right, font=\footnotesize] at (rel axis cs:1,1) 
{PRC-AUC: $0.414$ \\ $95\%$ CI: $[0.380, 0.448]$};
\node[anchor=north west, font=\small] at (rel axis cs:-0.15,1.12) {\textbf{d}};	

%
%
%
	\end{groupplot}
	\end{tikzpicture}
\tikzsetnextfilename{antibiotics_compare}
\begin{tikzpicture}
	\begin{axis}[
        			   ylabel={PRC-AUC $\uparrow$},
        			   xmin=-0.00, xmax=12.00, 
        			   ymin=-0.00, ymax=0.8,
        			   width=0.80\textwidth,
        			   height=0.3\textwidth,
        			   clip=false,
        			   x tick label style={align=center, text width=1.5cm},
        			   xtick style={draw=none},
        			   	every axis plot/.append style={
        			   	ybar,
        			   	bar width=18pt,
        			   	bar shift=0pt,
        			   	fill
        			   },
        		   		nodes near coords={
        		   		\pgfmathprintnumber[zerofill, precision=3]{\pgfplotspointmeta}
        	   			},
        		   		nodes near coords style={font=\scriptsize},
        		   		xtick={1.5, 4.5, 7.5, 10.5},
        		   		xticklabels={Antibiotics\\activity, HepG2\\cytotoxicity, HSkMC\\cytotoxicity, IMR-90\\cytotoxicity},
        		   		ytick={0.0, 0.2, 0.4, 0.6, 0.8},
        		   		yticklabels={0.0, 0.2, 0.4, 0.6, 0.8},
        		   		legend style={at={(1,1)}, anchor=north east, legend columns=1, /tikz/every even column/.append style={column sep=0.5cm}}, 
        		   		font=\footnotesize,
        		   		legend cell align={left},
        		   		legend image code/.code={
        		   			\draw [#1] (0cm,-0.1cm) rectangle (0.2cm,0.20cm);},
		]
        \addplot[ybar,c2]coordinates{(1.1,0.364)}; \addlegendentry{Chemprop};
        \addplot[ybar,c4]coordinates{(1.9,0.428)}; \addlegendentry{ChemFM-1B};
        
        \addplot[ybar,c2]coordinates{(4.1,0.176)}; 
        \addplot[ybar,c4]coordinates{(4.9,0.461)}; 
        
        \addplot[ybar,c2]coordinates{(7.1,0.168)}; 
		\addplot[ybar,c4]coordinates{(7.9,0.459)}; 
        
        \addplot[ybar,c2]coordinates{(10.1,0.335)}; 
		\addplot[ybar,c4]coordinates{(10.9,0.414)}; 
        \node[anchor=north west, font=\small] at (rel axis cs:-0.1,1.1) {\textbf{e}};	
	\end{axis}
\end{tikzpicture}
	
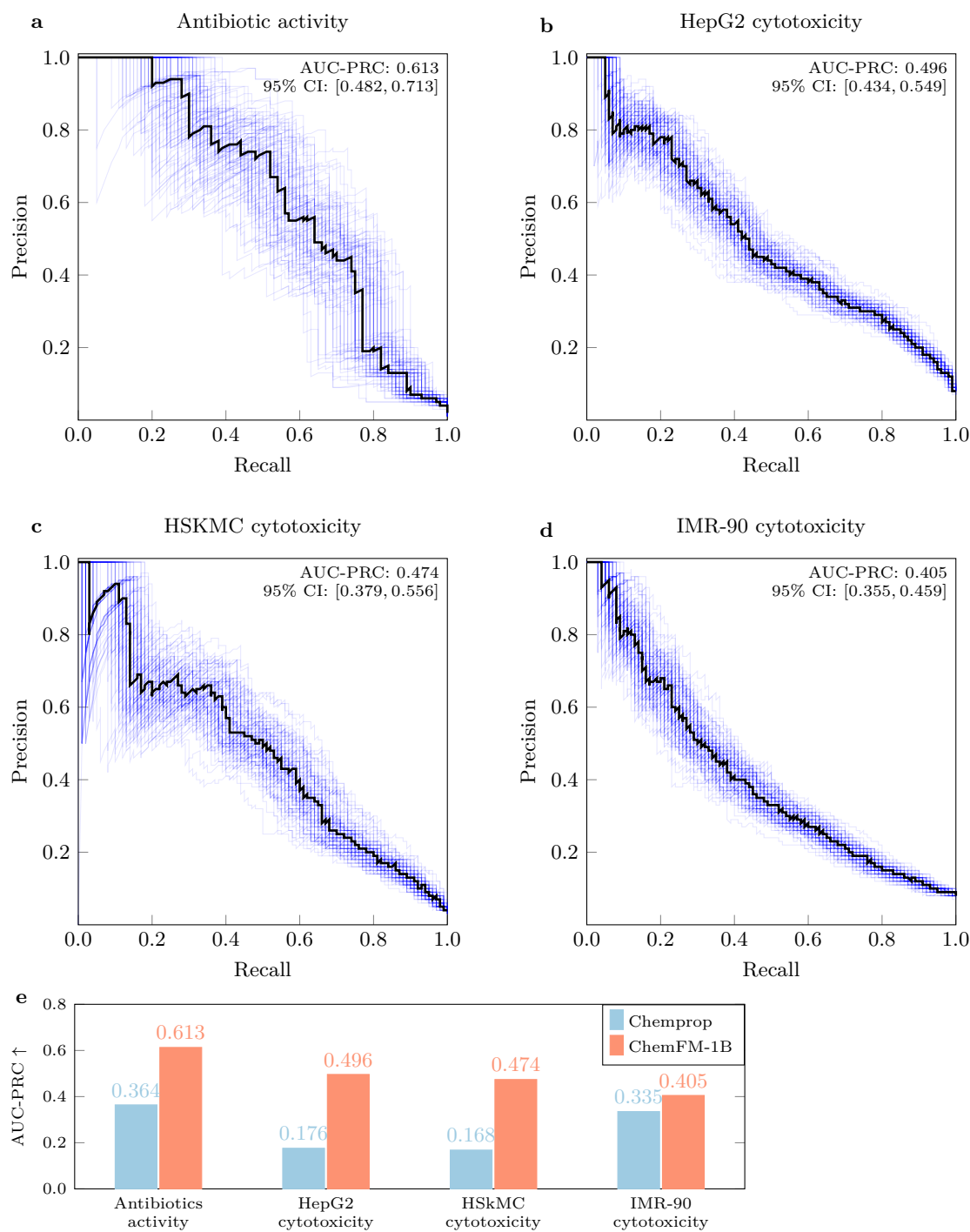
\captionof{figure}{
	\textbf{Precision-recall curve for ChemFM model for predicting antibiotic activity and human cell cytotoxicity.}
	\textbf{a}, Precision-recall curve for antibiotic activity prediction based on \textit{S. aureus} RN4220 growth inhibition.
	\textbf{b, c, d}, Precision-recall curves for human cell cytotoxicity prediction in human liver carcinoma cells (HepG2) (\textbf{b}), human primary skeletal muscle cells (HSkMC) (\textbf{c}), and human lung fibroblasts cells (IMR-90)  (\textbf{d}).
	The black lines represent precision-recall curves for the ChemFM models, while blue curves with $95\%$ confidence intervals (CI) show the variation from $100$ bootstrapping iterations. 
	The datasets are obtained from previous work\protect\citemethods{antibiotics_methods} by screening $\num{39312}$ compounds.
	In the original study, ten Chemprop\protect\citemethods{dmpnn_methods} models were trained and ensembled for each property.
	Here, we trained a single ChemFM model per property using the same datasets for direct performance comparison.
	\textbf{e}, Comparison of PRC-AUC values between ChemFM and Chemprop models for all four properties.}
	\label{fig:precision-recall-antibiotics}
\end{minipage}
\newpage

\noindent
\refstepcounter{supfigure}
\begin{minipage}{0.98\textwidth}
	\tikzsetnextfilename{data_efficiency}
	\definecolor{c2}{HTML}{9ecbe2}
	\definecolor{c4}{HTML}{fd9272}
\begin{tikzpicture}
	\begin{groupplot}[
		group style={
			group size=2 by 1,      
			horizontal sep=1.5cm      
		},
		width=8cm,
		height=6cm,
		xlabel={Percentage of training data},
		ylabel={PRC-AUC ($\uparrow$)},
		xtick={10, 20, 50, 100}, 
		grid=major,
		    legend style={
			at={(axis description cs:0,1), font=\small},
			anchor=north west,
			legend cell align={left},
		}
		]
		
		\nextgroupplot[
		ymin=0.5, ymax=0.8,      
		xmin=0, xmax=105,    
		]
		\addplot[mark=o,blue] coordinates {
			(10,0.5934305493670986)
			(20,0.6524737296120472)
			(50,0.6848537463703128)
			(100,0.739)
		};
		\addplot[mark=square,red] coordinates {
			(10,0.553)
			(20,0.601)
			(50,0.593)
			(100,0.686)
		};
		\legend{ChemFM-3B, Chemprop-RDKit}
		
		\nextgroupplot[
		ymin=0, ymax=0.65,   
		xmin=0, xmax=105, 
		xlabel={Percentage of training data},
		ylabel={Spearman ($\uparrow$)},
		]
		\addplot[mark=o,blue] coordinates {
			(10,0.34729745929346223)
			(20,0.4407433315717449)
			(50,0.47131015781941255)
			(100,0.551)
		};
		\addplot[mark=square,red] coordinates {
			(10,0.05)
			(20,0.142)
			(50,0.189)
			(100,0.329)
		};
		\legend{ChemFM-3B, DeepPurpose}
		
	\end{groupplot}
\end{tikzpicture}
	\captionof{figure}{
		\textbf{Comparison between ChemFM-3B and the previous best method across different percentages of the training dataset.}
		\textbf{a}, CYP2D6\_Substrate\_CarbonMangels classification task, compared with Chemprop-RDKit.
		\textbf{b},Half\_Life\_Obach regression task, compared with DeepPurpose.}
	\label{fig:data_efficiency}
\end{minipage}
\newpage
\noindent
\refstepcounter{suptable}
\begin{minipage}{1.0\textwidth}
	\captionof{table}{\textbf{Performance comparison on $12$ MoleculeNet benchmark datasets for molecular property prediction.}}
	\label{table:rst_prop_prediction}
	\footnotesize 
	\begin{tabular}{llccccccc}
				\toprule
		Category & Dataset & Task metric & MoleculeNet\citemethods{moleculenet_methods} & {ST\citemethods{smilestransformer_methods}} & Chemprop\citemethods{dmpnn_methods} & MMNB\citemethods{MMNB_methods} & {CF\citemethods{chemformer_methods}} & \makecell{ChemFM \\(1B/3B)} \\ 
		\midrule
		
		Pharmacokinetic & BBBP &  ROC-AUC $\uparrow$  & $0.690$ (Weave) &{$0.704$} & $0.738$ & $\underline{0.739}$ & - & $0.745/\mathbf{0.751}$\\
		\midrule
		Bioactivity & BACE & ROC-AUC $\uparrow$  & $0.806$ (Weave)& {$0.701$} & - & $\underline{0.835}$ & {-} & $0.857/\mathbf{0.869}$ \\
		& HIV & ROC-AUC $\uparrow$  & $0.763$ (GC) & {$\underline{0.807}$} & $0.729$ & $0.777$ & {-} & $0.785/\mathbf{0.807}$ \\
		& MUV & PRC-AUC $\uparrow$  & $\underline{0.109}$ (Weave) & {-} & $0.041$ & ${0.096}$ & {-} & $0.122/\mathbf{0.135}$ \\
		& PCBA & PRC-AUC $\uparrow$  & $0.136$ (GC) & {-} & $\underline{0.335}$ & ${0.276}$ & {-} & $0.322/\mathbf{0.346}$ \\
		\midrule
		Toxicity & Tox21 & ROC-AUC $\uparrow$  & $0.829$ (GC) & {$0.802$} & $\underline{0.851}$ & $0.845$ & {-} & $0.863/\mathbf{0.869}$ \\
		& SIDER & ROC-AUC $\uparrow$  & $0.638$ (GC) & {-} & $0.676$ & $\underline{0.680}$ & {-} & $0.702/\mathbf{0.709}$ \\
		& ClinTox & ROC-AUC $\uparrow$  & $0.832$ (Weave) & {$\underline{0.895}$} & $0.864$ & $0.888$ & {-} & $0.899/\mathbf{0.918}$ \\
		\midrule
		Physicochemical & ESOL & RMSE $\downarrow$  & $0.580$ (MPNN) & {$0.720$} & $\underline{0.555}$ & $0.575$ & {$0.633$} & $0.529/\mathbf{0.516}$ \\
		& FreeSolv & RMSE $\downarrow$  & $1.150$ (MPNN) & $1.650$ & {$\underline{1.075}$} & $1.155$ & {$1.230$} &  $0.906/\mathbf{0.830}$ \\
		& Lipop & RMSE $\downarrow$  & $0.655$ (GC) & {$0.921$} & $\underline{0.555}$ & $0.625$ & {$0.598$} & $0.547/\mathbf{0.545}$ \\
		\midrule
		\makecell[l]{Molecular\\ binding} & PDBbind & RMSE $\downarrow$  & $1.440$ (GC) & {-} & $1.391$ & $\underline{0.721}$ & {-} & $0.700/\mathbf{0.697}$ \\
\botrule
	\end{tabular}
\captionof*{tabledescription}{
All methods in this table were evaluated using the \textit{same} dataset splits from {Shen et al. }\citemethods{MMNB_methods}. 
\textbf{Bold} values indicate the best-performing models, 
while \underline{underline} values represent the best performance excluding ChemFM.
The ChemFM results are averaged over three runs with different dataset folds, while values for other models are sourced from {Shen et al. \citemethods{MMNB_methods}}. 
Metrics for classification tasks include ROC-AUC or PRC-AUC, while regression tasks are evaluated using RMSE. An upward arrow ($\uparrow$) indicates that higher values are better, while a downward arrow ($\downarrow$) indicates that lower values are better. 
ST denotes the SMILES Transformer~{smilestransformer}, and CF denotes the Chemformer~{chemformer}.
The brackets following each value in MoleculeNet denote the type of model used: Weave (Weave models), MPNN (Message Passing Neural Networks), and GC (Graph Convolutional Models).
	}

\end{minipage}


\newpage
\noindent
\refstepcounter{suptable}
\begin{minipage}{0.65\textwidth}
		\captionof{table}{
		\textbf{Performance comparison between AttentiveFP\protect\citemethods{attentiveFP_methods} and ChemFM-3B on four MoleculeNet benchmark datasets for molecular property prediction. }} \label{table:rst_prop_prediction_attentive_fp}
	    \tablebodyfont
		\begin{tabular}{lcccc}
			\toprule
			Dataset & Task metric & Split & AttentiveFP\citemethods{attentiveFP_methods} & ChemFM-3B \\ 
			\midrule
				Tox21 & ROC-AUC $\uparrow$ & Random  & $0.848\pm0.007$ & $\textbf{0.861}\pm{0.002}$ \\
				\midrule
				 ESOL & RMSE $\downarrow$ & Random  & $0.545\pm0.021$ & $\textbf{0.539}\pm0.033$ \\
				 \midrule
				 FreeSolv & RMSE $\downarrow$ & Random  & $1.145\pm0.183$ & $\mathbf{0.675}\pm0.050$ \\
				 \midrule
				 Lipophilicity & RMSE $\downarrow$ & Random  & $0.637\pm0.027$ & $\mathbf{0.543}\pm0.024$ \\
		\botrule
	\end{tabular}
\captionof*{tabledescription}{According to the evaluation in Table 2 of from {Shen et al. }\cite{MMNB}, MMNB generally outperformed AttentiveFP, while our ChemFM models performed better than MMNB across all benchmarks. 
	Therefore, for most datasets, we did not directly compare ChemFM with AttentiveFP, as it was reasonable to assume ChemFM's superiority.
	However, these four datasets represent cases where MMNB performed worse than AttentiveFP, making them essential for a direct comparison. 
	We reran AttentiveFP with three different random seeds and used identical data splits from AttentiveFP to fine-tune our ChemFM-3B model.
	No hyperparameter tuning was conducted for ChemFM-3B in these experiments; we used the hyperparameters optimized for the standard MoleculeNet benchmark datasets, as presented in Supplementary Table~\ref{table:hyperparameter_moleculenet}. \textbf{Bold} values indicate the best-performing models.	}
\end{minipage}

\newpage
\noindent
\refstepcounter{suptable}
\begin{minipage}{1.0\textwidth}
	\captionof{table}{\textbf{Performance comparison with Pretrain GNNs\protect\citemethods{pretraingnns_methods}, ChemBerta-2\protect\citemethods{chemberta2_methods}, $3$D InfoMax\protect\citemethods{3dinfomax_methods}, Mole-BERT\protect\citemethods{molebert_methods}, GraphMVP\protect\citemethods{graphmvp_methods}, and MoleculeSDE\protect\citemethods{moleculeSDE_methods}  on $11$ MoleculeNet benchmark datasets for molecular property prediction.}}		\label{table:rst_prop_prediction_molebert}
		\footnotesize
		\begin{tabular}{lccccccc}
			
			\toprule
			Dataset & {\makecell{Pretrain \\GNNs\citemethods{pretraingnns_methods}}} & {\makecell{Chem\\BERTa\citemethods{chemberta2_methods}}} & \makecell{$3$D \\ InfoMax\citemethods{3dinfomax_methods}} & Mole-BERT\citemethods{molebert_methods} & GraphMVP\citemethods{graphmvp_methods} & MoleculeSDE\citemethods{moleculeSDE_methods} &  \thead{ChemFM} \\ 
			\midrule
				%
				%
				BBBP   
				& {$0.688\pm0.008$} & {$0.728$} & $0.708\pm0.005$ & $0.719\pm0.016$ & $0.724\pm0.016$ & $\underline{0.732}\pm0.005$ & $\mathbf{0.733}\pm0.007$ \\
				Tox21 & {$0.767\pm0.004$} & {-} & $0.749\pm0.008$ & $0.768\pm0.005$ & $0.759\pm0.005$  & $\underline{0.768}\pm0.003$ & $\mathbf{0.795}\pm0.007$  \\
				ToxCast & {$0.642\pm0.005$} & {-} & $0.635\pm0.008$  & $0.643\pm0.002$ & $0.631\pm0.002$ & $\underline{0.652}\pm0.003$ &  
				$\mathbf{0.688}\pm0.004$ \\
				SIDER & {$0.610\pm0.007$} & {-} & $0.568 \pm0.021$ & $0.628\pm0.011$ & $\underline{0.639}\pm0.012$ & $0.608\pm0.004$ & $\mathbf{0.650}\pm0.014$  \\
				ClinTox & {$0.718\pm0.041$} &  {$0.563$} & $0.627\pm0.033$& $0.789\pm0.030$  & $0.791\pm0.028$ & $\underline{0.870}\pm0.005$ & $\mathbf{0.941}\pm0.017$  \\
				MUV & {$0.758\pm0.017$} & {-} & $0.762\pm0.014$ &$0.786\pm0.018$ & $0.777\pm0.006$ & $\underline{0.809\pm0.004}$ & $\mathbf{0.812}^\star\pm0.002$  \\
				HIV & {$0.773\pm0.010$} & {-} & $0.761\pm0.013$ &$0.782\pm0.008$ & $0.770\pm0.012$ & $\underline{0.788}\pm0.009$ & $\mathbf{0.793}\pm0.013$ \\
				BACE & {$0.796\pm0.012$} & {$0.799$} & $0.786\pm0.019$ & $0.808\pm0.014$ & $\underline{0.812}\pm0.009$& $0.804\pm0.009$ & $\mathbf{0.853}\pm0.005$  \\
				\midrule
				ESOL 
				& {-} & {$0.889$} &$\underline{0.894}\pm0.028$  & $1.015\pm0.030$ & $1.029\pm0.033$ & - & $\mathbf{0.844}\pm0.032$  \\
				FreeSolv & {-} & {-} & $\underline{2.337}\pm0.227$ & - & - & - & $\mathbf{2.130}\pm0.151$  \\
				Lipop & {-} &  {$0.798$} &$0.695\pm0.012 $& $\underline{0.676}\pm0.017$ & $0.681\pm0.010$ & - & $\mathbf{0.625}\pm0.010$ \\
		\botrule
	\end{tabular}
	\captionof*{tabledescription}{All methods in this table were evaluated using the \textit{same} datasets, 
		with a deterministic scaffold split applied to define the train/validation/test sets. 
		Results are averaged over three random training seeds, except for ChemBERTa-2, whose standard deviation were not reported in the original paper. 
		A horizontal line separates classification and regression tasks: classification tasks are evaluated using ROC-AUC (higher values are better), while regression tasks are evaluated using RMSE (lower values are better).
%
		%
		\textbf{Bold} values indicate the best-performing models, and \underline{underlined} values highlight the best performance excluding ChemFM. 
		%
		%
		All ChemFM results were fine-tuned from ChemFM-3B, except those marked with a $\star$, which were fine-tuned from ChemFM-1B.}
\end{minipage}

%
%
\newpage
\noindent
\refstepcounter{suptable}
\begin{minipage}{1.0\textwidth}
	\captionof{table}{\textbf{Performance comparison on 22 ADMET benchmark datasets with the previous best models for each molecular property prediction task.}} 		\label{table:rst_prop_prediction_admet}
		\footnotesize
		\begin{tabular}{lllll}
			\toprule
			Category & Dataset & Task metric & Previous best
			& ChemFM \\ 
			\midrule
			Absorption & Caco2\_Wang & MAE $\downarrow$ & $0.330\pm0.024$@Chemprop-RDKit\citemethods{chemprop-rdkit_methods} 
			& $\mathbf{0.322}^\star\pm0.026$ \\ 
			& Bioavailability\_Ma & ROC-AUC $\uparrow$ & $0.672\pm0.021$@DeepPurpose\citemethods{deeppurpose_methods} 
			& $\mathbf{0.715}\pm0.011$ \\ 
			& Lipophilicity\_AstraZeneca & MAE $\downarrow$ & $0.467\pm0.006$@Chemprop-RDKit\citemethods{chemprop-rdkit_methods} & $\mathbf{0.460}\pm0.006$\\ 
			& Solubility\_AqSolDB & MAE $\downarrow$ & $0.761\pm0.025$@Chemprop-RDKit\citemethods{chemprop-rdkit_methods} & $\mathbf{0.725}\pm0.011$\\ 
			& HIA\_Hou & ROC-AUC $\uparrow$ &$0.981\pm0.002$@Chemprop-RDKit\citemethods{chemprop-rdkit_methods} & $\mathbf{0.984}^\star\pm0.004$\\ 
			& Pgp\_Broccatelli & ROC-AUC $\uparrow$ & $0.929\pm0.006$@AttrMasking\citemethods{attrmasking_methods} & $\mathbf{0.931}\pm0.003$ \\
			\midrule
			Distribution & BBB\_Martins & ROC-AUC $\uparrow$ & $0.897\pm0.004$@ContextPred\citemethods{attrmasking_methods} & $\mathbf{0.908}\pm0.010$\\ 
			& PPBR\_AZ & MAE $\downarrow$ & $7.788\pm0.210$@Chemprop\cite{dmpnn} & $\mathbf{7.505}\pm0.073$\\ 
			& VDss\_Lombardo & Spearman $\uparrow$ & $0.561\pm0.025$@DeepPurpose\citemethods{deeppurpose_methods} & $\mathbf{0.662}\pm0.013$\\
			\midrule 
			Metabolism & CYP2C9\_Veith & PRC-AUC $\uparrow$ & $0.777\pm0.003$@Chemprop-RDKit\citemethods{chemprop-rdkit_methods}& $\mathbf{0.788}\pm0.005$ \\ 
			& CYP2D6\_Veith & PRC-AUC $\uparrow$ & $0.673\pm0.007$@Chemprop-RDKit\citemethods{chemprop-rdkit_methods} & $\mathbf{0.704}\pm0.003$\\ 
			& CYP3A4\_Veith & PRC-AUC $\uparrow$ & $0.876\pm0.003$@Chemprop-RDKit\citemethods{chemprop-rdkit_methods} & $\mathbf{0.878}\pm0.003$\\ 
			& CYP2C9\_Substrate\_CarbonMangels & PRC-AUC $\uparrow$ & $0.400\pm0.008$@Chemprop-RDKit\citemethods{chemprop-rdkit_methods} & $\mathbf{0.414}\pm0.027$ \\ 
			& CYP2D6\_Substrate\_CarbonMangels & PRC-AUC $\uparrow$ & $0.686\pm0.031$@Chemprop-RDKit\citemethods{chemprop-rdkit_methods} & $\mathbf{0.739}\pm0.024$\\   
			& CYP3A4\_Substrate\_CarbonMangels & ROC-AUC $\uparrow$ & $0.619\pm0.030$@Chemprop-RDKit\citemethods{chemprop-rdkit_methods} & 
			$\mathbf{0.654}\pm0.022$
			\\
			\midrule
			Excretion &Half\_Life\_Obach & Spearman $\uparrow$ &$0.329\pm0.083$@DeepPurpose\citemethods{deeppurpose_methods}  & $\mathbf{0.551}\pm0.020$\\ 
			& Clearance\_Hepatocyte\_AZ & Spearman $\uparrow$ & $0.439\pm0.026$@ContextPred\citemethods{attrmasking_methods} & $\mathbf{0.495}\pm0.030$\\ 
			& Clearance\_Microsome\_AZ & Spearman $\uparrow$ & $0.599\pm0.025$@Chemprop-RDKit\citemethods{chemprop-rdkit_methods} & 
			$\mathbf{0.611}\pm0.016$
			\\
			\midrule
			Toxicity & LD50\_Zhu & MAE $\downarrow$ & $0.606\pm0.024$@Chemprop\cite{dmpnn} & $\mathbf{0.541}\pm0.015$\\ 
			& hERG & ROC-AUC $\uparrow$ & $0.841\pm0.020$@DeepPurpose\citemethods{deeppurpose_methods} & $\mathbf{0.848}\pm0.009$\\ 
			& AMES & ROC-AUC $\uparrow$ & $0.850\pm0.004$@Chemprop-RDKit\citemethods{chemprop-rdkit_methods} & $\mathbf{0.854}\pm0.007$ \\ 
			& DILI & ROC-AUC $\uparrow$ & $0.919\pm0.008$@ContextPred\citemethods{attrmasking_methods} & $\mathbf{0.920}\pm0.012$\\ 
		\botrule
	\end{tabular}
	\captionof*{tabledescription}{	We used the default metrics and scaffold split methods, with the data pre-split by the benchmark.
		\textbf{Bold} values indicate the best-performing models. 
		Evaluation metrics include ROC-AUC, PRC-AUC, MAE, and Spearman’s rank correlation coefficient, with an upward arrow ($\uparrow$) indicating that higher values are better and a downward arrow ($\downarrow$) indicating that lower values are better. 
		All ChemFM results were fine-tuned from the ChemFM-3B model, except those marked with a $\star$, which were fine-tuned from the ChemFM-1B model.}
\end{minipage}

\newpage
\newpage
\noindent
\refstepcounter{suptable}
\begin{minipage}{0.85\textwidth}
	{
	\captionof{table}{{\textbf{Performance comparison on 22 ADMET benchmark datasets between ChemFM fine-tuned from pre-trained weights and from random initialization.}}} 		\label{table:without_pretraining}
	\footnotesize
		\begin{tabular}{lllll}
	\toprule
	Category & Dataset & Task metric & \makecell{ChemFM \\(w/o pretraining)}
	& ChemFM \\ 
	\midrule
	Absorption & Caco2\_Wang & MAE $\downarrow$ & $0.390\pm0.017$ 
	& $\mathbf{0.322}^\star\pm0.026$ \\ 
	& Bioavailability\_Ma & ROC-AUC $\uparrow$ & $0.575\pm0.031$
	& $\mathbf{0.715}\pm0.011$ \\ 
	& Lipophilicity\_AstraZeneca & MAE $\downarrow$ & $0.779\pm0.033$ & $\mathbf{0.460}\pm0.006$\\ 
	& Solubility\_AqSolDB & MAE $\downarrow$ & $0.857\pm0.019$ & $\mathbf{0.725}\pm0.011$\\ 
	& HIA\_Hou & ROC-AUC $\uparrow$ &$0.948\pm0.009$ & $\mathbf{0.984}^\star\pm0.004$\\ 
	& Pgp\_Broccatelli & ROC-AUC $\uparrow$ & $0.902\pm0.004$ & $\mathbf{0.931}\pm0.003$ \\
	\midrule
	Distribution & BBB\_Martins & ROC-AUC $\uparrow$ & $0.842\pm0.023$ & $\mathbf{0.908}\pm0.010$\\ 
	& PPBR\_AZ & MAE $\downarrow$ & $10.012\pm0.317$ & $\mathbf{7.505}\pm0.073$\\ 
	& VDss\_Lombardo & Spearman $\uparrow$ & $0.583\pm0.038$ & $\mathbf{0.662}\pm0.013$\\
	\midrule 
	Metabolism & CYP2C9\_Veith & PRC-AUC $\uparrow$ & $0.694\pm0.004$& $\mathbf{0.788}\pm0.005$ \\ 
	& CYP2D6\_Veith & PRC-AUC $\uparrow$ & $0.576\pm0.008$ & $\mathbf{0.704}\pm0.003$\\ 
	& CYP3A4\_Veith & PRC-AUC $\uparrow$ & $0.797\pm0.006$ & $\mathbf{0.878}\pm0.003$\\ 
	& CYP2C9\_Substrate\_CarbonMangels & PRC-AUC $\uparrow$ & $0.349\pm0.033$ & $\mathbf{0.414}\pm0.027$ \\ 
	& CYP2D6\_Substrate\_CarbonMangels & PRC-AUC $\uparrow$ & $0.658\pm0.030$ & $\mathbf{0.739}\pm0.024$\\   
	& CYP3A4\_Substrate\_CarbonMangels & ROC-AUC $\uparrow$ & $0.562\pm0.011$ & 
	$\mathbf{0.654}\pm0.022$
	\\
	\midrule
	Excretion &Half\_Life\_Obach & Spearman $\uparrow$ &$0.327\pm0.083$  & $\mathbf{0.551}\pm0.020$\\ 
	& Clearance\_Hepatocyte\_AZ & Spearman $\uparrow$ & $0.208\pm0.053$ & $\mathbf{0.495}\pm0.030$\\ 
	& Clearance\_Microsome\_AZ & Spearman $\uparrow$ & $0.283\pm0.027$ & 
	$\mathbf{0.611}\pm0.016$
	\\
	\midrule
	Toxicity & LD50\_Zhu & MAE $\downarrow$ & $0.667\pm0.026$ & $\mathbf{0.541}\pm0.015$\\ 
	& hERG & ROC-AUC $\uparrow$ & $0.820\pm0.024$ & $\mathbf{0.848}\pm0.009$\\ 
	& AMES & ROC-AUC $\uparrow$ & $0.767\pm0.005$ & $\mathbf{0.854}\pm0.007$ \\ 
	& DILI & ROC-AUC $\uparrow$ & $0.825\pm0.011$ & $\mathbf{0.920}\pm0.012$\\ 	
\botrule
\end{tabular}}
\end{minipage}
\newpage
\newpage
\noindent
\refstepcounter{suptable}
\begin{minipage}{0.78\textwidth}
	{
	\captionof{table}{{\textbf{Performance comparison on 11 MoleculeNet benchmark datasets between ChemFM models pre-trained on UniChem and ZINC20.}}}		\label{table:comparison_pretraining}
	\tablebodyfont
	\begin{tabular}{llccc}
	\toprule
	Category & Dataset & Task metric & \makecell{ChemFM-1B \\ (ZINC20)} & \makecell{ChemFM (1B/3B) \\ (UniChem)} \\ 
	\midrule
	
	Pharmacokinetic & BBBP &  ROC-AUC $\uparrow$  & $0.739$ & $\underline{0.745}/\mathbf{0.751}$\\
	\midrule
	Bioactivity & BACE & ROC-AUC $\uparrow$  & $0.457$ & $\underline{0.857}/\mathbf{0.869}$ \\
	& HIV & ROC-AUC $\uparrow$  & $0.721$ & $\underline{0.785}/\mathbf{0.807}$ \\
	& MUV & PRC-AUC $\uparrow$  & $0.033$ & $\underline{0.122}/\mathbf{0.135}$ \\
	& PCBA & PRC-AUC $\uparrow$  & $0.209$ & $\underline{0.322}/\mathbf{0.346}$ \\
	\midrule
	Toxicity & Tox21 & ROC-AUC $\uparrow$  & $0.816$ & $\underline{0.863}/\mathbf{0.869}$ \\
	& SIDER & ROC-AUC $\uparrow$  & $0.592$ & $\underline{0.702}/\mathbf{0.709}$ \\
	& ClinTox & ROC-AUC $\uparrow$  & $\underline{0.900}$ & $0.899/\mathbf{0.918}$ \\
	\midrule
	Physicochemical & ESOL & RMSE $\downarrow$  & $0.541$ & $\underline{0.529}/\mathbf{0.516}$ \\
	& FreeSolv & RMSE $\downarrow$  & $\underline{0.905}$ & $0.906/\mathbf{0.830}$ \\
	& Lipophilicity & RMSE $\downarrow$  & $0.626$ & $\underline{0.547}/\mathbf{0.545}$ \\
	\midrule
	Molecular binding & PDBbind-Full & RMSE $\downarrow$  & $0.744$ & $\underline{0.700}/\mathbf{0.697}$ \\
	\botrule
\end{tabular}
}
	\end{minipage}

\newpage
\noindent
\refstepcounter{suptable}
\begin{minipage}{0.68\textwidth}
	{
	\captionof{table}{{\textbf{Performance comparison of ChemFM-3B and baseline methods on odor prediction\protect\citemethods{odor_methods} and chromatographic retention time prediction\protect\citemethods{smrt_methods}.}}}		\label{table:additional_property_prediction}
	\tablebodyfont
	\begin{tabular}{lccc}
	\toprule
	Task & Metric & Model & Performance \\
	\midrule
	\multirow{2}{*}{Odor prediction} & \multirow{2}{*}{ROC-AUC $\uparrow$} 
	& ChemFM-3B & $\mathbf{0.902}$ \\
	& & MPNN (OpenPOM) & 0.887 \\
	\midrule
	\multirow{2}{*}{Retention time } & \multirow{2}{*}{MAE $\downarrow$} 
	& ChemFM-3B & $\mathbf{39}$ \\
	& & \makecell{(Fingerprints + Descriptors) \\+ Regression NN} & 57 \\
	\botrule
\end{tabular}
	\captionof*{tabledescription}{{Odor prediction results are reported as the mean ROC-AUC over 5-fold random cross-validation, using identical splits as in the baseline.
		Retention time prediction results are reported as mean absolute error (MAE) in second  from a single run using the same $75\%/25\%$ train/test random split as in the baseline.
		\textbf{Bold} values indicate the best-performing models. 
		}}
	}
\end{minipage}

\newpage
\begin{sup_table}[!ht]
	\caption{\textbf{Summary of $12$ MoleculeNet benchamark datasets evaluated in this work.}}
		\footnotesize
		\begin{tabular}[t]{p{0.12\linewidth}lp{0.2\linewidth}lllll}
			\toprule
			Category & Dataset & Dataset description & \makecell[lt]{Number of \\ molecules} & \makecell[lt]{Number of \\ tasks} & \makecell[lt]{Task \\ type} & \makecell[lt]{Task \\ metric} & \makecell[lt]{Split \\ method} \\ 
			\midrule
			Pharmacokinetic & \makecell[lt]{BBBP} & predicts binary labels for blood-brain barrier permeability   & $\num{2039}$ & $1$ & C & ROC-AUC  & Scaffold\\
				\midrule
				Bioactivity & BACE & predicts binary labels for inhibitors of human $\beta$-secretase 1 (BACE1) &  $\num{1513}$ &  1 & C& ROC-AUC  &  Scaffold\\
				\cmidrule{2-8}
				& HIV & predicts binary labels for inhibitors of HIV replication  & $\num{41127}$ & 1 & C & ROC-AUC & Scaffold \\
				\cmidrule{2-8}
				& MUV &  predicts binary labels for a series of bioactivities, which is used for validating virtual screening methods & $\num{93087}$ & 17 & C & PRC-AUC & Random \\
				\cmidrule{2-8}
				& PCBA &  predicts binary labels for bioactivities across multiple assays from the PubChem  & $\num{437929}$ &  128 & C
				& PRC-AUC & Random  \\
				\midrule
				Toxicity & Tox21 & predicts molecular toxicity with binary labels from assays developed by the Tox21 challenge.  & $\num{7831}$ & 12 & C & ROC-AUC & Random \\
				\cmidrule{2-8}
				& SIDER & predicts adverse drug reactions with binary labels based on marketed drugs and their recorded side effects & $\num{1427}$ & 27 & C & ROC-AUC & Random \\
				\cmidrule{2-8}
				& ClinTox & predicts drug toxicity with binary labels comparing FDA-approved drugs and those failing clinical trials due to toxicity  & $\num{1427}$ & 2 & C & ROC-AUC & Random  \\
				\midrule
				Physicochemical & ESOL & predicts water solubility with continuous labels measured in $\log \unit[per-mode=symbol]{\mole\per\liter}$ & $\num{1128}$ & 1 & R & RMSE & Random \\
				\cmidrule{2-8}
				& FreeSolv & predicts hydration free energy with continuous labels measured in $\unit[per-mode=symbol]{\kilo\cal\per\mole}$ for experimental and calculated free energies & $\num{642}$ & 1 & R & RMSE & Random  \\
				\cmidrule{2-8}
				& Lipophilicity & predicts lipophilicity with continuous labels for experimental logP values, important for absorption and distribution in drugs & $\num{4200}$ & 1 & R & RMSE & Random   \\
				\midrule
				Molecular binding & PDBbind-Full & predicts continuous labels of experimentally measured binding affinities for protein-ligand complexes & $\num{9880}$ & 1 & R & RMSE & Time \\
		\botrule
	\end{tabular}
	\tablebodyfont
	\footnotetext{
	We used the recommended task metrics and split methods, 
	with additional details and dataset sources provided in Wu et al. \citemethods{moleculenet_methods} and {Shen et al. }\citemethods{MMNB_methods}.
	The task type ``C'' indicates classification task, while ``R'' indicates regression task.
	For classification tasks, the evaluation metrics are primarily Receiver Operating Characteristic Area Under the Curve (ROC-AUC) or Precision-Recall Curve Area Under the Curve (PRC-AUC). For regression tasks, the evaluation is based on Root Mean Squared Error (RMSE).}
	\label{table:dataset_molecule_net}
\end{sup_table}

\noindent
\refstepcounter{suptable}
\begin{minipage}{0.81\textwidth}
	\footnotesize
	\captionof{table}{\textbf{Hyperparameters for fine-tuning MoleculeNet benchamark datasets for molecular property prediction.}}
	\label{table:hyperparameter_moleculenet}
		\begin{tabular}{lcccccc}
			\toprule
			Dataset & BBBP & BACE & HIV & MUV & PCBA & Tox21 \\
			\midrule
			Optimizer & \multicolumn{6}{c}{AdamW}\\
			\midrule
			LR scheduler & \multicolumn{6}{c}{Cosine}\\
			\midrule
			Warmup ratio & \multicolumn{6}{c}{$0.05$} \\
			\midrule
			Minimum LR & \multicolumn{6}{c}{$0.1\times\text{LR}$ } \\
			\midrule
			Attention dropout & $0.2$/$0.2$ & $0.4$/$0.2$ & $0.2$/$0.0$ & $0.0$/$0.0$ & $0.0$/$0.2$ & $0.0$/$0.0$\\
			\midrule
			Learning rate (LR) & \makecell[c]{$\num{4e-4}$/\\$\num{2e-4}$} & \makecell[c]{$\num{8e-4}$/\\$\num{2e-4}$} & \makecell[c]{$\num{8e-5}$/\\$\num{8e-5}$} & \makecell[c]{$\num{2e-4}$/\\$\num{8e-5}$} & \makecell[c]{$\num{1e-4}$/\\$\num{1e-4}$} & \makecell[c]{$\num{4e-4}$/\\$\num{8e-5}$}\\
			\midrule
			Epochs & $20$/$20$ & $20$/$20$ & $10$/$10$ & $5$/$5$ & $5$/$10$ & $5$/$5$\\
			\midrule
			Batch size & $16$/$16$ & $64$/$16$ & $8$/$8$ & $16$/$8$ & $8$/$8$ & $32$/$8$\\
			\midrule
			Weight decay & $0.01$/$0.05$ & $0.01$/$0.01$ & $0.1$/$0.05$ & $0.1$/$0.01$ & $0.05$/$0.01$ & $0.05$/$0.01$\\
			\midrule
			LoRA $\alpha$ & \multicolumn{6}{c}{$1.0$ }\\
			\midrule
			LoRA rank & $16$/$4$ & $32$/$32$ & $2$/$1$ & $1$/$16$ & $16$/$8$ & $32$/$32$\\	
			\midrule
			LoRA dropout & $0.2$/$0.2$ & $0.6$/$0.6$ & $0.2$/$0.4$ & $0.4$/$0.1$ & $0.4$/$0.1$ & $0.6$/$0.4$\\
			\midrule
			\makecell[lt]{Number of \\trainable parameters} &
			\makecell[ct]{$12.6$M/\\$6.5$M} & \makecell[ct]{$25.2$M/\\$51.9$M} & \makecell[ct]{$1.6$M/\\$1.6$M} & \makecell[ct]{$0.8$M/\\$26.0$M} & \makecell[ct]{$12.6$M/\\$13.0$M} &
			\makecell[ct]{$25.2$M/\\$51.9$M}
			\\
			\botrule
	\end{tabular}
\end{minipage}


\noindent
\begin{minipage}{0.81\textwidth}
\footnotesize
		\begin{tabular}{lcccccc}
			\toprule
			Dataset & SIDER & ClinTox & ESOL & FreeSolv & Lipophilicity & PDBbind-Full \\
			\midrule
			Optimizer & \multicolumn{6}{c}{AdamW}\\
			\midrule
			LR scheduler & \multicolumn{6}{c}{Cosine}\\
			\midrule
			Warmup ratio & \multicolumn{6}{c}{$0.05$} \\
			\midrule
			Minimum LR & \multicolumn{6}{c}{$0.1\times\text{LR}$ } \\
			\midrule
			Attention dropout & $0.2$/$0.2$ & $0.2$/$0.0$ & $0.0$/$0.0$ & $0.2$/$0.0$ & $0.0$/$0.0$ & $0.0$/$0.2$\\
			\midrule
			Learning rate (LR) & \makecell[c]{$\num{4e-4}$/\\$\num{4e-4}$} & \makecell[c]{$\num{1e-4}$/\\$\num{8e-5}$} & \makecell[c]{$\num{8e-4}$/\\$\num{2e-4}$} & \makecell[c]{$\num{8e-4}$/\\$\num{8e-4}$} & \makecell[c]{$\num{2e-4}$/\\$\num{8e-5}$} & \makecell[c]{$\num{4e-4}$/\\$\num{4e-4}$}\\
			\midrule
			Epochs & $20$/$10$ & $20$/$20$ & $50$/$50$ & $50$/$50$ & $20$/$20$ & $5$/$5$\\
			\midrule
			Batch size & $16$/$8$ & $8$/$8$ & $32$/$8$ & $8$/$8$ & $8$/$8$ & $64$/$8$\\
			\midrule
			Weight decay & $0.05$/$0.05$ & $0.1$/$0.1$ & $0.05$/$0.01$ & $0.1$/$0.01$ & $0.05$/$0.01$ & $0.01$/$0.1$\\
			\midrule
			LoRA $\alpha$ & \multicolumn{6}{c}{$1.0$ }\\
			\midrule
			LoRA rank & $8$/$4$ & $32$/$16$ & $8$/$32$ & $4$/$4$ & $8$/$32$ & $32$/$2$\\	
			\midrule
			LoRA dropout & $0.6$/$0.1$ & $0.1$/$0.2$ & $0.4$/$0.6$ & $0.4$/$0.4$ & $0.4$/$0.4$ & $0.2$/$0.1$\\
			\midrule
			\makecell[lt]{Number of \\trainable parameters} &
			\makecell[ct]{$6.3$M/\\$6.5$M} & \makecell[ct]{$25.2$M/\\$26.0$M} & \makecell[ct]{$6.3$M/\\$51.9$M} & \makecell[ct]{$3.2$M/\\$6.5$M} 
			& \makecell[ct]{$6.3$M/\\$51.9$M} & \makecell[ct]{$25.2$M/\\$3.2$M}
			\\
			\botrule
		\end{tabular}
\captionof*{tabledescription}{
			For each hyperparameter, the value before the slash corresponds to ChemFM-1B, and the value after the slash corresponds to ChemFM-3B.}
\end{minipage}

\newpage
\refstepcounter{suptable}
\noindent
{
	\footnotesize
\begin{longtable}[l]{p{0.10\linewidth}lp{0.35\linewidth}lll}
			\caption{\textbf{Summary of $22$ ADMET benchamark datasets evaluated in this work.}}
		\label{table:dataset_admet}\\\toprule
		Category & Dataset & Dataset description & \makecell[lt]{Number of \\ molecules}  & \makecell[lt]{Task \\ type} & \makecell[lt]{Task \\ metric}  \\ 
		\endfirsthead
		\caption*{\textbf{\textbf{\tablename\ \thetable{} Continued:} Summary of $22$ ADMET benchamark datasets evaluated in this work.}}\\\toprule
		Category & Dataset & Dataset description & \makecell[lt]{Number of \\ molecules}  & \makecell[lt]{Task \\ type} & \makecell[lt]{Task \\ metric}  \\ 
		\midrule 
		\endhead	
		\multicolumn{6}{r}{{Continued on next page}} \\ 
		\endfoot
		\bottomrule
		\endlastfoot
		
			\midrule
			Absorption & \makecell[lt]{Caco2\_Wang} & predicts drug permeability, measured in $\unit[per-mode=symbol]{\centi\meter\per\second}$, using the Caco-2 cell line as an in vitro model to simulate human intestinal tissue permeability   & $\num{906}$ & R & MAE \\
			\cmidrule{2-6}
			& \makecell[lt]{Bioavailability\_\\Ma} & predicts oral bioavailability with binary labels, indicating the rate and extent a drug becomes available at its site of action   & $\num{640}$ & C & ROC-AUC \\
			\cmidrule{2-6}
			& \makecell[lt]{Lipophilicity\_\\AstraZeneca} & predicts lipophilicity with continuous labels, measured as a log-ratio, indicating a drug's ability to dissolve in lipid environments   & $\num{4200}$ & R & MAE \\
			\cmidrule{2-6}
			& \makecell[lt]{Solubility\_\\AqSolDB} & predicts aqueous solubility with continuous labels, measured in $\unit{\log \mole\per\liter}$, indicating a drug's ability to dissolve in water   & $\num{9982}$ & R & MAE \\
			\cmidrule{2-6}
			& \makecell[lt]{HIA\_Hou} & predicts human intestinal absorption (HIA) with binary labels, indicating a drug's ability to be absorbed into the bloodstream   & $\num{578}$ & C & ROC-AUC \\
			\cmidrule{2-6}
			& \makecell[lt]{Pgp\_Broccatelli} & predicts P-glycoprotein (Pgp) inhibition with binary labels, indicating a drug's potential to alter bioavailability and overcome multidrug resistance  & $\num{1212}$ & C & ROC-AUC \\
			\midrule
			Distribution 
			& \makecell[lt]{BBB\_Martins} & predicts blood-brain barrier permeability with binary labels, indicating a drug's ability to penetrate the barrier to reach the brain  & $\num{1915}$ & C & ROC-AUC \\
			\cmidrule{2-6}
			& \makecell[lt]{PPBR\_AZ} & predicts plasma protein binding rate with continuous labels, indicating the percentage of a drug bound to plasma proteins in the blood    & $\num{1797}$ & R & MAE \\
			\cmidrule{2-6}
			& \makecell[lt]{VDss\_Lombardo} & predicts the volume of distribution at steady state (VDss) in $\unit[per-mode=symbol]{\liter\per\kilogram}$, indicating drug concentration in tissues versus blood     & $\num{1130}$ & R & Spearman \\
			\midrule
			Metabolism
			& \makecell[lt]{CYP2C9\_Veith} & predicts CYP2C9 inhibition with binary labels, indicating the drug's ability to inhibit the CYP2C9 enzyme involved in metabolism  & $\num{12092}$ & C & PRC-AUC \\
			\cmidrule{2-6}
			& \makecell[lt]{CYP2D6\_Veith} & predicts CYP2D6 inhibition with binary labels, indicating the drug's potential to inhibit the CYP2D6 enzyme involved in metabolism  & $\num{13130}$ & C & PRC-AUC \\
			\cmidrule{2-6}
			& \makecell[lt]{CYP3A4\_Veith} & predicts CPY3A4 inhibition with binary labels, indicating the drug's ability to inhibit the CPY3A4 enzyme involved in metabolism  & $\num{12328}$ & C & PRC-AUC \\
			\cmidrule{2-6}
			& \makecell[lt]{CYP2C9\_Substrate\_\\CarbonMangels} & predicts whether a drug is a substrate of the CYP2C9 enzyme with binary labels, indicating its potential to be metabolized   & $\num{666}$ & C & PRC-AUC \\
			\cmidrule{2-6}
			& \makecell[lt]{CYP2D6\_Substrate\_\\CarbonMangels} & predicts whether a drug is a substrate of the CYP2D6 enzyme with binary labels, indicating its potential to be metabolized   & $\num{664}$ & C & PRC-AUC \\
			\cmidrule{2-6}
			& \makecell[lt]{CYP3A4\_Substrate\_\\CarbonMangels} & predicts whether a drug is a substrate of the CYP3A4 enzyme with binary labels, indicating its potential to be metabolized   & $\num{667}$ & C & ROC-AUC \\
			\midrule
			Excretion & \makecell[lt]{Half\_Life\_Obach} & predicts the half-life duration of a drug, measured in hours, indicating the time for its concentration to reduce by half   & $\num{667}$ & R & Spearman \\
			\cmidrule{2-6}
			& \makecell[lt]{Clearance\_Hepatocyte\_\\AZ} & predicts drug clearance, measured in 
			$\unit{\mu\liter\per(\minute\times 10^6 \text{cells})}$, from hepatocyte experiments, indicating the rate at which the drug is removed from body   & $\num{1020}$ & R & Spearman \\
			\cmidrule{2-6}
			& \makecell[lt]{Clearance\_Microsome\_\\AZ} & predicts drug clearance, measured in 
			$\unit{\milli\liter\per(\minute\times\gram)}$, from microsome experiments, indicating the rate at which the drug is removed from body   & $\num{1102}$ & R & Spearman \\
			\midrule
			\pagebreak
			Toxicity
			& \makecell[lt]{LD50\_Zhu} & predicts the acute toxicity of a drug, measured as the dose leading to lethal effects in $\unit{\log \kilogram\per\mole}$  & $\num{7385}$ & R & MAE \\
			\cmidrule{2-6}
			& \makecell[lt]{hERG} & predicts whether a drug blocks the hERG channel, which is crucial for heart rhythm, potentially leading to adverse effects   & $\num{648}$ & C & ROC-AUC \\
			\cmidrule{2-6}
			& \makecell[lt]{AMES} & predicts whether a drug is mutagenic with binary labels, indicating its ability to induce genetic alterations   & $\num{7255}$ & C & ROC-AUC \\
			\cmidrule{2-6}
			& \makecell[lt]{DILI} & predicts whether a drug can cause liver injury with binary labels, indicating its potential for hepatotoxicity  & $\num{475}$ & C & ROC-AUC \\			

\end{longtable}
\captionof*{tabledescription}{All datasets are single-task and use the default metrics and scaffold split methods, with the data pre-split by the benchmark.  
			Additional details and dataset sources are provided in Huang et al\citemethods{admet_methods}. 
			Task type ``C'' refers to classification tasks, while ``R'' refers to regression tasks. 
			For classification tasks, the primary evaluation metric is Receiver Operating Characteristic Area Under the Curve (ROC-AUC ) or Precision-Recall Curve Area Under the Curve (PRC-AUC). 
			For regression tasks, the evaluation is based on Mean Absolute Error (MAE) or Spearman's rank correlation coefficient.}
}

\newpage
\noindent
\refstepcounter{suptable}
\begin{minipage}{0.89\textwidth}
	\captionof{table}{\textbf{Hyperparameters for fine-tuning ADMET benchamark datasets for molecular property prediction.}}
	\label{table:hyperparameter_admet}
		\footnotesize
		\begin{tabular}{lcccccc}
			\toprule
			Dataset & \makecell[c]{Caco2\_ \\Wang $\star$} & \makecell[c]{Bioavailability\_\\Ma} & \makecell[c]{Lipophilicity\_\\AstraZeneca} & \makecell[c]{Solubility\_\\AqSolDB} & \makecell[c]{HIA\_\\Hou $\star$} & \makecell[c]{Pgp\_\\Broccatelli} \\
			\midrule
			Optimizer & \multicolumn{6}{c}{AdamW}\\
			\midrule
			LR scheduler & \multicolumn{6}{c}{Cosine}\\
			\midrule
			Warmup ratio & \multicolumn{6}{c}{$0.05$} \\
			\midrule
			Minimum LR & \multicolumn{6}{c}{$0.1\times\text{LR}$ } \\
			\midrule
			Attention dropout & $0.2$ & $0.2$ & $0.0$ & $0.0$ & $0.2$ & $0.2$\\
			\midrule
			Learning rate (LR) & $\num{2e-4}$ & $\num{4e-4}$ & $\num{2e-4}$ & $\num{4e-4}$ & $\num{2e-4}$ & $\num{4e-5}$\\
			\midrule
			Epochs & $50$ & $20$ & $20$ & $20$ & $50$ & $50$\\
			\midrule
			Batch size & $16$ & $16$ & $8$ & $8$ & $8$ & $8$\\
			\midrule
			Weight decay & $0.1$ & $0.05$ & $0.1$ & $0.1$ & $0.01$ & $0.1$\\
			\midrule
			LoRA $\alpha$ & \multicolumn{6}{c}{$1.0$ }\\
			\midrule
			LoRA rank & $16$ & $8$ & $8$ & $8$ & $2$ & $1$\\	
			\midrule
			LoRA dropout & $0.4$ & $0.6$ & $0.6$ & $0.2$ & $0.4$ & $0.4$\\
			\midrule
			\makecell[lt]{Number of \\trainable parameters} 
			& \makecell[cc]{$12.6$M} & \makecell[cc]{$13.0$M}
			& \makecell[cc]{$13.0$M} & \makecell[cc]{$13.0$M}
			& \makecell[cc]{$1.6$M} & \makecell[cc]{$1.6$M}
			\\
			\botrule
		\end{tabular}
%

		\begin{tabular}[t]{lcccccc}
			\toprule
			Dataset & \makecell[c]{BBB\_ \\Martins} & \makecell[c]{PPBR\_\\AZ} & \makecell[c]{VDss\_\\Lombardo} & \makecell[c]{CYP2C9\_\\Veith} & \makecell[c]{CYP2D6\_\\Veith} & \makecell[c]{CYP2A4\_\\Veith} \\
			\midrule
			Optimizer & \multicolumn{6}{c}{AdamW}\\
			\midrule
			LR scheduler & \multicolumn{6}{c}{Cosine}\\
			\midrule
			Warmup ratio & \multicolumn{6}{c}{$0.05$} \\
			\midrule
			Minimum LR & \multicolumn{6}{c}{$0.1\times\text{LR}$ } \\
			\midrule
			Attention dropout & $0.2$ & $0.0$ & $0.2$ & $0.0$ & $0.2$ & $0.1$ \\
			\midrule
			Learning rate (LR) & $\num{4e-4}$ & $\num{2e-5}$ & $\num{2e-4}$ & $\num{2e-5}$ & $\num{2e-5}$ & $\num{7e-5}$ \\
			\midrule
			Epochs & $5$ & $20$ & $20$ & $20$ & $20$ & $20$ \\
			\midrule
			Batch size & $16$ & $8$ & $8$ & $16$ & $16$ & $8$ \\
			\midrule
			Weight decay & $0.05$ & $0.01$ & $0.05$ & $0.05$ & $0.1$ & $0.1$ \\
			\midrule
			LoRA $\alpha$ & \multicolumn{6}{c}{$1.0$ }\\
			\midrule
			LoRA rank & $8$ & $16$ & $16$ & $8$ & $16$ & $16$ \\	
			\midrule
			LoRA dropout & $0.4$ & $0.1$ & $0.2$ & $0.6$ & $0.6$ & $0.2$ \\
			\midrule
			\makecell[lt]{Number of \\trainable parameters} 
			& \makecell[cc]{$13.0$M} & \makecell[cc]{$26.0$M}
			& \makecell[cc]{$26.0$M} & \makecell[cc]{$13.0$M}
			& \makecell[cc]{$26.0$M} & \makecell[cc]{$26.0$M}	
			\\
			\botrule
		\end{tabular}

		\begin{tabular}[t]{lcccc}
			\toprule
			Dataset & \makecell[c]{CYP2C9\_Substrate\_\\CarbonMangels} & \makecell[c]{CYP2D6\_Substrate\_\\CarbonMangels} & \makecell[c]{CYP3A4\_Substrate\_\\CarbonMangels} & \makecell[c]{Half\_Life\_\\Obach} \\
			\midrule
			Optimizer & \multicolumn{4}{c}{AdamW}\\
			\midrule
			LR scheduler & \multicolumn{4}{c}{Cosine}\\
			\midrule
			Warmup ratio & \multicolumn{4}{c}{$0.05$} \\
			\midrule
			Minimum LR & \multicolumn{4}{c}{$0.1\times\text{LR}$ } \\
			\midrule
			Attention dropout & $0.2$ & $0.2$ & $0.4$ & $0.2$  \\
			\midrule
			Learning rate (LR) & $\num{2e-4}$ & $\num{2e-4}$ & $\num{1e-4}$ & $\num{8e-4}$  \\
			\midrule
			Epochs & $20$ & $20$ & $20$ & $20$  \\
			\midrule
			Batch size & $16$ & $16$ & $16$ & $16$  \\
			\midrule
			Weight decay & $0.05$ & $0.01$ & $0.05$ & $0.05$  \\
			\midrule
			LoRA $\alpha$ & \multicolumn{4}{c}{$1.0$ }\\
			\midrule
			LoRA rank & $16$ & $1$ & $32$ & $4$  \\	
			\midrule
			LoRA dropout & $0.4$ & $0.4$ & $0.4$ & $0.1$  \\
			\midrule
			\makecell[lt]{Number of \\trainable parameters} 
			& \makecell[cc]{$26.0$M} & \makecell[cc]{$1.6$M}
			& \makecell[cc]{$52.0$M} & \makecell[cc]{$6.5$M}
			\\
			\botrule
		\end{tabular}
\begin{adjustbox}{minipage=3.0cm, right}
	continued on next page
\end{adjustbox}
\end{minipage}
\newpage
\noindent
\begin{minipage}{0.87\textwidth}
	\captionof*{table}{\textbf{\tablename\ \thetable{} Continued:} \textbf{Hyperparameters for fine-tuning ADMET benchamark datasets for molecular property prediction.}} 
		\footnotesize
		\begin{tabular}[t]{lccccccc}
			\toprule
			Dataset & \makecell[c]{Clearance\_\\Hepatocyte\_AZ} & \makecell[c]{Clearance\_\\Hepatocyte\_AZ} & \makecell[c]{LD50\_\\Zhu} & \makecell[c]{hERG} & \makecell[c]{AMES} & \makecell[c]{DILI}\\
			\midrule
			Optimizer & \multicolumn{6}{c}{AdamW}\\
			\midrule
			LR scheduler & \multicolumn{6}{c}{Cosine}\\
			\midrule
			Warmup ratio & \multicolumn{6}{c}{$0.05$} \\
			\midrule
			Minimum LR & \multicolumn{6}{c}{$0.1\times\text{LR}$ } \\
			\midrule
			Attention dropout & $0.0$ & $0.0$ & $0.0$ & $0.2$ & $0.2$ & $0.3$ \\
			\midrule
			Learning rate (LR) & $\num{1e-4}$ & $\num{2e-4}$ & $\num{4e-4}$ & $\num{1e-4}$ & $\num{4e-5}$ & $\num{4e-4}$ \\
			\midrule
			Epochs & $20$ & $20$ & $20$ & $20$ & $20$ & $10$ \\
			\midrule
			Batch size & $8$ & $8$ & $16$ & $16$ & $16$ & $16$ \\
			\midrule
			Weight decay & $0.1$  & $0.01$ & $0.1$ & $0.1$ & $0.01$ & $0.01$ \\
			\midrule
			LoRA $\alpha$ & \multicolumn{6}{c}{$1.0$ }\\
			\midrule
			LoRA rank & $1$ & $2$ & $8$ & $8$ & $4$ & $8$ \\	
			\midrule
			LoRA dropout & $0.4$  & $0.1$ & $0.4$ & $0.5$ & $0.4$ & $0.3$ \\
			\midrule
			\makecell[lt]{Number of \\trainable parameters} 
			& \makecell[cc]{$1.6$M} & \makecell[cc]{$3.2$M}
			& \makecell[cc]{$13.0$M} & \makecell[cc]{$13.0$M}
			& \makecell[cc]{$6.5$M} & \makecell[cc]{$13.0$M}
			\\
			\botrule
		\end{tabular}
	\captionof*{tabledescription}{The hyperparameters listed are for the ChemFM-3B model, except those marked with a $\star$, which are for the ChemFM-1B model. }
\end{minipage}
\newpage
\noindent
\refstepcounter{suptable}
\begin{minipage}{0.85\textwidth}
	\tablebodyfont
	\captionof{table}{\textbf{Reasons for excluding certain methods from comparison in ADMET benchmark.}} \label{tab:exclusion_reasons}
   \begin{tabular}{l p{10cm}}
	\hline
	\textbf{Method} & \textbf{Reason for Exclusion} \\ 
	\midrule
	ZairaChem & Uses both training and validation datasets for training in all runs\footnotemark[1]. \\ \midrule
	MapLight series & Uses both training and validation datasets for training in all runs\footnotemark[1].  \\ \midrule
	Random Forest & Uses both training and validation datasets for training in all runs\footnotemark[1].  \\ \midrule
	SimGCN & Different hyperparameters are used across different datasets, and no hyperparameter search criteria are found or stated\footnotemark[2]. \\ \midrule
	CFA & Optimized using the test dataset. \\ \midrule
	DeepMol (AutoML) & Optimized using the test dataset. \\ \midrule
	BaseBoosting & Uses both training and validation datasets for training in all runs\footnotemark[1]. \\ \midrule
	XGBoost & Uses both training and validation datasets for training in all runs\footnotemark[1].  \\ \midrule 
	MolMapNet-D & Uses both training and validation datasets for training in all runs\footnotemark[1]. \\ \midrule
	Basic ML & Uses both training and validation datasets for training in all runs\footnotemark[1]. \\ \midrule
	RFStacker & Uses both training and validation datasets for training in all runs\footnotemark[1]. \\ \midrule
	Innoplexus ADME & Uses both training and validation datasets for training in all runs\footnotemark[1]. \\ \midrule
	Euclia ML model & Uses both training and validation datasets for training in all runs\footnotemark[1]. \\ \midrule
	ContextPred\footnotemark[3] & Data information leakage during pre-training\footnotemark[4]. \\ \midrule
	AttrMasking\footnotemark[3] & Data information leakage during pre-training\footnotemark[4]. \\ \midrule
	Lantern series\footnotemark[5]  & Different hyperparameters are used across different datasets, and no hyperparameter search criteria are found or stated\footnotemark[2]. \\
	\botrule
\end{tabular}
\captionof*{tabledescription}{{1. Lack of variations in the training dataset, and since the dataset sizes are normally less than 1000, including the validation dataset can considerably improve performance.}\\
2. Different hyperparameters are used across different datasets. It could be optimized using the test dataset, but this is not confirmed. \\
3. The information leakage for ContextPred and AttrMasking occurred only on CYP-based benchmarks. \\
4. The method provider confirmed the data information leakage in this GitHub issue: \url{{https://github.com/mims-harvard/TDC/issues/166}}. \\
5. Lantern series methods are only applied to the BBB\_Martins dataset.
}
\end{minipage}
\newpage
\noindent
\refstepcounter{suptable}
\begin{minipage}{0.68\textwidth}
	\captionof{table}{\textbf{Hyperparameters for fine-tuning ChemFM-1B for predicting antibiotic activity and human cell cytotoxicity.}}
	\label{table:hyperparameter_antibiotics}
	\tablebodyfont
	\begin{tabular}{lcccc}
		\toprule
		Hyperparameters & \makecell[c]{Antibiotic \\ activity} & \makecell[c]{HepG2 \\ cytotoxicity} & \makecell[c]{HSkMC \\ cytotoxicity} & \makecell[c]{IMR-90 \\ cytotoxicity} \\
		\midrule
		Optimizer & \multicolumn{4}{c}{AdamW} \\
		\midrule
		LR scheduler & \multicolumn{4}{c}{Cosine} \\
		\midrule
		Warmup ratio & \multicolumn{4}{c}{$0.05$} \\
		\midrule
		Minimum LR & \multicolumn{4}{c}{$0.1\times\text{LR}$} \\
		\midrule
		Attention dropout & \multicolumn{4}{c}{$0.1$} \\
		\midrule
		Learning rate (LR) & \multicolumn{4}{c}{$\num{2e-4}$}\\
		\midrule
		Epochs & \multicolumn{4}{c}{$5$} \\
		\midrule
		Batch size & \multicolumn{4}{c}{$8$}\\
		\midrule
		Weight decay & \multicolumn{4}{c}{$0.05$}  \\
		\midrule
		LoRA $\alpha$ & \multicolumn{4}{c}{$1.0$}\\
		\midrule
		LoRA rank & \multicolumn{4}{c}{$2$} \\
		\midrule
		LoRA dropout & \multicolumn{4}{c}{$0.1$} \\
		\midrule
		\makecell[lt]{Number of \\trainable parameters} & \multicolumn{4}{c}{$1.6$M} \\
		\bottomrule
	\end{tabular}
\end{minipage}

\newpage
\setcounter{suptable}{0}
\setcounter{supfigure}{0}
\subsection{Supplementary information for conditional molecular generation}

\noindent
\refstepcounter{supfigure}
\begin{minipage}{1.0\textwidth}
	\tikzsetnextfilename{cond_gen_prop_1}
	\begin{tikzpicture}
		\begin{groupplot}[group style={group size=2 by 2, 		vertical sep=40pt,
		horizontal sep=60pt,}, 	width=0.49\textwidth,
	height=0.49\textwidth,
	every axis/.append style={
		font=\small
	}]
\input{figures/results_mol_gen_prop_distribution_1d_1}

\input{figures/results_mol_gen_prop_distribution_1d_2}		
\input{figures/results_mol_gen_prop_distribution_1d_3}	
\input{figures/results_mol_gen_prop_distribution_1d_4}
\end{groupplot}
	\end{tikzpicture}
\\
\begin{adjustbox}{minipage=3.5cm, right}
	continued on next page
\end{adjustbox}
\\
\captionof{figure}{\textbf{Distribution of properties for generated molecules on the GuacaMol dataset\protect\citemethods{guacamol_methods}.} 
		Distribution of molecular properties for generated molecules conditioned on different property combinations. Each subplot corresponds to a specific combination of properties, with the target values indicated in the axis labels.
	}
\label{fig:rst_mol_gen_prop_dist_1d}
\end{minipage}

\newpage
\noindent
\refstepcounter{supfigure}
\begin{minipage}{1.0\textwidth}
	\tikzsetnextfilename{cond_gen_prop_2}

\captionof*{figure}{\textbf{\figurename\ \thefigure{} Continued:} \textbf{Distribution of properties for generated molecules on the GuacaMol dataset\protect\citemethods{guacamol_methods}.} 
	Distribution of molecular properties for generated molecules conditioned on different property combinations. Each subplot corresponds to a specific combination of properties, with the target values indicated in the axis labels.
	}
\end{minipage}

\newpage
\noindent
\refstepcounter{supfigure}
\begin{minipage}{1.0\textwidth}
	\tikzsetnextfilename{cond_gen_scaffold_prop_1}
	\begin{tikzpicture}
		\begin{groupplot}[group style={group size=2 by 2, 		vertical sep=40pt,
				horizontal sep=60pt,}, 	width=0.49\textwidth,
			height=0.49\textwidth,
			every axis/.append style={
				font=\small
			}]
			\input{figures/results_mol_gen_scaffold_prop_distribution_1d_1}
			\input{figures/results_mol_gen_scaffold_prop_distribution_1d_2}		
			\input{figures/results_mol_gen_scaffold_prop_distribution_1d_3}	
			\input{figures/results_mol_gen_scaffold_prop_distribution_1d_4}	
			
		\end{groupplot}
	\end{tikzpicture}
\\
\begin{adjustbox}{minipage=3.5cm, right}
	continued on next page
\end{adjustbox}
\\
	\captionof{figure}{\textbf{Distribution of properties for generated molecules on the MOSES dataset\protect\citemethods{moses_methods}.} 
Distribution of molecular properties for generated molecules conditioned on five different scaffold and property combinations. Each subplot corresponds to a specific combination, with the target values indicated in the axis labels.
	}
	\label{fig:rst_mol_gen_scaffold_prop_dist_1d}
\end{minipage}

\newpage
\noindent
\refstepcounter{supfigure}
\begin{minipage}{1.0\textwidth}
	\tikzsetnextfilename{cond_gen_scaffold_prop_2}

	\captionof*{figure}{\textbf{\figurename\ \thefigure{} Continued:} \textbf{Distribution of properties for generated molecules on the MOSES dataset\protect\citemethods{moses_methods}.} 
		Distribution of molecular properties for generated molecules conditioned on five different scaffold and property combinations. Each subplot corresponds to a specific combination, with the target values indicated in the axis labels.
	}
\end{minipage}

\newpage
\noindent
\refstepcounter{suptable}
\begin{minipage}{0.945\textwidth}
{
	\captionof{table}{	{{\textbf{Performance comparison for conditional molecule generation on the GuacaMol\protect\citemethods{guacamol_methods} dataset.}}}} 	 \label{table:result_conditional_generation_full}
	\tablebodyfont
\begin{tabular}{cccccc}
	\toprule
	Property & Model & Validity $\uparrow$ & Uniqueness $\uparrow$ & Novelty $\uparrow$ & \makecell[c]{Mean average  deviation \\ (MAD) $\downarrow$} \\   
	
	\midrule
	\multirow{5}{*}{logP}  
	&  cRNN  & $0.931$ & $0.930$ & $0.926$ &$0.202$ \\ 
	\cmidrule{2-6}
	&  MolGPT     & $0.971$            &  $0.969$         &  $0.947$             &   $0.230$ \\
	\cmidrule{2-6}
	&  \makecell{ChemFM-3B \\ (w/o pretraining)}  & $0.952$ & $0.952$ & $0.944$ &$0.257$ \\
	\cmidrule{2-6}
	&  ChemFM-3B    & $\mathbf{0.981}$            &  $\mathbf{0.981}$         &  $\mathbf{0.966}$             &   $\mathbf{0.182}$  \\
	\midrule
	\multirow{5}{*}{TPSA}      
	&  cRNN  & $0.923$ & $0.923$ & $0.917$ &$2.629$ \\
	\cmidrule{2-6}
	&  MolGPT     & $0.971$            &  $0.969$         &  $0.945$             &   $3.562$ \\
	\cmidrule{2-6}
	&  \makecell{ChemFM-3B \\ (w/o pretraining)}  & $0.953$ & $0.953$ & $0.942$ &$4.055$ \\
	\cmidrule{2-6}
	&  ChemFM-3B    & $\mathbf{0.979}$            &  $\mathbf{0.979}$         &  $\mathbf{0.963}$             &   $\mathbf{2.466}$  \\
	\midrule
	\multirow{5}{*}{SAS}       
	&  cRNN  & $0.935$ & $0.934$ & $0.922$ &$0.185$ \\
	\cmidrule{2-6}
	&  MolGPT     & $0.978$            &  $0.974$         &  $0.941$             &   $0.133$ \\
	\cmidrule{2-6}
	&  \makecell{ChemFM-3B \\ (w/o pretraining)}  & $0.957$ & $0.956$ & $0.939$ &$0.167$ \\
	\cmidrule{2-6}
	&  ChemFM-3B    & $\mathbf{0.986}$            &  $\mathbf{0.985}$         &  $\mathbf{0.957}$             &   $\mathbf{0.126}$  \\
	\midrule
	\multirow{5}{*}{QED}      
	&  cRNN  & $0.933$ & $0.933$ & $0.925$ &$0.058$ \\
	\cmidrule{2-6}
	&  MolGPT     & $0.974$            &  $0.971$         &  $0.940$             &   $0.056$ \\
	\cmidrule{2-6}
	&  MolGPT-3B  & $0.954$ & $0.954$ & $0.943$ &$0.059$ \\
	\cmidrule{2-6}
	&  ChemFM-3B    & $\mathbf{0.982}$            &  $\mathbf{0.982}$         &  $\mathbf{0.963}$             &   $\mathbf{0.045}$  \\
	\midrule
	\multirow{5}{*}{SAS + logP}      
	&  cRNN  & $0.932$ & $0.931$ & $0.926$ &$0.209/0.201$ \\
	\cmidrule{2-6}
	&  MolGPT     & $0.972$            &  $0.963$         &  $0.947$             &   $0.147 / 0.253$ \\
	\cmidrule{2-6}
	&  \makecell{ChemFM-3B \\ (w/o pretraining)}  & $0.955$ & $0.953$ & $0.943$ &$0.163/0.271$ \\
	\cmidrule{2-6}
	&  ChemFM-3B    & $\mathbf{0.980}$            &  $\mathbf{0.975}$         &  $\mathbf{0.960}$             &   $\mathbf{0.137 / 0.195}$  \\
	\midrule
	\multirow{5}{*}{SAS + TPSA}      
	&  cRNN  & $0.923$ & $0.921$ & $0.917$ &$0.218/3.032$ \\
	\cmidrule{2-6}
	&  MolGPT     & $0.971$            &  $0.960$         &  $0.944$             &   $0.155 / 3.785$ \\
	\cmidrule{2-6}
	&  \makecell{ChemFM-3B \\ (w/o pretraining)}  & $0.959$ & $0.954$ & $0.944$ &$0.172/4.148$ \\
	\cmidrule{2-6}
	&  ChemFM-3B    & $\mathbf{0.980}$            &  $\mathbf{0.971}$         &  $\mathbf{0.956}$             &   $\mathbf{0.138} / \mathbf{2.659}$ \\
	\botrule
	\multirow{5}{*}{TPSA + logP}      
	&  cRNN  & $0.903$ & $0.901$ & $0.898$ &$2.845/0.212$ \\
	\cmidrule{2-6}
	&  MolGPT     & $0.964$            &  $0.958$         &  $0.947$             &   $3.715 / 0.243$ \\
	\cmidrule{2-6}
	&  \makecell{ChemFM-3B \\ (w/o pretraining)}  & $0.951$ & $0.949$ & $0.943$ &$4.036/0.255$ \\
	\cmidrule{2-6}
	&  ChemFM-3B    & $\mathbf{0.973}$            &  $\mathbf{0.970}$         &  $\mathbf{0.962}$             &   $\mathbf{2.415 / 0.184}$  \\
	\botrule
	\multirow{5}{*}{TPSA + logP + SAS}      
	&  cRNN  & $0.937$ & $0.928$ & $0.922$ &$3.232/0.226/0.261$ \\
	\cmidrule{2-6}
	&  MolGPT     & $0.972$            &  $0.942$         &  $0.931$             &   $3.797 / 0.268 / 0.180$  \\
	\cmidrule{2-6}
	&  \makecell{ChemFM-3B \\ (w/o pretraining)}  & $0.965$ & $0.941$ & $0.932$ &$3.626/0.256/0.186$ \\
	\cmidrule{2-6}
	&  ChemFM-3B    & $\mathbf{0.975}$            &  $\mathbf{0.946}$         &  $\mathbf{0.936}$             &   $\mathbf{2.289}/ \mathbf{0.191} / \mathbf{0.166}$  \\					
	\bottomrule
\end{tabular}
}
		\captionof*{tabledescription}{{
			Molecules were generated according to desired property values, with performance compared across ChemFM-3B, ChemFM-3B without pre-training (both using a single model), and cRNN~\citemethods{crnn_methods} and MolGPT~\citemethods{molgpt_methods} (each requiring $8$ separate models).
			 Metrics include validity, uniqueness, novelty, and mean absolute deviation (MAD) between the conditioned and actual properties of the generated molecules. 
			\textbf{Bold} values indicate the best performance for each metric. 
			It should be noted that validity, uniqueness, and novelty are computed against the total number of generated molecules, rather than only the valid ones, to more accurately reflect model performance (\hyperref[sec:conditonal_generation_evaluation]{Methods}).}}
\end{minipage}

\newpage
\noindent
\refstepcounter{suptable}
{
\footnotesize
\begin{longtable}[l]{ccccccccc}
	\caption{\textbf{Performance comparison on standard benchmarks for conditional molecule generation on the MOSES dataset\protect\citemethods{moses_methods}.}}
	\label{table:rst_conditional_molecular_generation}\\\toprule
	Property & Scaffold  & Model &\makecell[c]{Generation \\ count} & \makecell[c]{Valid \\ molecules $\uparrow$} & \makecell[c]{Unique \\ molecules $\uparrow$} & \makecell[c]{Novel \\ molecules $\uparrow$} & \makecell[c]{Same scaffold \\ molecules $\uparrow$} & \makecell[c]{MAD $\downarrow$}  \\
	\midrule 	
	\endfirsthead
	\caption*{\textbf{\textbf{\tablename\ \thetable{} Continued:} Performance comparison on standard benchmarks for conditional molecular generation on the MOSES dataset.}}\\\toprule
	Property & Scaffold  & Model &\makecell[c]{Generation \\ count} & \makecell[c]{Valid \\ molecules $\uparrow$} & \makecell[c]{Unique \\ molecules $\uparrow$} & \makecell[c]{Novel \\ molecules $\uparrow$} & \makecell[c]{Same scaffold \\ molecules $\uparrow$} & \makecell[c]{MAD $\downarrow$}  \\
	\midrule 
	\endhead	
	 \multicolumn{9}{r}{{Continued on next page}} \\ 
	\endfoot
	\bottomrule
	\endlastfoot	
	\multirow{13}{*}{logP} 
& \multirow{2}{*}{(a)} & MolGPT & $\num{30000}$ & $\num{29679}$ & $\num{7132}$ & $\num{7132}$ & $\num{7132}$ & $\num{0.123}$\\
&                      & ChemFM & $\num{30000}$ & $\num{29538}$ & $\num{10117}$ & $\num{10117}$ & $\num{10117}$ & $\num{0.087}$\\
\cmidrule{2-9}
& \multirow{2}{*}{(b)} & MolGPT & $\num{30000}$ & $\num{26880}$ & $\num{15669}$ & $\num{15669}$ & $\num{15669}$ & $\num{0.126}$\\
&                      & ChemFM & $\num{30000}$ & $\num{27606}$ & $\num{20011}$ & $\num{20011}$ & $\num{19519}$ & $\num{0.097}$\\
\cmidrule{2-9}
& \multirow{2}{*}{(c)} & MolGPT & $\num{30000}$ & $\num{28536}$ & $\num{10537}$ & $\num{10537}$ & $\num{10531}$ & $\num{0.127}$\\
&                      & ChemFM & $\num{30000}$ & $\num{29077}$ & $\num{11398}$ & $\num{11398}$ & $\num{11392}$ & $\num{0.077}$\\
\cmidrule{2-9}
& \multirow{2}{*}{(d)} & MolGPT & $\num{30000}$ & $\num{29643}$ & $\num{8886}$ & $\num{8886}$ & $\num{8885}$ & $\num{0.136}$\\
&                      & ChemFM & $\num{30000}$ & $\num{29832}$ & $\num{11559}$ & $\num{11559}$ & $\num{11558}$ & $\num{0.093}$\\
\cmidrule{2-9}
& \multirow{2}{*}{(e)} & MolGPT & $\num{30000}$ & $\num{29666}$ & $\num{2334}$ & $\num{2334}$ & $\num{2328}$ & $\num{0.111}$\\
&                      & ChemFM & $\num{30000}$ & $\num{29629}$ & $\num{3521}$ & $\num{3521}$ & $\num{3521}$ & $\num{0.084}$\\
\cmidrule{2-9}
& \multirow{2}{*}{Total} & MolGPT & $\num{150000}$ & $\num{144404}$ & $\num{44558}$ & $\num{44558}$ & $\num{44545}$ & $\num{0.125}$\\
&                      & ChemFM & $\num{150000}$ & \boldmath$\num{145682}$ & \boldmath$\num{56606}$ & \boldmath$\num{56606}$ & \boldmath$\num{56107}$ & \boldmath$\num{0.087}$\\
\midrule
	\multirow{13}{*}{SAS} 
& \multirow{2}{*}{(a)} & MolGPT & $\num{30000}$ & $\num{29253}$ & $\num{13663}$ & $\num{13663}$ & $\num{13663}$ & $\num{0.122}$\\
&                      & ChemFM & $\num{30000}$ & $\num{26899}$ & $\num{16922}$ & $\num{16922}$ & $\num{16921}$ & $\num{0.106}$\\
\cmidrule{2-9}
& \multirow{2}{*}{(b)} & MolGPT & $\num{30000}$ & $\num{26719}$ & $\num{11846}$ & $\num{11846}$ & $\num{11843}$ & $\num{0.140}$\\
&                      & ChemFM & $\num{30000}$ & $\num{27088}$ & $\num{15801}$ & $\num{15801}$ & $\num{15337}$ & $\num{0.129}$\\
\cmidrule{2-9}
& \multirow{2}{*}{(c)} & MolGPT & $\num{30000}$ & $\num{25269}$ & $\num{10803}$ & $\num{10803}$ & $\num{10791}$ & $\num{0.143}$\\
&                      & ChemFM & $\num{30000}$ & $\num{27685}$ & $\num{14535}$ & $\num{14534}$ & $\num{14513}$ & $\num{0.140}$\\
\cmidrule{2-9}
& \multirow{2}{*}{(d)} & MolGPT & $\num{30000}$ & $\num{29011}$ & $\num{12253}$ & $\num{12253}$ & $\num{12253}$ & $\num{0.138}$\\
&                      & ChemFM & $\num{30000}$ & $\num{29264}$ & $\num{15710}$ & $\num{15710}$ & $\num{15691}$ & $\num{0.113}$\\
\cmidrule{2-9}
& \multirow{2}{*}{(e)} & MolGPT & $\num{30000}$ & $\num{28540}$ & $\num{4087}$ & $\num{4087}$ & $\num{4065}$ & $\num{0.102}$\\
&                      & ChemFM & $\num{30000}$ & $\num{29644}$ & $\num{5195}$ & $\num{5195}$ & $\num{5192}$ & $\num{0.093}$\\
\cmidrule{2-9}
& \multirow{2}{*}{Total} & MolGPT & $\num{150000}$ & $\num{138792}$ & $\num{52652}$ & $\num{52652}$ & $\num{52615}$ & $\num{0.129}$\\
&                      & ChemFM & $\num{150000}$ & \boldmath$\num{140580}$ & \boldmath$\num{68163}$ & \boldmath$\num{68162}$ & \boldmath$\num{67654}$ & \boldmath$\num{0.123}$\\
	\midrule
	\multirow{13}{*}{TPSA} 
& \multirow{2}{*}{(a)} & MolGPT & $\num{30000}$ & $\num{29809}$ & $\num{8134}$ & $\num{8134}$ & $\num{8132}$ & $\num{1.887}$\\
&                      & ChemFM & $\num{30000}$ & $\num{29726}$ & $\num{10777}$ & $\num{10777}$ & $\num{10777}$ & $\num{1.581}$\\
\cmidrule{2-9}
& \multirow{2}{*}{(b)} & MolGPT & $\num{30000}$ & $\num{26509}$ & $\num{14519}$ & $\num{14519}$ & $\num{14518}$ & $\num{2.948}$\\
&                      & ChemFM & $\num{30000}$ & $\num{27666}$ & $\num{17262}$ & $\num{17259}$ & $\num{16698}$ & $\num{2.291}$\\
\cmidrule{2-9}
& \multirow{2}{*}{(c)} & MolGPT & $\num{30000}$ & $\num{29522}$ & $\num{10240}$ & $\num{10240}$ & $\num{10240}$ & $\num{2.002}$\\
&                      & ChemFM & $\num{30000}$ & $\num{28617}$ & $\num{10742}$ & $\num{10742}$ & $\num{10734}$ & $\num{1.392}$\\
\cmidrule{2-9}
& \multirow{2}{*}{(d)} & MolGPT & $\num{30000}$ & $\num{29655}$ & $\num{10264}$ & $\num{10264}$ & $\num{10263}$ & $\num{2.702}$\\
&                      & ChemFM & $\num{30000}$ & $\num{29783}$ & $\num{11546}$ & $\num{11546}$ & $\num{11542}$ & $\num{2.056}$\\
\cmidrule{2-9}
& \multirow{2}{*}{(e)} & MolGPT & $\num{30000}$ & $\num{28716}$ & $\num{1873}$ & $\num{1873}$ & $\num{1867}$ & $\num{3.785}$\\
&                      & ChemFM & $\num{30000}$ & $\num{29575}$ & $\num{3838}$ & $\num{3838}$ & $\num{3835}$ & $\num{3.240}$\\
\cmidrule{2-9}
& \multirow{2}{*}{Total} & MolGPT & $\num{150000}$ & $\num{144211}$ & $\num{45030}$ & $\num{45030}$ & $\num{45020}$ & $\num{2.651}$\\
&                      & ChemFM & $\num{150000}$ & \boldmath$\num{145367}$ & \boldmath$\num{54165}$ & \boldmath$\num{54162}$ & \boldmath$\num{53586}$ & \boldmath$\num{2.114}$\\
	\midrule
	\multirow{13}{*}{QED} 
& \multirow{2}{*}{(a)} & MolGPT & $\num{30000}$ & $\num{29618}$ & $\num{12931}$ & $\num{12931}$ & $\num{12931}$ & $\num{0.043}$\\
&                      & ChemFM & $\num{30000}$ & $\num{29332}$ & $\num{16723}$ & $\num{16723}$ & $\num{16723}$ & $\num{0.044}$\\
\cmidrule{2-9}
& \multirow{2}{*}{(b)} & MolGPT & $\num{30000}$ & $\num{25516}$ & $\num{15659}$ & $\num{15659}$ & $\num{15646}$ & $\num{0.040}$\\
&                      & ChemFM & $\num{30000}$ & $\num{27421}$ & $\num{21597}$ & $\num{21597}$ & $\num{21002}$ & $\num{0.036}$\\
\cmidrule{2-9}
& \multirow{2}{*}{(c)} & MolGPT & $\num{30000}$ & $\num{29055}$ & $\num{13108}$ & $\num{13108}$ & $\num{13105}$ & $\num{0.034}$\\
&                      & ChemFM & $\num{30000}$ & $\num{28914}$ & $\num{14643}$ & $\num{14643}$ & $\num{14633}$ & $\num{0.033}$\\
\cmidrule{2-9}
& \multirow{2}{*}{(d)} & MolGPT & $\num{30000}$ & $\num{29812}$ & $\num{12978}$ & $\num{12978}$ & $\num{12978}$ & $\num{0.064}$\\
&                      & ChemFM & $\num{30000}$ & $\num{29797}$ & $\num{14368}$ & $\num{14368}$ & $\num{14362}$ & $\num{0.043}$\\
\cmidrule{2-9}
& \multirow{2}{*}{(e)} & MolGPT & $\num{30000}$ & $\num{27457}$ & $\num{2918}$ & $\num{2918}$ & $\num{2909}$ & $\num{0.072}$\\
&                      & ChemFM & $\num{30000}$ & $\num{29330}$ & $\num{5127}$ & $\num{5127}$ & $\num{5116}$ & $\num{0.095}$\\
\cmidrule{2-9}
& \multirow{2}{*}{Total} & MolGPT & $\num{150000}$ & $\num{141458}$ & $\num{57594}$ & $\num{57594}$ & $\num{57569}$ & $\num{0.051}$\\
&                      & ChemFM & $\num{150000}$ & \boldmath$\num{144794}$ & \boldmath$\num{72458}$ & \boldmath$\num{72458}$ & \boldmath$\num{71836}$ & \boldmath$\num{0.050}$\\
	\midrule  
\pagebreak
\multirow{13}{*}{\shortstack{TPSA + \\ $\log$P}} 
& \multirow{2}{*}{(a)} & MolGPT & $\num{40000}$ & $\num{38440}$ & $\num{8919}$ & $\num{8919}$ & $\num{8919}$ & $\num{3.125}/\num{0.238}$\\
&                      & ChemFM & $\num{40000}$ & $\num{38845}$ & $\num{13536}$ & $\num{13536}$ & $\num{13536}$ & $\num{3.268}/\num{0.206}$\\
\cmidrule{2-9}
& \multirow{2}{*}{(b)} & MolGPT & $\num{40000}$ & $\num{33172}$ & $\num{16609}$ & $\num{16609}$ & $\num{16578}$ & $\num{3.912}/\num{0.173}$\\
&                      & ChemFM & $\num{40000}$ & $\num{36388}$ & $\num{20940}$ & $\num{20940}$ & $\num{20251}$ & $\num{2.767}/\num{0.124}$\\
\cmidrule{2-9}
& \multirow{2}{*}{(c)} & MolGPT & $\num{40000}$ & $\num{38190}$ & $\num{13901}$ & $\num{13901}$ & $\num{13881}$ & $\num{3.492}/\num{0.125}$\\
&                      & ChemFM & $\num{40000}$ & $\num{37873}$ & $\num{13828}$ & $\num{13828}$ & $\num{13816}$ & $\num{2.779}/\num{0.110}$\\
\cmidrule{2-9}
& \multirow{2}{*}{(d)} & MolGPT & $\num{40000}$ & $\num{36010}$ & $\num{11104}$ & $\num{11104}$ & $\num{11102}$ & $\num{4.749}/\num{0.165}$\\
&                      & ChemFM & $\num{40000}$ & $\num{39085}$ & $\num{11394}$ & $\num{11394}$ & $\num{11391}$ & $\num{2.520}/\num{0.179}$\\
\cmidrule{2-9}
& \multirow{2}{*}{(e)} & MolGPT & $\num{40000}$ & $\num{36122}$ & $\num{3966}$ & $\num{3965}$ & $\num{3942}$ & $\num{3.651}/\num{0.230}$\\
&                      & ChemFM & $\num{40000}$ & $\num{35762}$ & $\num{6950}$ & $\num{6950}$ & $\num{6904}$ & $\num{5.104}/\num{0.172}$\\
\cmidrule{2-9}
& \multirow{2}{*}{Total} & MolGPT & $\num{200000}$ & $\num{181934}$ & $\num{54499}$ & $\num{54498}$ & $\num{54422}$ & $\num{3.771}/\num{0.186}$\\
&                      & ChemFM & $\num{200000}$ & \boldmath$\num{187953}$ & \boldmath$\num{66648}$ & \boldmath$\num{66648}$ & \boldmath$\num{65898}$ & \boldmath$\num{3.266}$/\boldmath$\num{0.159}$\\
\midrule
\multirow{13}{*}{\shortstack{SAS + \\ $\log$P}} 
& \multirow{2}{*}{(a)} & MolGPT & $\num{40000}$ & $\num{36497}$ & $\num{11866}$ & $\num{11866}$ & $\num{11859}$ & $\num{0.154}/\num{0.238}$\\
&                      & ChemFM & $\num{40000}$ & $\num{32753}$ & $\num{15401}$ & $\num{15401}$ & $\num{15401}$ & $\num{0.124}/\num{0.159}$\\
\cmidrule{2-9}
& \multirow{2}{*}{(b)} & MolGPT & $\num{40000}$ & $\num{30249}$ & $\num{10962}$ & $\num{10962}$ & $\num{10958}$ & $\num{0.156}/\num{0.164}$\\
&                      & ChemFM & $\num{40000}$ & $\num{35512}$ & $\num{15937}$ & $\num{15937}$ & $\num{15565}$ & $\num{0.166}/\num{0.133}$\\
\cmidrule{2-9}
& \multirow{2}{*}{(c)} & MolGPT & $\num{40000}$ & $\num{36784}$ & $\num{12745}$ & $\num{12745}$ & $\num{12734}$ & $\num{0.157}/\num{0.157}$\\
&                      & ChemFM & $\num{40000}$ & $\num{35456}$ & $\num{14073}$ & $\num{14073}$ & $\num{14001}$ & $\num{0.164}/\num{0.141}$\\
\cmidrule{2-9}
& \multirow{2}{*}{(d)} & MolGPT & $\num{40000}$ & $\num{37290}$ & $\num{11616}$ & $\num{11616}$ & $\num{11613}$ & $\num{0.136}/\num{0.192}$\\
&                      & ChemFM & $\num{40000}$ & $\num{38417}$ & $\num{13995}$ & $\num{13995}$ & $\num{13966}$ & $\num{0.113}/\num{0.204}$\\
\cmidrule{2-9}
& \multirow{2}{*}{(e)} & MolGPT & $\num{40000}$ & $\num{39243}$ & $\num{4361}$ & $\num{4361}$ & $\num{4262}$ & $\num{0.125}/\num{0.167}$\\
&                      & ChemFM & $\num{40000}$ & $\num{38666}$ & $\num{7059}$ & $\num{7059}$ & $\num{7050}$ & $\num{0.120}/\num{0.190}$\\
\cmidrule{2-9}
& \multirow{2}{*}{Total} & MolGPT & $\num{200000}$ & $\num{180063}$ & $\num{51550}$ & $\num{51550}$ & $\num{51426}$ & $\num{0.145}$/$\num{0.184}$\\
&                      & ChemFM & $\num{200000}$ & \boldmath$\num{180804}$ & \boldmath$\num{66465}$ & \boldmath$\num{66465}$ & \boldmath$\num{65983}$ & \boldmath$\num{0.137}$/\boldmath$\num{0.166}$\\
\midrule
\multirow{13}{*}{\shortstack{TPSA + \\ SAS}} 
& \multirow{2}{*}{(a)} & MolGPT & $\num{40000}$ & $\num{37901}$ & $\num{16217}$ & $\num{16217}$ & $\num{16214}$ & $\num{3.417}/\num{0.175}$\\
&                      & ChemFM & $\num{40000}$ & $\num{34935}$ & $\num{16994}$ & $\num{16994}$ & $\num{16994}$ & $\num{2.990}/\num{0.131}$\\
\cmidrule{2-9}
& \multirow{2}{*}{(b)} & MolGPT & $\num{40000}$ & $\num{32343}$ & $\num{12808}$ & $\num{12808}$ & $\num{12744}$ & $\num{4.633}/\num{0.189}$\\
&                      & ChemFM & $\num{40000}$ & $\num{34274}$ & $\num{15585}$ & $\num{15585}$ & $\num{15117}$ & $\num{4.136}/\num{0.181}$\\
\cmidrule{2-9}
& \multirow{2}{*}{(c)} & MolGPT & $\num{40000}$ & $\num{31731}$ & $\num{13522}$ & $\num{13522}$ & $\num{13512}$ & $\num{3.960}/\num{0.181}$\\
&                      & ChemFM & $\num{40000}$ & $\num{36445}$ & $\num{15526}$ & $\num{15525}$ & $\num{15482}$ & $\num{3.640}/\num{0.181}$\\
\cmidrule{2-9}
& \multirow{2}{*}{(d)} & MolGPT & $\num{40000}$ & $\num{35947}$ & $\num{13793}$ & $\num{13793}$ & $\num{13793}$ & $\num{4.066}/\num{0.165}$\\
&                      & ChemFM & $\num{40000}$ & $\num{38614}$ & $\num{16790}$ & $\num{16790}$ & $\num{16740}$ & $\num{3.343}/\num{0.133}$\\
\cmidrule{2-9}
& \multirow{2}{*}{(e)} & MolGPT & $\num{40000}$ & $\num{39196}$ & $\num{5170}$ & $\num{5170}$ & $\num{5120}$ & $\num{3.292}/\num{0.149}$\\
&                      & ChemFM & $\num{40000}$ & $\num{38941}$ & $\num{6010}$ & $\num{6010}$ & $\num{6005}$ & $\num{3.443}/\num{0.119}$\\
\cmidrule{2-9}
& \multirow{2}{*}{Total} & MolGPT & $\num{200000}$ & $\num{177118}$ & $\num{61510}$ & $\num{61510}$ & $\num{61383}$ & $\num{3.840}/\num{0.171}$\\
&                      & ChemFM & $\num{200000}$ & \boldmath$\num{183209}$ & \boldmath$\num{70905}$ & \boldmath$\num{70904}$ & \boldmath$\num{70338}$ & \boldmath$\num{3.504}$/\boldmath$\num{0.148}$\\
\midrule
\multirow{20}{*}{\shortstack{TPSA + \\ $\log$P + \\ SAS}} 
& \multirow{2}{*}{(a)} & MolGPT & $\num{80000}$ & $\num{66217}$ & $\num{14699}$ & $\num{14699}$ & $\num{14699}$ & $\num{5.125}/\num{0.539}/\num{0.348}$\\
&                      & ChemFM & $\num{80000}$ & $\num{56565}$ & $\num{19454}$ & $\num{19454}$ & $\num{19454}$ & $\num{4.334}/\num{0.378}/\num{0.220}$\\
\cmidrule{2-9}
& \multirow{2}{*}{(b)} & MolGPT & $\num{80000}$ & $\num{63655}$ & $\num{17279}$ & $\num{17279}$ & $\num{17257}$ & $\num{5.096}/\num{0.225}/\num{0.227}$\\
&                      & ChemFM & $\num{80000}$ & $\num{60128}$ & $\num{27610}$ & $\num{27610}$ & $\num{26850}$ & $\num{4.207}/\num{0.187}/\num{0.258}$\\
\cmidrule{2-9}
& \multirow{2}{*}{(c)} & MolGPT & $\num{80000}$ & $\num{59363}$ & $\num{16086}$ & $\num{16086}$ & $\num{16068}$ & $\num{5.650}/\num{0.256}/\num{0.220}$\\
&                      & ChemFM & $\num{80000}$ & $\num{64752}$ & $\num{22422}$ & $\num{22422}$ & $\num{22315}$ & $\num{4.681}/\num{0.285}/\num{0.220}$\\
\cmidrule{2-9}
& \multirow{2}{*}{(d)} & MolGPT & $\num{80000}$ & $\num{50119}$ & $\num{12081}$ & $\num{12081}$ & $\num{12080}$ & $\num{5.145}/\num{0.332}/\num{0.243}$\\
&                      & ChemFM & $\num{80000}$ & $\num{67716}$ & $\num{18970}$ & $\num{18970}$ & $\num{18895}$ & $\num{4.337}/\num{0.483}/\num{0.197}$\\
\cmidrule{2-9}
& \multirow{2}{*}{(e)} & MolGPT & $\num{80000}$ & $\num{74433}$ & $\num{7228}$ & $\num{7228}$ & $\num{7111}$ & $\num{5.749}/\num{0.384}/\num{0.233}$\\
&                      & ChemFM & $\num{80000}$ & $\num{73882}$ & $\num{8858}$ & $\num{8858}$ & $\num{8787}$ & $\num{6.079}/\num{0.304}/\num{0.196}$\\
\cmidrule{2-9}
& \multirow{2}{*}{Total} & MolGPT & $\num{400000}$ & $\num{313787}$ & $\num{67373}$ & $\num{67373}$ & $\num{67215}$ & $\num{5.370}$/{$\num{0.352}$}/$\num{0.255}$\\
&                      & ChemFM & $\num{400000}$ & \boldmath$\num{323043}$ & \boldmath$\num{97314}$ & \boldmath$\num{97314}$ & \boldmath$\num{96301}$ & {\boldmath$\num{4.780}$}/\boldmath$\num{0.329}$/\boldmath$\num{0.217}$\\
\end{longtable}
\captionof*{tabledescription}{
Generations were conditioned on five test scaffolds and various property combinations, comparing the performance of ChemFM-3B, which uses a single model, to MolGPT\protect\citemethods{molgpt_methods}, which uses $8$ separate models. 
Generation performance is reported as the number of valid, unique, and novel molecules. A valid molecule is defined as 1) syntactically correct and 2) having a Tanimoto similarity greater than $0.8$ to the conditioned scaffold.
For scaffold matching, we report the number of molecules that match the conditioned scaffold and the mean absolute deviation (MAD) between the conditioned and actual property values. The generations are based on five test scaffolds: (a) \texttt{O=C(Cc1ccccc1)NCc1ccccc1}, (b) \texttt{c1cnc2[nH]ccc2c1}, (c) \texttt{c1ccc(-c2ccnnc2)cc1}, (d) \texttt{c1ccc(-n2cnc3ccccc32)cc1}, and (e) \texttt{O=C(c1cc[nH]c1)N1CCN(c2ccccc2)CC1}.}
}

\newpage
\setcounter{suptable}{0}
\setcounter{supfigure}{0}
\subsection{Supplementary information for reaction prediction tasks}
\noindent
\refstepcounter{suptable}
\begin{minipage}{.66\textwidth}
	\captionof{table}{\textbf{Comprehensive comparison of synthesis and retro-synthesis reaction prediction performance across standard USPTO benchmarks.}} \label{table:rst_rct_prediction_detail}
	\tablebodyfont
		\begin{tabular}{lllccc}
			\toprule
			Task category & Dataset & Model & Top-$1$ & Top-$3$ & Top-$5$ \\ 
			\midrule
			Synthesis & USPTO-MIT
			                             & WLDN\citemethods{usptomit_methods} & $74.0$ & - & -  \\
			&							 & MT\citemethods{mt_methods} & $88.6$ & $93.5$ & $94.2$ \\
			&                            & AT\citemethods{at_methods} & $\underline{90.4}$ & - & $\underline{96.5}$  
			\\
			&							 & MEGAN\citemethods{megan_methods} & $86.3$ & $92.4$ & $94.0$ \\
			&                            & R-SMILES\citemethods{rsmiles_methods} & $90.0$ & $\underline{95.6}$ & 96.4 
			\\
			&                            & ChemFM & $\mathbf{90.5}$ & $\mathbf{95.7}$ & $\mathbf{96.6}$
			
			\\                                                
			\midrule
			Retro-synthesis & USPTO-50K & 
										  AT\citemethods{at_methods}       & $53.5$ &  -     &  $81.0$  \\
			&                            &MEGAN\citemethods{megan_methods} & $48.1$ & $70.7$ & $78.4$ \\
			&							 &GraphRetro\citemethods{graphretro_methods} & $53.7$ & $68.3$ & $72.2$ \\
			&                            &GLN\citemethods{gln_methods} & $52.5$ & $69.0$ & $75.6$ \\
			&                            &RetroXpert\citemethods{retroxpert_methods} & $50.4$ & $61.1$ & $62.3$ \\
			&                            & GTA\citemethods{gta_methods} & $51.1$ & - & - \\
			&                            &RetroPrime\citemethods{retroprime_methods} & $51.4$ & $70.8$ & $74.0$ \\
			&                            &LocalRetro\citemethods{localretro_methods}    & $53.4$ & $77.5$ & ${85.9}$ \\
			&                            &Chemformer\citemethods{chemformer_methods} & 54.3 & 62.3 & 63.0
			\\
			&                            &Retroformer\citemethods{retroformer_methods} & 53.2 & 71.1 & 76.6\\
			&                            &Graph2Edits\citemethods{graph2edits_methods} & 55.1 & 77.3 & 83.4\\
			&                            &G$^2$Retro\citemethods{g2retro_methods} & 54.1 & 74.1 & 81.2\\
			&                            &R-SMILES\citemethods{rsmiles_methods} & $\underline{56.0}$ & $\underline{79.0}$ & $\underline{86.1}$
			\\
			&                            & ChemFM & $58.0$ & $\mathbf{80.0}$ &
			 $\mathbf{86.3}$ \\
			 &                            & ChemFM$^\star$ & $\mathbf{59.7}$ & $79.2$ & $84.2$ 				
			\\
			\cmidrule(r){2-6}
			& USPTO-MIT & 
			                               LocalRetro\citemethods{localretro_methods} & 54.1 & 73.7 & 79.4 \\
			&                            & RetroTRAE\citemethods{retroTRAE_methods} & 58.3 & - & - \\
			&                            & R-SMILES\citemethods{rsmiles_methods} & $\underline{60.3}$ & $\underline{77.9}$ & $\underline{82.8}$ 
			\\
			&                            & ChemFM & $61.6$ & $\mathbf{78.7}$ & $\mathbf{83.0}$ \\
			&                      	& ChemFM$^\star$ & $\mathbf{62.4}$ & $78.5$ & $82.5$ 						
			\\
			\cmidrule(r){2-6}
			& USPTO-Full & 
			
			                               AT\citemethods{at_methods}       & $46.2$ &  -     &  -  \\
			&                            & MEGAN\citemethods{megan_methods} & 33.6 & - & - \\
			&                            & GLN\citemethods{gln_methods} & 39.3 & - & - \\
			&                            & RetroPrime\citemethods{retroprime_methods} & $44.1$ & $59.1$ & $62.8$ \\
			&                            & LocalRetro\citemethods{localretro_methods}    & $39.1$ & $53.3$ & $58.4$ \\	
			&                            & RetroXpert\citemethods{retroxpert_methods} & 49.4 & 63.6 & 67.6 \\
			&                            & GTA\citemethods{gta_methods} & 46.6 & - & - \\
			&                            & Substructure\citemethods{substructure_methods} & 48.2 & - & - \\
			&                            & R-SMILES\citemethods{rsmiles_methods} & $\underline{48.9}$ & $\underline{66.6}$ & $\underline{72.0}$ \\
			&                            & ChemFM & $\mathbf{51.7}$ & $\mathbf{68.0}$ & $\mathbf{72.5}$ 
			\\	
			\botrule
		\end{tabular}
\captionof*{tabledescription}{
This table compares the top-1, top-3, and top-5 accuracies (in percentages) of our ChemFM model against various models from the literature on the USPTO-MIT, USPTO-50K, and USPTO-Full datasets.
\textbf{Bold} values indicate the best performance, and \underline{underlined} values represent the best performance other than ChemFM.
A hyphen ``-'' signifies that the result was not reported in the corresponding paper.
All results from other methods are obtained directly from their respective publications, except for the R-SMILES results, which were replicated using publicly available models.
ChemFM$^\star$ denotes ChemFM with further pre-training, which achieve better top-$1$ results but show a decrease in top-$3$ and $5$ performance.
}
\end{minipage}
\newpage
\refstepcounter{suptable}
\begin{minipage}{1.0\textwidth}
	\footnotesize{
	\captionof{table}{\textbf{Summary of USPTO benchmark datasets evaluated in this work.}} 	\label{table:dataset_uspto}

		\begin{tabular}[t]{llllp{0.25\linewidth}}
			\toprule
			 Dataset & \makecell[lt]{Number of reactions \\ (training)} & \makecell[lt]{Number of reactions \\ (validation)} & \makecell[lt]{Number of reactions \\ (testing)} & Dataset description \\ 
			\midrule
			USPTO-Full\citemethods{uspto_full_methods} & \num{768630} & \num{96071} & \num{96023} & 
			Largest dataset, containing reactions extracted from patents (1976-2016), covering a wide range of chemical reaction types.\\
%
			\midrule
			USPTO-MIT\citemethods{usptomit_methods} & \num{411685} & \num{30182} & \num{40265} & 
			Refined subset of USPTO-Full, with duplicates and erroneous reactions removed. Commonly used for forward reaction prediction benchmarks.
			\\
			\midrule
			USPTO-50K\citemethods{uspto50k_methods} & \num{40003} & \num{5001} & \num{5007} & 
			Smaller curated dataset of 50K reactions across 10 types, commonly used for benchmarking retro-synthesis prediction tasks. \\
			\bottomrule
		\end{tabular}
	}
\end{minipage}


\newpage
\noindent
\refstepcounter{suptable}
\begin{minipage}{0.83\textwidth}
	\captionof{table}{\textbf{Hyperparameters for fine-tuning ChemFM-3B for reaction prediction tasks.}}
	\label{table:hyperparameter_reaction}
	\tablebodyfont
	\begin{tabular}{lcccc}
		\toprule
		Hyperparameters & \makecell[c]{USPTO-MIT \\ (Synthesis)} & \makecell[c]{USPTO-50K\\ (Retro-synthesis)} & \makecell[c]{USPTO-MIT\\ (Retro-synthesis)} & \makecell[c]{USPTO-Full\\ (Retro-synthesis)} \\
		\midrule
		Optimizer & \multicolumn{4}{c}{AdamW} \\
		\midrule
		LR scheduler & \multicolumn{4}{c}{Cosine} \\
		\midrule
		Warmup ratio & \multicolumn{4}{c}{$0.05$} \\
		\midrule
		Minimum LR & \multicolumn{4}{c}{$0.1\times\text{LR}$} \\
		\midrule
		Attention dropout & $0.1$ & $0.1$ & $0.1$ & $0.1$ \\
		\midrule
		Learning rate (LR) & $\num{1e-4}$ & $\num{1e-4}$ & $\num{1e-4}$ &  $\num{1e-4}$\\
		\midrule
		Epochs & $10$ & $10$ & $10$ & $10$ \\
		\midrule
		Batch size & $64$ & $64$  & $64$ & $64$\\
		\midrule
		Weight decay & $0.01$ & $0.01$  &  $0.01$  & $0.01$  \\
		\midrule
		LoRA $\alpha$ & - & $1.0$ & $1.0$ & $1.0$\\
		\midrule
		LoRA rank & - & $32$ & $64$ & $64$ \\
		\midrule
		LoRA dropout & - & $0.1$ & $0.1$ &  $0.1$ \\
		\midrule
		\makecell[lt]{Number of \\trainable parameters} & $3.0$B & $54$M & $106$M & $106$M \\
		\bottomrule
	\end{tabular}
	\captionof*{tabledescription}{
		For synthesis prediction using USPTO-MIT, we perform full-parameter fine-tuning, while all retro-synthesis experiments use LoRA. 
		The number of epochs refers to training on the augmented dataset.
		Due to computational resource constraints, we do not perform hyperparameter searching. Two or three sets of hyperparameters are chosen based on prior experience, and we report the best one. Results were found to be quite stable within a reasonable hyperparameter range. A hyphen ``-'' indicates that the hyperparameter is not applicable.
	}
\end{minipage}

\clearpage
\bibliographystylemethods{naturemag}
\bibliographymethods{sn-bibliography-method}

\end{appendices}

\end{document}